\documentclass[extra,mreferee]{gji}
\usepackage{timet}
\usepackage{amsmath}
\usepackage{bm}
\usepackage{amsfonts}
\usepackage{hyperref}
\usepackage{graphicx}
\usepackage{subcaption}
\hypersetup{
    colorlinks=true,
    citecolor=blue
}
\usepackage[dvipsnames]{xcolor}

\usepackage{xr}
\makeatletter
\newcommand*{\addFileDependency}[1]{
\typeout{(#1)}
\@addtofilelist{#1}
\IfFileExists{#1}{}{\typeout{No file #1.}}
}\makeatother
\newcommand*{\myexternaldocument}[1]{%
\externaldocument{#1}%
\addFileDependency{#1.tex}%
\addFileDependency{#1.aux}%
}
\myexternaldocument{supp}

\title[Model of Ionospheric and Magnetospheric Magnetic Fields]{A decade of the fast-varying ionospheric and magnetospheric magnetic fields from ground and multi-satellite observations}
\author[Min and Grayver]{Jingtao Min$^1$, Alexander Grayver$^2$ \\ 
    $^1$ Institute of Geophysics, ETH Zurich, 8092 Zurich, Switzerland\\ 
    $^2$ Institute of Geophysics and Meteorology, University of Cologne, 50923 Cologne, Germany}
\date{\today}

\begin{document}

\maketitle

\begin{summary}
The time-varying geomagnetic field is a superposition of contributions from multiple internal and external current systems. A major source of geomagnetic variations at periods less than a few years are current systems external to the solid Earth, namely the ionospheric and magnetospheric currents, as well as associated induced currents. The separation of these three sources is mathematically underdetermined using either ground or satellite measurements alone, but becomes tractable when the two datasets are combined.
Based on this concept, we developed a new geomagnetic field modelling approach that allows us to simultaneously characterise the mid-latitude ionospheric, magnetospheric and the internal induced magnetic fields using ground and satellite observations for all local times and magnetic conditions, and without prescribing any harmonic behaviour on these current systems in time, as is typical in other models.
By applying this new method to a 10-year dataset of ground observatory and multi-satellite measurements from 2014 to 2023, we obtained the time series of the spherical harmonic coefficients of the ionospheric, magnetospheric and induced fields. 
These new time series allow the study of complex non-periodic dynamics of the external magnetic fields during global geomagnetic storms, as well as periodicities in the magnetospheric coefficients linked to solar activities and periodic ionospheric magnetic fields linked to lunar daily variations, contributing to a more complete picture of the dynamics of the external currents and magnetosphere-ionosphere interactions, and facilitating more accurate space weather nowcast and forecast. 
Finally, the new approach allows for a better characterisation of internal induced field sources, leading to higher quality electromagnetic transfer functions. 
\end{summary}


\section{Introduction}

Time variations of the magnetic field observed on the ground or at a spacecraft represent a superposition of contributions from multiple time-varying current systems.
The electric current generated by the dynamo action in the fluid outer core is responsible for sustaining the main field over geological time scales, and is also the major contributor to magnetic field variations at long periods, known as secular variations \citep{backus1996foundations}.
Meanwhile, the extraneous current systems in the ionosphere and the magnetosphere generate time-varying magnetic fields, which then induce currents in the electrically conductive solid Earth \citep{schuster_diurnal_1889, Olsen1999Review}.
Characterising the ionospheric and magnetospheric contributions to the observed magnetic field is of interest for several reasons.
First, magnetic fields of ionospheric and magnetospheric origins are used to quantitatively study respective electric current systems, and serve as a proxy for understanding the underlying dynamics in these regions \citep{yamazaki_sq_2017, Tsyganenko2019, laundal_electrojet_2021}.
On the other hand, the relation between these magnetic fields of external origin and their induced counterparts in the Earth interior can be used to probe the electrical conductivity in the Earth’s subsurface \citep{Olsen1999a, Kelbert2009, kuvshinov2012deep, grayver_unravelling_2024}.
More accurate space physics and subsurface conductivity models are then essential for a practically useful space weather hazard evaluation, as the latter is driven by the dynamics in the extraneous current systems \citep{toth_space_2005, pulkkinen_community-wide_2013} but sensitive to the subsurface electrical conductivity as well \citep{Juusola2020, kelbert2020modified}. Last but not least, better characterisation of the ionospheric and magnetospheric systems also improves the core field modelling \citep{finlay2017challenges}, and indirectly leads to improvements in crust and ocean field models \citep{olsen_lcs-1_2017, grayver_tides_2024}.

Ionospheric and magnetospheric contributions in the geomagnetic field models have long been constrained by using the magnetic field measurements from ground observatories \citep{sugiura_hourly_1964,pulkkinen2003ionospheric,olsen_chaos-4_2014}. While this conventional source of geomagnetic observations provides high-quality validated long time series \citep{intermagnet_2013,Macmillan2013}, there are some shortcomings. First, their distribution over the globe is very non-uniform and the number of observatories does not grow. This prohibits the accurate retrieval of geomagnetic field features at higher spherical harmonic degrees, and poses a limit to the spatial resolution of the geomagnetic field models, especially over the oceans. Another fundamental limitation is related to separation of the external field sources. Since all magnetic field contributions of external origin are external to the ground observatories, it is mathematically impossible to separate the ionospheric signal from the magnetospheric signal, unless restrictive prior constraints are imposed. Previous attempts to study the individual extraneous magnetic signals mostly rely on temporal intervals, frequency bands or spatial modes where the contribution of one of the current systems is assumed to be dominant \textit{a priori}. For instance, the magnetospheric contribution is often studied by working with the Dst index \citep{sugiura_hourly_1964} or its analogues \citep[e.g. the RC-index,][]{Finlay2015}, which is assumed to reflect the axisymmetric part of the magnetospheric ring current \citep{Maus2004, Olsen2005}, although it most certainly contains part of the ionospheric field as well. The large-scale ionospheric Sq fields were studied by selecting the data during magnetic quiet times and parametrising the temporal evolution using a few space and time harmonics \citep{winch_spherical_1981,Schmucker1999,yamazaki_sq_2017}. In studies by \cite{Egbert2021, Zenhaeusern2021}, authors adopted a more advanced spatial basis derived from a physics-driven model of the ionospheric dynamics called TIE-GCM. While these models provide a more detailed spatial description of magnetic fields, and allow one to perform physics-based interpolation in regions with lack of observations, their temporal content is limited and mostly focuses on modelling dynamics within the daily band. In addition, it is not trivial to extend these models to include largely non-periodic magnetospheric fields. Further, more work is needed to be able to reconcile this approach with satellite observations.

Noting the problems and limitations with geomagnetic ground observatories, recent studies increasingly incorporate satellite measurements in geomagnetic field modelling.
Continuous magnetic field observations from Low-Earth-Orbit (LEO) satellites have enabled significant advances in geomagnetic field modelling, leading to improved spatial resolution of various magnetic field models, including those for the slow-varying or static core \citep{alken_international_2021}, crustal \citep{maus_fifth-generation_2007, olsen_lcs-1_2017} and oceanic \citep{sabaka_cm6_2020, grayver_tides_2024} fields. This also holds for models of ionospheric and magnetospheric fields \citep{lesur_second_2010, chulliat_first_2016, sabaka_comprehensive_2018, finlay_chaos-7_2020, baerenzung_kalmag_2020}. However, the second problem regarding separation of external field sources is still mostly at large. Most external field models attempt to achieve this by reducing the spatial and/or temporal complexity of the external field model parametrisations. For instance, by taking only the dark-side magnetic observations, these models attribute the magnetic field external to observatories and satellites to the magnetosphere, assuming ionospheric magnetic fields can be neglected at night \citep{lesur_second_2010, finlay_chaos-7_2020, baerenzung_kalmag_2020}. This approach limits the usable data to only the dark-side field measurements, typically discarding more than half of the available data. In addition, even the dark-side data may contain ionospheric contributions. Apart from the potential dynamics in the F region during the night, the transient behaviour of electromagnetic induction also means that the dark-side data contain internal signals that are induced by the day-time ionosphere and persist into the night \citep{Maus2004, Grayver2021}. Another common practice is to use only observations taken during quiet magnetic conditions. For instance, a dedicated magnetic field model of the ionosphere and equatorial electrojet (EEJ) by \cite{chulliat_first_2016} delivers a model of the mid-latitude ionosphere during quiet phases. Although the model has a high spatial resolution, the temporal dynamics of the ionosphere is approximated by a small set of daily and seasonal time harmonics. Other temporal components are approximated by modulating the predicted field with an external time-varying proxy (e.g. F10.7 flux). More elaborate Comprehensive Inversion (CI) approach tries to parametrise the ionospheric magnetic field together with other components, including a lowest-degree magnetospheric field \citep{sabaka_comprehensive_2018}. While the parametrisation of ionospheric currents in CI is quite elaborate and fits observations at mid and equatorial latitudes well, a small set of discrete points in the frequency domain at annual, semiannual and diurnal bands also means that other temporal variations are not captured. Similarly, the model is constructed using only quiet magnetic periods. Further, these models are difficult to update on the fly as small volumes of new observations are added in time, limiting their applications for operational space-weather or for studying individual events, such as geomagnetic storms.

To overcome the limitations of current models, and to better exploit the increasing number of satellite observations delivered by dedicated missions and platform magnetometers, we present here a novel approach to simultaneously characterise the ionospheric, magnetospheric and internally induced magnetic fields using all available ground and space observations and high temporal resolution. Our approach utilises the geometrical configuration of ground and low Earth orbit (LEO) geomagnetic observations. Indeed, the ground observatories alone cannot distinguish between the ionospheric and magnetospheric contributions, as both are external to the observations. The introduction of satellite observations in the region between the ionosphere and magnetospheric currents, however, renders the problem feasible. We construct model in short time-bins, allowing for a continuous Swarm-era model, including all local times and magnetic activity conditions. In addition to ionospheric and magnetospheric components, we also estimate time series of the spherical harmonic coefficients of the internally induced field, rendering the model output also relevant for mantle induction studies.

The rest of the article is organised as follows. We introduce the model setup in Section \ref{sec:method}, focusing on how our formulation would allow simultaneous co-estimation of external sources using short time bins instead of predefined time harmonics. This is followed by a description of the data pre-processing (Section \ref{sec:sec-data}) and model parametrisation (Section \ref{sec:sec-model-est}). In Section \ref{sec:results} we present the results from processing the 10-year time series with our designated methodology. The results provide some new inputs for electromagnetic induction studies, and insights into ionospheric and magnetospheric dynamics, which we will discuss in Section \ref{sec:discussions}.

\section{Methodology}\label{sec:method}

The time-dependent magnetic field at location $\mathbf{r}$ and time $t$ is described by the Maxwell's equations, which read
\begin{equation}
\begin{aligned}
    \nabla\cdot \mathbf{B}(\mathbf{r}, t) &= 0, \\
    \nabla\times \mathbf{B}(\mathbf{r}, t) &= \mu \left(\mathbf{j}(\mathbf{r}, t) + \varepsilon \frac{\partial \mathbf{E}(\mathbf{r}, t)}{\partial t}\right), 
\end{aligned}
\end{equation}
where $\mathbf{j}$ is the total current density, $\mu$ is the magnetic permeability, $\varepsilon$ is the electric permittivity, and $\mathbf{B}$ and $\mathbf{E}$ are the magnetic and electric fields, respectively.
Unless otherwise stated, the physical variables are given in SI units.
In an electrically insulating medium under magneto-quasistatic approximation (MQS) \citep{larsson_electromagnetics_2007}, the right hand side of the second equation vanishes identically, yielding an irrotational magnetic field, in which case the magnetic field is a potential field, expressed through a scalar potential $V$,
\begin{equation}
    \mathbf{B}(\mathbf{r}, t) = - \nabla V(\mathbf{r}, t).
\end{equation}
Combined with the solenoidal property of the magnetic field, the scalar potential is shown to satisfy the Laplace equation
\begin{equation}
    \nabla^2 V = 0.
\end{equation}
In a spherical shell, this leads to the following form of general solutions
\begin{equation}\label{eqn:proto-gauss}
\begin{aligned}
    V(\mathbf{r}, t) = a \sum_{n=1}^{+\infty} \sum_{m=0}^n \Big[&\left(g_{nm}(t) \cos m\phi + h_{nm}(t) \sin m\phi \right) \left(\frac{r}{a}\right)^{-(n+1)} \\
    + &\left(q_{nm}(t) \cos m\phi + s_{nm}(t) \sin m\phi \right) \left(\frac{r}{a}\right)^n \Big] P_n^m(\cos\theta),
\end{aligned}
\end{equation}
where $(r,\theta,\phi)$ are the spherical radius, the colatitude and the azimuth, $P_n^m(\cos\theta)$ is the associated Legendre polynomial of degree $n$ and order $m$, and $a$ is a scaling factor for radius. The time-dependent coefficients $g_{nm}$ and $h_{nm}$ represent the contribution to the magnetic field of internal origin (i.e. from electric current sources within the inner boundary of the shell), and $q_{nm}$ and $s_{nm}$ represent the contributions of external origin (i.e. from electric current sources outside the outer boundary of the shell).

In geomagnetism, the spherical harmonic (SH) representation (\ref{eqn:proto-gauss}) applies to any observations made in the electric current free region surrounding the Earth.
In this case, the radius scaling factor $a$ is conveniently chosen to be the mean radius of the Earth, and $P_n^m(\cos\theta)$ is chosen to be Schmidt quasi-normalised associated Legendre polynomial \citep{winch_geomagnetism_2005}. Coefficients $g_n^m$, $h_n^m$, $q_n^m$ and $s_n^m$ are referred to as the real Gauss coefficients. 

The SH representation has long been used for constructing geomagnetic field models from ground-based observations \citep{gauss1839}. Since the ground observations are made in the electrically insulating neutral atmosphere, the potential formulation of the magnetic field in a spherical shell is valid. In recent decades, an increasing number of studies also included satellite observations in the same SH representation \citep{langel_initial_1980, bloxham_simultaneous_1989, langlais_igrf_2000, Olsen2006, baerenzung_kalmag_2020}. This approach is justified since satellite observations taken at $400-800$ km altitude are away from major source regions: the magnetospheric current systems \citep{ganushkina_current_2018}, and the E-region dynamo in the ionosphere \citep{yamazaki_sq_2017}. 

\begin{figure}
    \centering
    \includegraphics[width=0.95\linewidth]{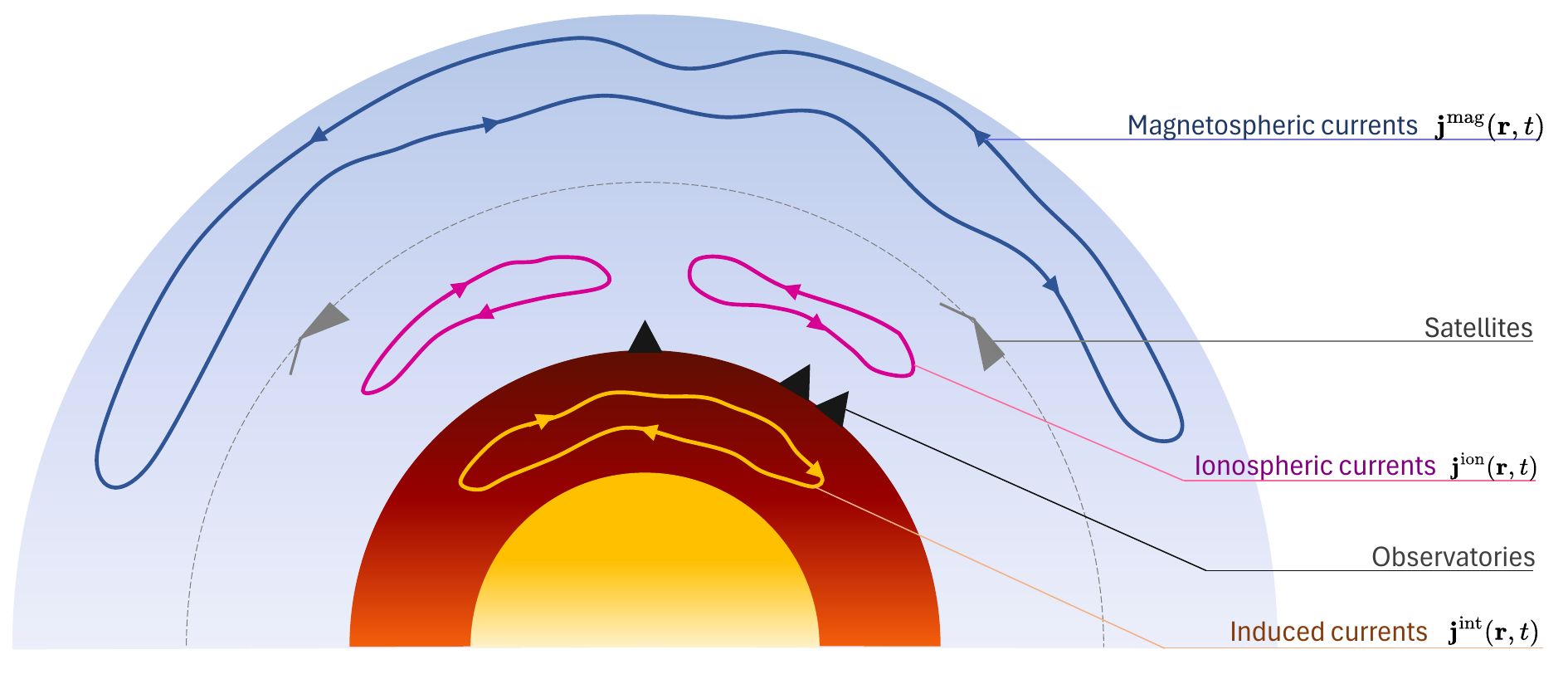}
    \caption{Geometric configuration of the model, showing time dependent spatially varying external and internal electric currents at middle geomagnetic latitudes (that is, excluding the polar regions).}
    \label{fig:model-config}
\end{figure}

Note, however, because of the presence of ionospheric currents, for instance those generated in the E-region dynamo, the ground observatories and the satellites do not "see" the same set of source configurations. 
Taking this into account, we introduce two sets of Gauss coefficients. For the ground observations, the SH representation reads
\begin{equation}\label{eqn:gauss-obs}
\begin{aligned}
    V_\mathrm{obs}(\mathbf{r}, t) = a \sum_{n=1}^{+\infty} \sum_{m=0}^n \Big[&\left(g_{nm}^\mathrm{obs}(t) \cos m\phi + h_{nm}^\mathrm{obs}(t) \sin m\phi \right) \left(\frac{r}{a}\right)^{-(n+1)} \\
    + &\left(q_{nm}^\mathrm{obs} (t)\cos m\phi + s_{nm}^\mathrm{obs}(t) \sin m\phi \right) \left(\frac{r}{a}\right)^n \Big] P_n^m(\cos\theta),
\end{aligned}
\end{equation}
while for the satellite observations we have
\begin{equation}\label{eqn:gauss-sat}
\begin{aligned}
    V_\mathrm{sat}(\mathbf{r}, t) = a \sum_{n=1}^{+\infty} \sum_{m=0}^n \Big[&\left(g_{nm}^\mathrm{sat}(t) \cos m\phi + h_{nm}^\mathrm{sat}(t) \sin m\phi \right) \left(\frac{r}{a}\right)^{-(n+1)} \\
    + &\left(q_{nm}^\mathrm{sat}(t) \cos m\phi + s_{nm}^\mathrm{sat}(t) \sin m\phi \right) \left(\frac{r}{a}\right)^n \Big] P_n^m(\cos\theta).
\end{aligned}
\end{equation}
In this study, we are interested in the magnetic fields that are ultimately attributed to extraneous current systems. For this purpose, we consider three major electrical current systems contributing to the magnetic field: 1) the magnetospheric currents, the most dominant of which is the ring current, 2) the ionospheric currents, mostly from the E-region dynamo, and 3) the corresponding induced signals originating in the solid Earth or in the ocean, and hence internal to all observations. As we are most interested in the large-scale features in the magnetic field, we shall restrict the ionospheric currents to the mid-latitude currents. This excludes the equatorial and polar current systems, which require dedicated parametrisations. Meanwhile, the core field with its secular variations and the lithospheric field are considered removed from the system. The geometric setup of our model is illustrated in Fig. \ref{fig:model-config}.

Under these considerations, the observatory Gauss coefficients of internal origin represent contributions from electric currents induced in the Earth's interior or in the ocean by nature of electromagnetic induction. We write these coefficients alternatively as
\begin{equation}\label{eqn:gauss-decomp-obs-int}
    g_{nm}^\mathrm{obs} = g_{nm}^\mathrm{int}, \quad h_{nm}^\mathrm{obs} = h_{nm}^\mathrm{int}.
\end{equation}
The satellite Gauss coefficients of external origin represent contributions from magnetospheric currents, and are written as
\begin{equation}\label{eqn:gauss-decomp-sat-ext}
    q_{nm}^\mathrm{sat} = q_{nm}^\mathrm{mag}, \quad s_{nm}^\mathrm{sat} = s_{nm}^\mathrm{mag}.
\end{equation}
The observatory Gauss coefficients of external origin contain contributions from the magnetospheric currents as well as from the ionospheric currents. They can thus be decomposed as
\begin{equation}\label{eqn:gauss-decomp-obs-ext}
    q_{nm}^\mathrm{obs} = q_{nm}^\mathrm{ion} + q_{nm}^\mathrm{mag}, \quad s_{nm}^\mathrm{obs} = s_{nm}^\mathrm{ion} + s_{nm}^\mathrm{mag}.
\end{equation}
The same holds for the satellite Gauss coefficients of internal origin. They describe the field generated by both ionospheric currents and induced currents, and can be decomposed as
\begin{equation}\label{eqn:gauss-decomp-sat-int}
    g_{nm}^\mathrm{sat} = g_{nm}^\mathrm{ion} + g_{nm}^\mathrm{int}, \quad h_{nm}^\mathrm{sat} = h_{nm}^\mathrm{ion} + h_{nm}^\mathrm{int}.
\end{equation}
Note that the coefficients $g_{nm}^\mathrm{int}$ and $h_{nm}^\mathrm{int}$ in Eq. (\ref{eqn:gauss-decomp-sat-int}) are the same as in (\ref{eqn:gauss-decomp-obs-int}). Since the induced currents reside within the Earth, and are internal to both the ground and the satellite observations, the field generated at the surface can be upward continued as the same potential field to a satellite altitude. Similarly, the coefficients $q_{nm}^\mathrm{mag}$ and $h_{nm}^\mathrm{mag}$ in Eqs. (\ref{eqn:gauss-decomp-obs-ext}) and (\ref{eqn:gauss-decomp-sat-ext}) are also the same. The only component different is the ionospheric component, described by $q_{nm}^\mathrm{ion}$ and $s_{nm}^\mathrm{ion}$ in the ground setting, and $g_{nm}^\mathrm{ion}$ and $h_{nm}^\mathrm{ion}$ in the satellite setting.
It is generally not possible to establish a one-to-one relation between the fields generated by the ionosphere within the ionosphere and outside the ionosphere, without knowing \textit{a priori} the spatial distribution of the ionospheric currents. However, assuming that the electric current in the ionosphere occupies a domain with limited radial extent, one can adopt the thin-sheet approximation, whereby the ionospheric current is approximated as a spherical sheet of radius $a+h$. This approximation does not only associate a unique sheet current to an internal or an external magnetic field, but also provides the continuity of the radial magnetic field across the thin sheet:
\begin{equation}
    B_r|_{r\rightarrow (a+h)^+} = B_r|_{r\rightarrow (a+h)^-}.
\end{equation}
Using $B_r = -\partial_r V_\mathrm{obs}$ for $r<a+h$ and $B_r = -\partial_r V_\mathrm{sat}$ for $r>a+h$, a one-to-one relation between the ionospheric Gauss coefficients can be established \citep[see also][]{sabaka_comprehensive_2002}
\begin{equation}\label{eqn:gaussc-ionos-link-sheet}
\begin{aligned}
    n\left(\frac{a+h}{a}\right)^{n-1} q_{nm}^\mathrm{ion} &= - (n+1)\left(\frac{a+h}{a}\right)^{-(n+2)} g_{nm}^\mathrm{ion},\\
    n\left(\frac{a+h}{a}\right)^{n-1} s_{nm}^\mathrm{ion} &= - (n+1)\left(\frac{a+h}{a}\right)^{-(n+2)} h_{nm}^\mathrm{ion}. 
\end{aligned}
\end{equation}
Combining Eqs. (\ref{eqn:gauss-decomp-obs-int}-\ref{eqn:gauss-decomp-sat-int}) with (\ref{eqn:gaussc-ionos-link-sheet}), we can rewrite the SH representation for ground observations (\ref{eqn:gauss-obs}) as
\begin{equation}\label{eqn:model-V-repr-obs}
\begin{aligned}
    V_\mathrm{obs}(\mathbf{r},t) = a &\sum_{n=1}^{+\infty} \sum_{m=0}^n \Big[\left(g_{nm}^\mathrm{int}(t) \cos m\phi + h_{nm}^\mathrm{int}(t) \sin m\phi \right) \left(\frac{r}{a}\right)^{-(n+1)} \\
    + &\left(\left(q_{nm}^\mathrm{ion}(t) + q_{nm}^\mathrm{mag}(t)\right) \cos m\phi + \left(s_{nm}^\mathrm{ion}(t) + s_{nm}^\mathrm{mag}(t)\right) \sin m\phi \right) \left(\frac{r}{a}\right)^n \Big] P_n^m(\cos\theta),
\end{aligned}
\end{equation}
and the SH representation for satellites (\ref{eqn:gauss-sat}) as
\begin{equation}\label{eqn:model-V-repr-sat}
\begin{aligned}
    V_\mathrm{sat}(\mathbf{r}, t) = a &\sum_{n=1}^{+\infty} \sum_{m=0}^n \Bigg[\Bigg(\left(g_{nm}^\mathrm{int}(t) - \frac{n}{n+1} \left(\frac{a+h}{a}\right)^{2n+1} q_{nm}^\mathrm{ion}(t)\right) \cos m\phi \\
    + &\left(h_{nm}^\mathrm{int}(t) - \frac{n}{n+1} \left(\frac{a+h}{a}\right)^{2n+1} s_{nm}^\mathrm{ion}(t) \right) \sin m\phi \Bigg) \left(\frac{r}{a}\right)^{-(n+1)} \\
    + &\left(q_{nm}^\mathrm{mag}(t) \cos m\phi + s_{nm}^\mathrm{mag}(t) \sin m\phi \right) \left(\frac{r}{a}\right)^n \Bigg] P_n^m(\cos\theta).
\end{aligned}
\end{equation}
These representations reduce the magnetic potential to the source-specific Gauss coefficients $(g_{nm}^\mathrm{int}, h_{nm}^\mathrm{int})$, $(q_{nm}^\mathrm{ion}, s_{nm}^\mathrm{ion})$, $(q_{nm}^\mathrm{mag}, s _{nm}^\mathrm{mag})$, hereinafter referred to as internally induced coefficients, ionospheric coefficients and magnetospheric coefficients, respectively. These Gauss coefficients can be estimated by minimising the data misfit of the combined satellite and ground observatory dataset. The compilation and pre-processing of the dataset and the model estimation process are detailed in Sections \ref{sec:sec-data} and \ref{sec:sec-model-est}, respectively.

\newcommand{\refmod}{\ref{eqn:model-V-repr-obs}-\ref{eqn:model-V-repr-sat}}

\section{Data}\label{sec:sec-data}

We compiled a joint dataset of ground observatory and satellite observations from data provided by the European Space Agency's (ESA) VirES for Swarm service\footnote{VirES for Swarm service, https://vires.services/}. The dataset spans a total length of nearly 10 years, from January 2014 to December 2023. The compilation and pre-processing of the data is largely facilitated by the Python interface \textit{viresclient} \citep{smith_esa-viresvires-python-client_2024}. 

While Eqs. (\refmod) provides magnetic field models that are continuous in both space and time, we implement a time-discrete version by estimating the Gauss coefficients within non-overlapping time bins. For this purpose, the ground and satellite observations must be both arranged into separate time bins, within which the corresponding SH coefficients are assumed to be constant. This inevitably introduces a trade-off between spatial coverage and the temporal resolution of the satellite data. We found that a time bin of $3$ hours provides a fair balance between the spatial and temporal resolution. The model parametrisation can be easily adjusted depending on future applications and availability of new satellite missions and observatories. 

\subsection{Satellite data}

We used vector magnetic field measurements from the Swarm satellites (Swarm data product \texttt{SW\_MAGx\_LR\_1B} where $\mathtt{x} \in \{\mathtt{A,B,C}\}$) within the designated time period \citep{olsen_swarm_2013}. For better local time coverage, the Swarm data are further supplemented by the vector magnetic field measurements from CryoSat-2 (Swarm data product \texttt{CS\_OPER\_MAG}) from 2014 to 2023 \citep{olsen_cryosat2_2020}, as well as those from Grace-FO satellites (Swarm data product \texttt{GFx\_OPER\_FGM\_ACAL\_CORR} where $\mathtt{x} \in \{\mathtt{1,2}\}$) from 2018 to 2023 \citep{stolle_gracefo_2021}. All satellite measurements are down-sampled to a one-minute sampling interval. As we are interested in the magnetic fields that are ultimately attributed to the extraneous sources, the core field (product \texttt{SW\_MCO\_SHA\_2C}) and the static lithospheric field (product \texttt{SW\_MLI\_SHA\_2C}) as given by the Comprehensive Inversion chain \citep{sabaka_cm6_2020} are subtracted from all magnetic field measurements.

We then filter out satellite observations taken at locations with an absolute quasi-dipole (QD) latitude \citep{richmond_ionospheric_1995} below $5^\circ$ or above $56^\circ$. This is to avoid the effects of equatorial and polar electrojets, which are localised current systems requiring dedicated parametrisations and cannot be well described by low-degree spherical harmonics. 
Contrary to approaches deriving full geomagnetic models with a focus on core models, such as the CHAOS model \citep{finlay_chaos-7_2020} or the Comprehensive Model \citep{sabaka_cm6_2020}, we do not filter observations based on solar zenith angle or magnetic indices, nor do we parametrise the ionospheric field using specific diurnal/seasonal time harmonics. 
Since the spatial and temporal behaviour of the external field is of prime interest here, we impose no constraint on the magnetic conditions and inject no prior knowledge on the periodicity of reconstructed sources. Instead, the parameter estimation is carried out independently for each time window, and the temporal behaviour will be analysed in a completely \textit{a posteriori} fashion. A significant advantage of such model is the ease of its update, which can be performed on the fly in real-time (limited only by the geomagnetic data latency).

\subsection{Ground observatory data}

\begin{figure}
    \centering
    \includegraphics[width=0.98\linewidth]{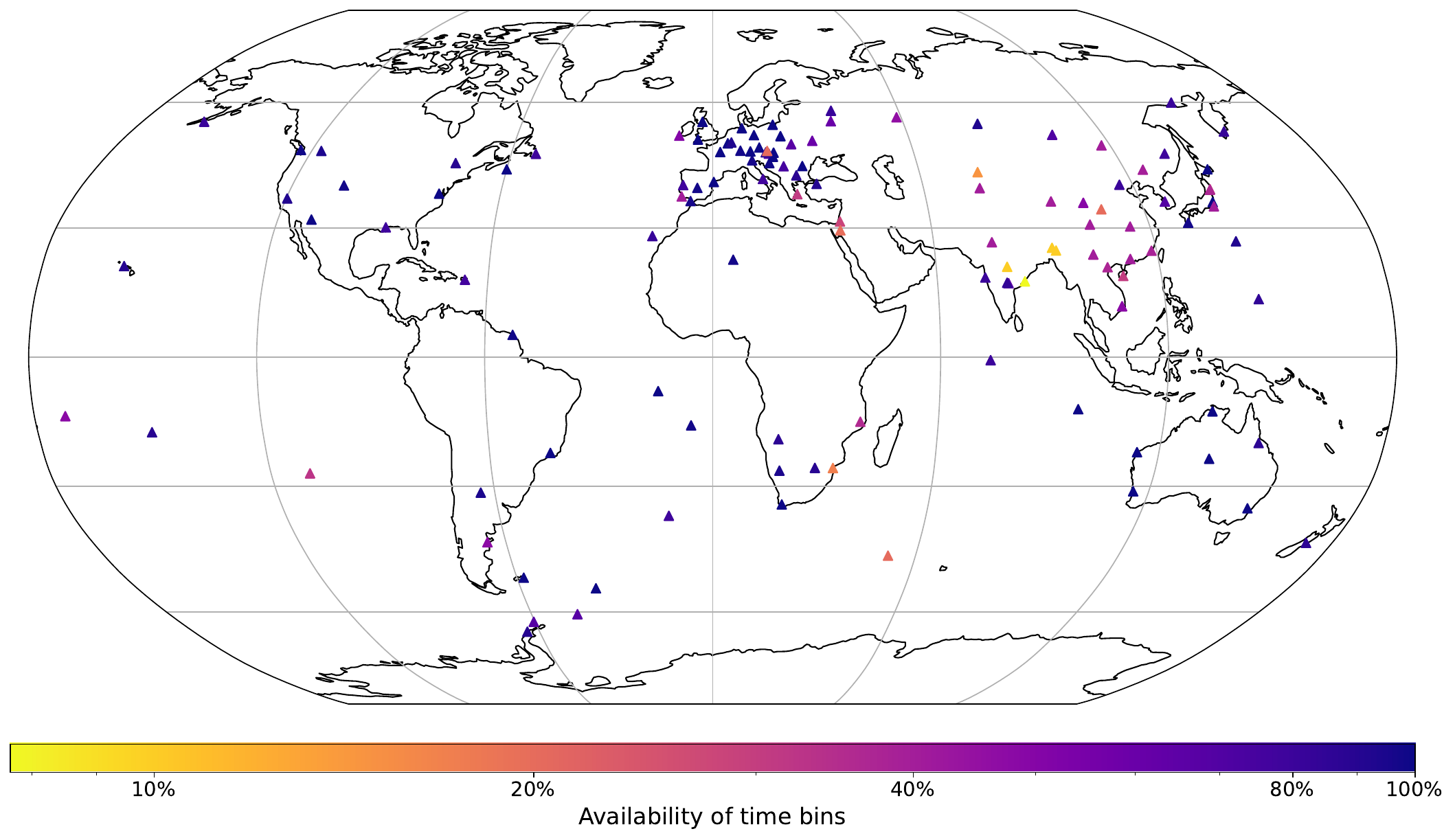}
    \caption{Distribution of ground observatories within $5^\circ$N - $56^\circ$N and $5^\circ$S - $56^\circ$S in magnetic latitude. The colour of the markers indicates the percentage of time bins where the observatory data are available in the period of 2013-2023.}
    \label{fig:obs-distribution-ntbin}
\end{figure}

We compiled the vector magnetic field data measured by ground observatories from the INTERMAGNET and World Data Centre for Geomagnetism (WDC) within the same time period \citep{irds-2020}. The hourly mean time series are binned in $3$-hr time bins for consistency with the satellite data.
The same filtering based on QD latitude ($5^\circ - 56^\circ$ north and south) has been applied.
The distribution of ground observatories used in this study is shown in Fig. \ref{fig:obs-distribution-ntbin}. 

\subsubsection{Observatory bias}

One of the difficulties in dealing with ground observatory data lies in the observatory biases. They appear as large offsets in the components of the vector magnetic field, which are not explained by the model, and at the same time are not well correlated even between relatively nearby stations. They are interpreted to represent short-wavelength, localised anomaly crustal fields that are not captured by a lithospheric field model  \citep{langel_new_methods_1982, sabaka_comprehensive_2002}. In addition, observatory-specific baseline errors introduce biases that cannot be represented in magnetic field models \citep{lesur_estimating_2017}. 

While these observatory biases are not described by the magnetic field model to be derived, they nevertheless appear as intercepts in the regression and hence affect the parameter estimation process. For this reason, they can be seen as \textit{nuisance parameters} in our model. Provided an estimate for such biases already exists, one can circumvent this issue by using the time differential ground observations for data fitting \citep{Finlay2015}. Determining the observatory bias from observations, on the other hand, is more challenging. One simple solution is to take the arithmetic mean of each vector component at a ground observatory over magnetic quiet times \citep{olsen_chaos-4_2014}. This serves as a pragmatic approach, effectively removing the means and leaving only zero-mean time series from which external and the induced fields are estimated, but lacks physical reasoning. From the data perspective, it has long been recognised that this bias cannot be determined unless satellite data unaffected by such bias are used (\citeauthor{langel_new_methods_1982} \citeyear{langel_new_methods_1982}; for an explanation, see Appendix \ref{sec:app-bias-obs}). Even when the satellite data are used, the estimation only works when the region between satellite and ground is void of electric currents, implying that no field originates in the ionosphere below satellites \citep{langel_new_methods_1982}, or when the ionosphere is parametrised using time harmonics \citep{sabaka_comprehensive_2002}. Relaxing the temporal prior knowledge on the ionosphere as we do in this study results in a non-trivial nullspace when the bias is included (Appendix \ref{sec:app-bias-full}).

Since the observatory bias cannot be co-estimated in our model with a piece-wise constant temporal parametrisation of the ionosphere, we derive these biases in a separate step. We took ground observatory and satellite data from magnetic quiet times (defined as times when $K_p \leq 2$ and $|dDst/dt| \leq 3\mathrm{nT/h}$) with the Sun at least $10^\circ$ below the horizon. Data fulfilling these conditions are referred to as \textit{quiet night} data. The quiet night data are then fitted using a magnetic field model with only magnetospheric field and the internally induced field, and the observatory bias consisting of one time independent vector per observatory. Additional vector biases are introduced for observatories where large offsets are seen in different segments of observation. Due to the limited spatial coverage of the filtered data, the magnetospheric field and the internally induced field are parametrised only up to SH degree and order one. Although the dark-side data are not fully free of ionospheric and its corresponding induced field contributions, such effects are minimal due to much weaker ionospheric currents \citep{price_new_1963}. Using a parametrisation without the ionosphere on the quiet night data is a necessary compromise since the addition of the ionospheric source would lead to an intrinsic non-uniqueness in the bias estimation problem (Appendix \ref{sec:app-bias-full}). In addition, note that the nullspace of the forward problem with the ionosphere is characterised by a static field. Therefore, the omission of the ionosphere in the bias estimation can only introduce a small static offset for the entire data span (10 years in our model), which does not interfere with the temporal variability of the field estimates.

The bias values are then determined using a computation- and memory-efficient linear least squares algorithm for block-diagonal matrix with intercept (Appendix \ref{sec:app-bias-diag}).
The estimated observatory biases are shown in supplementary Fig. \ref{fig:obs-bias-est-map}. 
The observatory biases derived in this study agree well with the CI-derived biases, and both are close to the quiet night mean values (Figs. \ref{fig:obs-bias-est}, \ref{fig:obs-bias-est-diffmean}, Table \ref{tab:bias-rmsd}).  
The reason to carry out this estimation instead of using the CI-derived biases is to retain the consistency with our data set, and with our model which does not prescribe a periodic behaviour of the ionospheric field, whereas CI does. These vector biases are then removed from the ground observatory measurements.

\begin{figure}
    \centering
    \includegraphics[width=.95\linewidth]{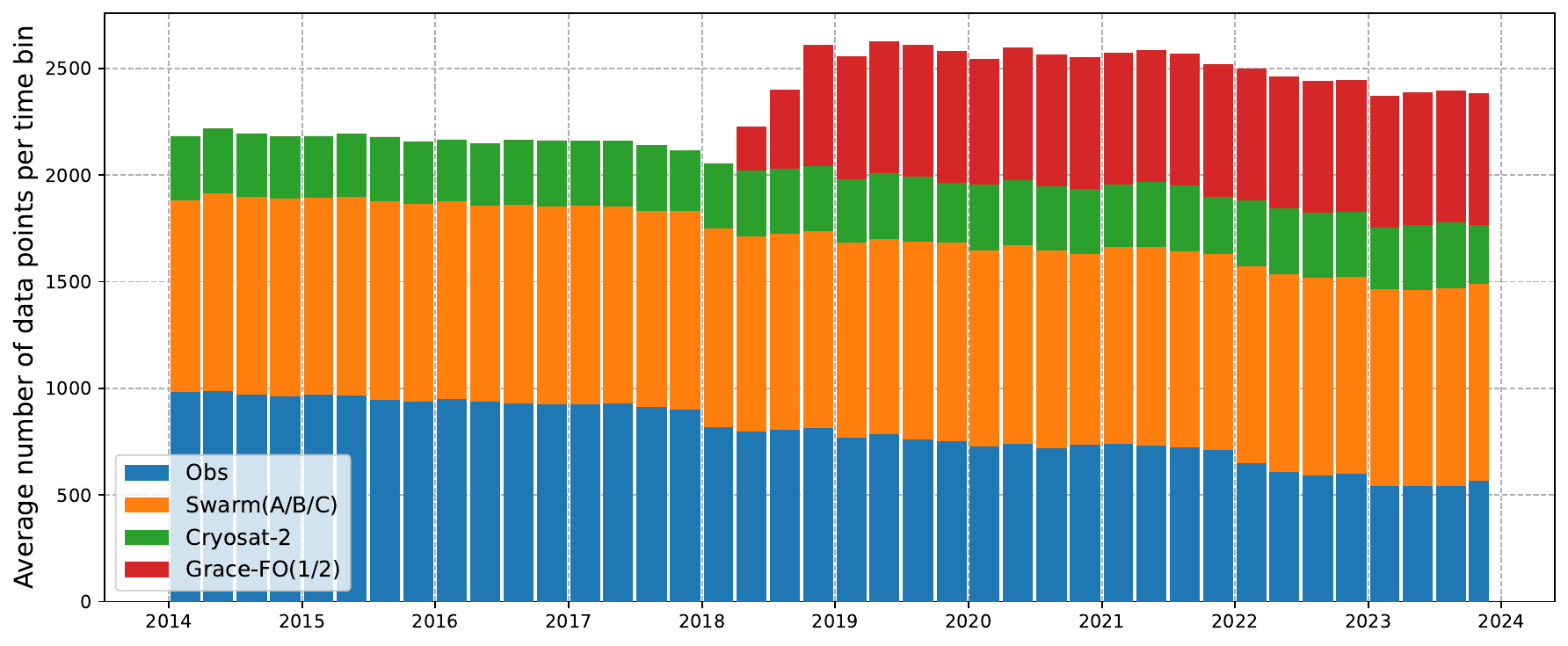}
    \caption{Average number of data points per time bin as a function of time. Each vector component is treated as an independent datum.}
    \label{fig:ndata-time-distribution}
\end{figure}

The pre-processing of the data yields a total number of 7.7 million ground observatory measurements, and 15.0 million satellite measurements of the three-component vector magnetic field.
An average time bin contains 800 magnetic field vector component data sampled by (on average) 89 ground observatories, and 1558 vector component data sampled at 519 satellite locations. These numbers vary in time depending on the availability of satellites and observatories, as is shown in Fig. \ref{fig:ndata-time-distribution}.

\section{Model construction and selection}\label{sec:sec-model-est}

The magnetic field model is parametrised in three sets of source-specific Gauss coefficients: $g_{nm}^\mathrm{int}(t)$ and $h_{nm}^\mathrm{int}(t)$ for the internally induced field, $q_{nm}^\mathrm{ion}(t)$ and $s_{nm}^\mathrm{ion}(t)$ for the ionospheric field, and $q_{nm}^\mathrm{mag}(t)$ and $s_{nm}^\mathrm{mag}(t)$ for the magnetospheric field (Eq. \refmod). Denoting these Gauss coefficients in the $i$-th time bin jointly as $\mathbf{m}_i$, and the observatory and satellite data jointly as $\mathbf{d}_i$, the model estimation procedure in each time bin can be formulated as solving the following optimisation problem
\begin{equation}\label{eqn:opt-form}
    \hat{\mathbf{m}}_i = \arg\min_{\mathbf{m}_i} L \left(\mathbf{W}_i (\mathbf{d}_i - \mathbf{G}_i \mathbf{m}_i) \right)
\end{equation}
where $\mathbf{W}_i$ is the data weight matrix for the data in the $i$-th time bin, $\mathbf{G}_i$ is the Gauss matrix, and $L(\cdot)$ is a generic misfit function. We assume independent sampling of each data point and assign uniform variance to each observation, hence $\mathbf{W} = \mathbf{I}$. The Gauss matrix $\mathbf{G}_i$ is the design matrix for the regression problem, and can be obtained by evaluating the gradients of the potentials in Eqs. (\refmod) at observation locations.

Since no temporal behaviour is imposed on any of the SH modes, all the modes and their coefficients with a spherical degree no greater than the truncation degree are treated equally in Eqs. (\refmod). Since spherical harmonics form an irreducible representation of the 3-D rotation group, there is no difference between using different coordinate systems in our formulation. We chose to work in geocentric geographic coordinates to simplify integration with the existing dataset, as well as with other ground-based observations that are naturally provided in this coordinate system.

The truncation degree for the spherical harmonics is selected based on two arguments. First, the local time coverage of the satellites gives a theoretical upper bound on the maximum azimuthal wavenumber. Our dataset contains data from six satellites in polar orbits, which, with an approximate orbital period of 90-100 minutes, cover 24 local times within a $3$-hr time bin (each satellite covers two local times per orbit, and completes two orbits per time bin). However, among these satellites, two pairs always share the same local times as they are very close in orbits (Swarm A and C, Grace-FO 1 and 2), and Swarm B satellite also evolves to the same orbital plane as Swarm A and C around 2021. Therefore, the satellites at best sample the magnetic field at 12 local times within one time bin. 
According to the Nyquist sampling theorem, the field that can be reconstructed without aliasing is limited to azimuthal wavenumber $m < 6$, even with uniform sampling.
Beyond this azimuthal wavenumber, even if the combined external field can be estimated from the observatory dataset, it can no longer be convincingly separated into ionospheric and magnetospheric contributions without being aliased by the satellites. As we are using $N=M$, this limits the truncation SH degree to $N < 6$.

Given $N < 6$, the specific parametrisation is further selected using $K$-fold cross-validation (CV), a technique that provides an estimate on the generalisation error of the model. In a $K$-fold CV, the full dataset to be used for parameter estimation is split into $K$ parts. Model parameters are estimated in each fold using $K-1$ parts of the datasets, while the remaining part serves as the validation set for testing how the estimated model performs on unseen data. The average of the $K$ validation errors then yields the proxy for the generalisation error \citep{arlot_survey_2010}. While a pure extension of the model parametrisation will always reduce the total data misfit (provided global minimum is reached), the generalisation error will stagnate or deteriorate when the model is being increasingly fitted to the noise rather than the true signal, and hence provides means to detect overfitting.

\begin{figure}
    \centering
    \includegraphics[width=0.8\linewidth]{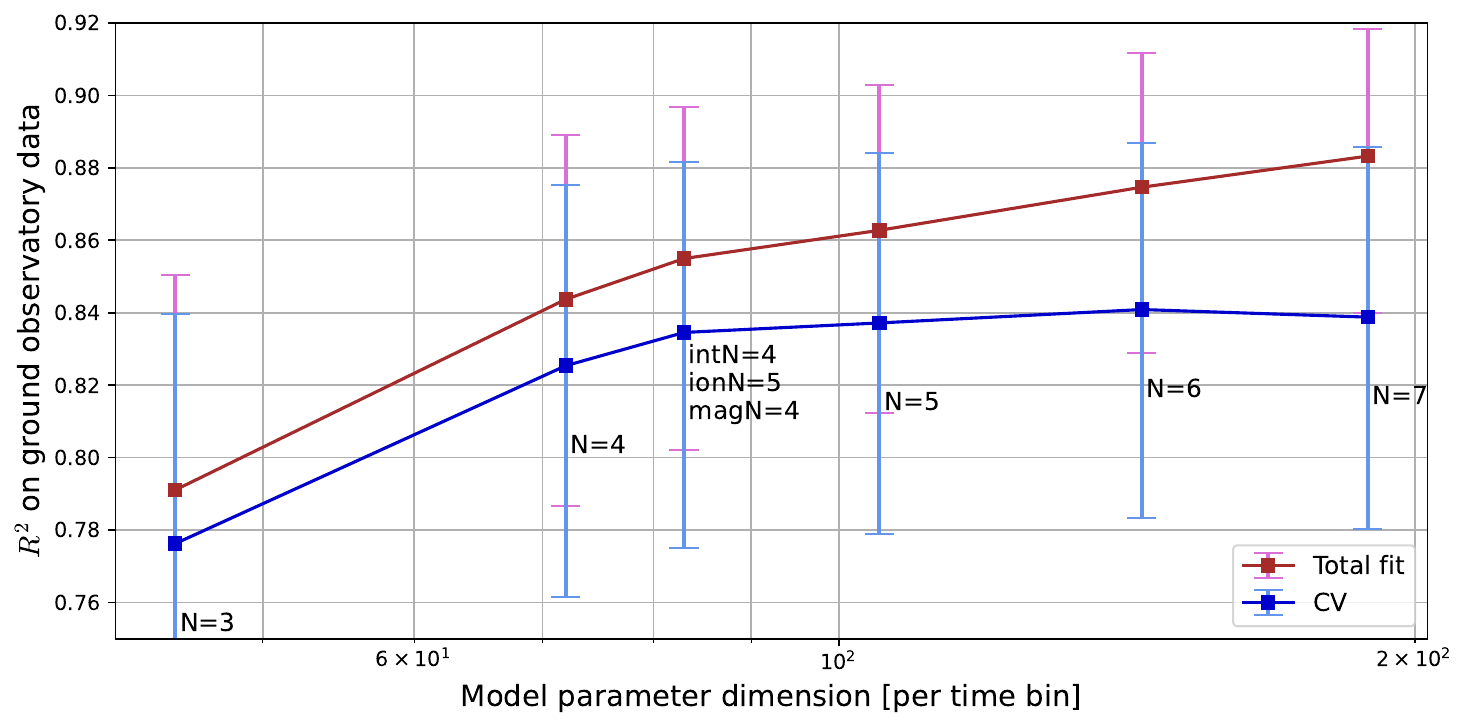}
    \caption{The coefficient of determination $R^2$ from fitting the whole ground observatory dataset (red) and from CV (blue) for models estimated using SH truncation degrees from $3$ to $7$. The square markers show the median of the $R^2$ scores, while the upper and lower error bar limits show the 1st and 3rd quantiles, respectively.}
    \label{fig:CV-N3-6-obs}
\end{figure}

\begin{figure}
    \centering
    \includegraphics[width=0.8\linewidth]{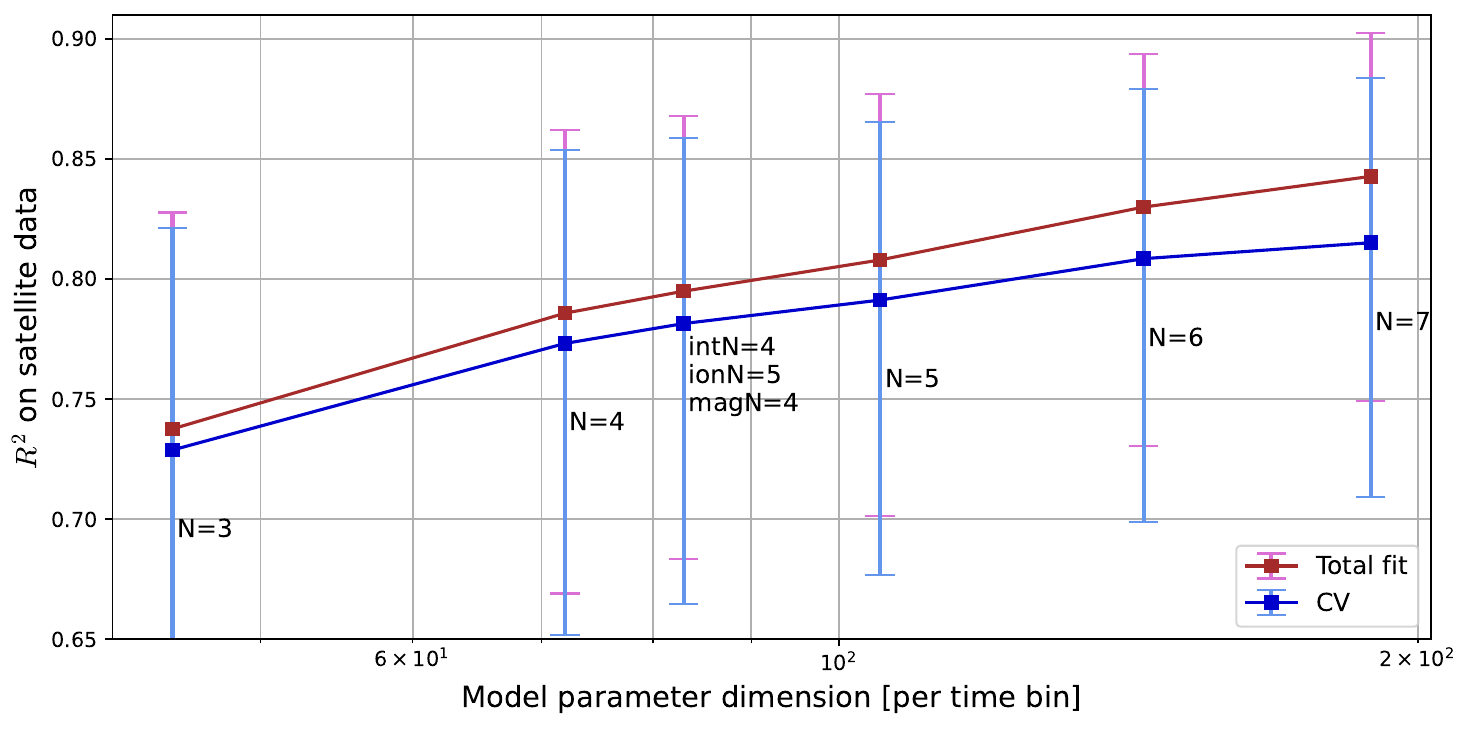}
    \caption{Same as Fig. \ref{fig:CV-N3-6-obs}, but for the satellite dataset.}
    \label{fig:CV-N3-6-sat}
\end{figure}

We performed a 5-fold CV on the combined dataset in each time bin, for SH truncation degrees ranging from $3$ to $7$. The performance of the model is evaluated using the coefficient of determination, defined as 
\begin{equation}
    R^2 = 1 - \frac{\|\mathbf{d} - \hat{\mathbf{d}}\|_2^2}{\|\mathbf{d}\|_2^2 - N_d \overline{d}^2}
\end{equation}
where $\mathbf{d}$ is the observed data vector, $\hat{\mathbf{d}}$ is data predicted by a model, $\overline{d}$ is the arithmetic mean of $\mathbf{d}$, and $N_d = \dim \mathbf{d}$ the dimension of the data. For each of the model parametrisations, the coefficients of determination $R^2$ are computed on the validation set of the ground observatory and satellite datasets separately and averaged over the CV folds. Statistics of the CV $R^2$ score shows that while the estimated model seems to perform better as the truncation degree increases on the satellite dataset (blue line in Fig. \ref{fig:CV-N3-6-sat}), the CV score for the ground observatory dataset stagnates at $N=5$ and then decreases for $N=7$ (blue line in Fig. \ref{fig:CV-N3-6-obs}). This observation is consistent with the theoretical argument that SH modes with degrees $>5$ cannot be resolved with the current data. 
A more careful investigation shows that a model with induced and magnetospheric fields parametrised to $N=4$, but ionospheric field extended to $N=5$ already explains most of the variance in the data and is located on the CV score plateau of $N=5$. Compared to this model, the 25\% more free parameters in the $N=5$ model only yields a marginal increase of 0.002 in the CV score. This implies that the extra degrees of freedom in the $N=5$ model does not yield a significant gain in explaining the data, while increasing the risk of overfitting. 
Based on these observations, we adopted the parametrisation in which the induced and magnetospheric fields are expanded to $N=4$, and the ionospheric field is expanded to $N=5$.
For each time bin, Gauss coefficients $\mathbf{m}_i$ are estimated by solving Eq. (\ref{eqn:opt-form}) using a robust Huber loss as the loss function $L(\cdot)$.

\section{Results}\label{sec:results}

\subsection{The role of ionosphere in the model}\label{sec:results-ionos-cv}

By estimating coefficients in each time bin, we obtained the 10-year time series of Gauss coefficients describing the magnetospheric field, the ionospheric field, and the induced counterpart. The introduction of the ionospheric field in our approach, as described in Section \ref{sec:method}, is the major reason all three fields can be characterised individually. While the inclusion of the ionosphere is dictated by reality, it results in additional model parameters and requires the sheet current approximation. Hence, a statistical experiment is needed to justify that adding the ionosphere indeed results in a better model.

To this end, we constructed another model with the same model parametrisation, but without the ionosphere layer. This model effectively attributes all satellite and observatory data variations to a common internal source (internally induced currents) and a common external source (magnetospheric currents). 
Comparing the results from the \textit{reduced} model (model without the ionosphere) to the results we obtained using the complete model with the ionosphere provides a data-based criterion for justifying the addition of the ionospheric component into our model.

\begin{figure}
    \centering
    \includegraphics[width=\linewidth]{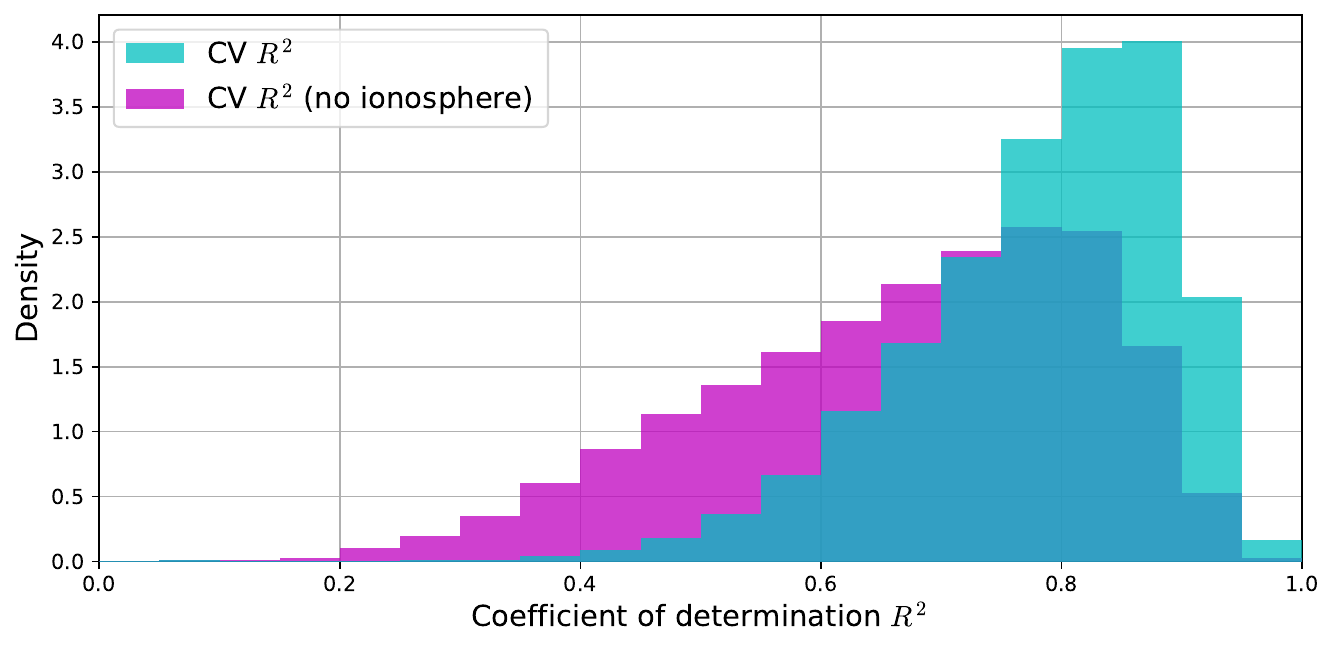}
    \caption{Histogram of the coefficient of determination $R^2$ based on the $5$-fold cross-validation.}
    \label{fig:R2-CV-bilayer-hist}
\end{figure}

Naturally, introducing an ionospheric field is able to explain more variance in the data simply by expanding the parameter space. The data fit alone is, hence, not a good indicator of the quality of a model, as the model can also achieve lower misfits by fitting noise.
To objectively compare the quality of the model and rule out the possibility of overfitting, we once again employ the CV approach.
Using a 5-fold CV, we see that there is a drastic improvement in the CV averaged $R^2$ score (Fig. \ref{fig:R2-CV-bilayer-hist}) when an ionosphere layer is included.
The parametrisation with an ionosphere is thus not only better at explaining the data variance overall, but can explain the unseen observations better. This strongly indicates that the model parametrisation with an ionosphere provides better descriptions of the geomagnetic field.

\begin{figure}
    \centering
    \includegraphics[width=\linewidth]{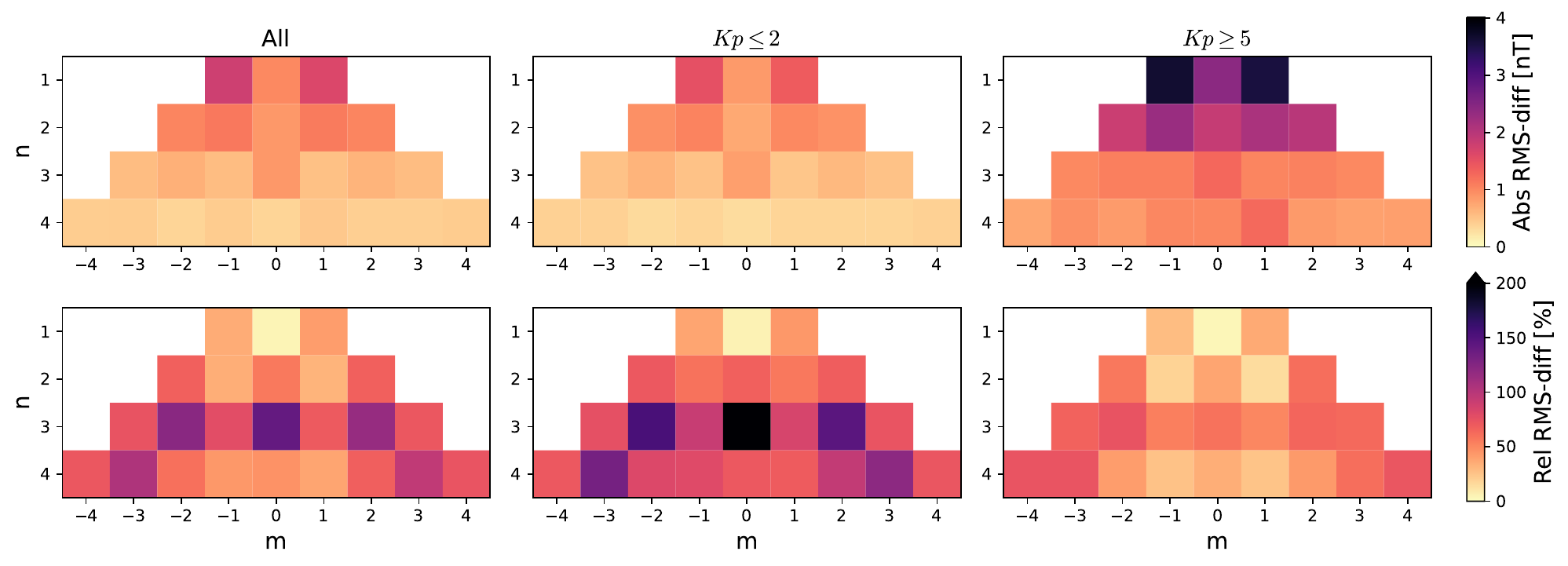}
    \caption{RMS difference for magnetospheric Gauss coefficients estimated using the models with and without the ionosphere. The rows show absolute (upper) and relative (lower) RMS differences, respectively. The columns show the RMS difference for the complete time series (left), the $Kp\leq 2$ subset (middle), and the $Kp \geq 5$ subset (right), respectively.}
    \label{fig:RMS-ionos-mag}
\end{figure}

As is evident from different $R^2$ values, introducing the ionosphere in our model parametrisation certainly changes the data fit. It is however unclear how the introduction of the ionospheric field affects the existing field estimates - that is, how the estimates of the induced and magnetospheric magnetic fields are affected by the introduction of the ionosphere. To further quantify the impact of the ionospheric layer on the magnetospheric field coefficient estimates, we turn to two statistical overviews of our field estimates. First, as an overview of the difference in the Gauss coefficients, we compute the root-mean-square (RMS) differences between the magnetospheric field coefficient time series estimated with and without the ionosphere, defined as
\begin{equation}
    \mathrm{RMSD}_{nm}^\mathrm{mag} = \left\{\begin{aligned}
        \sqrt{\frac{1}{N_t} \sum_{i=1}^{N_t} \left[q_{nm}^\mathrm{mag}(t_i) - \widetilde{q}_{nm}^\mathrm{mag}(t_i)\right]^2},\quad m \geq 0 \\ 
        \sqrt{\frac{1}{N_t} \sum_{i=1}^{N_t} \left[s_{n|m|}^\mathrm{mag}(t_i) - \widetilde{s}_{n|m|}^\mathrm{mag}(t_i)\right]^2},\quad m < 0
    \end{aligned}\right.
\end{equation}
where $N_t$ is the total number of time bins.
The absolute RMS differences, together with their relative values using the full model estimates as a reference, are shown in Fig. \ref{fig:RMS-ionos-mag} for different spherical harmonics. The notation $\widetilde{f}$ here denotes the field coefficients $f$ estimated in the reduced model, without an ionosphere. In the notation for the RMS differences, a negative $m$ is used to denote sine modes, whereas the non-negative $m$ is reserved for cosine modes.
For $q_{10}^\mathrm{mag}$, the RMS difference with and without an ionosphere is $2.4$nT during storm time ($Kp \geq 5$), and as low as $0.86$nT during quiet time ($Kp \leq 2$). Due to the strong magnetospheric $P_1^0$ signal, these differences translate to merely $3.8\%$ and $4.7\%$ during storm time and quiet time, respectively.
The similarity of the two model estimates also extends to other spherical harmonics, such as the $q_{21}^\mathrm{mag}$ component. For this mode, the estimates with and without an ionosphere have an RMS difference of $2.1$nT during storm time and $1.0$nT during quiet time. Although the RMS differences are always smaller during quiet times, we observe that the relative differences are typically higher (Fig. \ref{fig:RMS-ionos-mag}), due to the overall weaker magnetospheric field and greater contributions from the ionospheric field.

Next, we turn to the distribution of the Gauss coefficient pair estimated with and without an ionosphere. In other words, we are interested in the distribution of vectors $(f_{nm}, \Tilde{f}_{nm})$, where $f$ denotes a Gauss coefficient estimated using our complete model (\refmod), and $\Tilde{f}$ again denotes the same quantity estimated using the reduced model without the ionosphere.

\begin{figure}
    \centering
    \includegraphics[width=\linewidth]{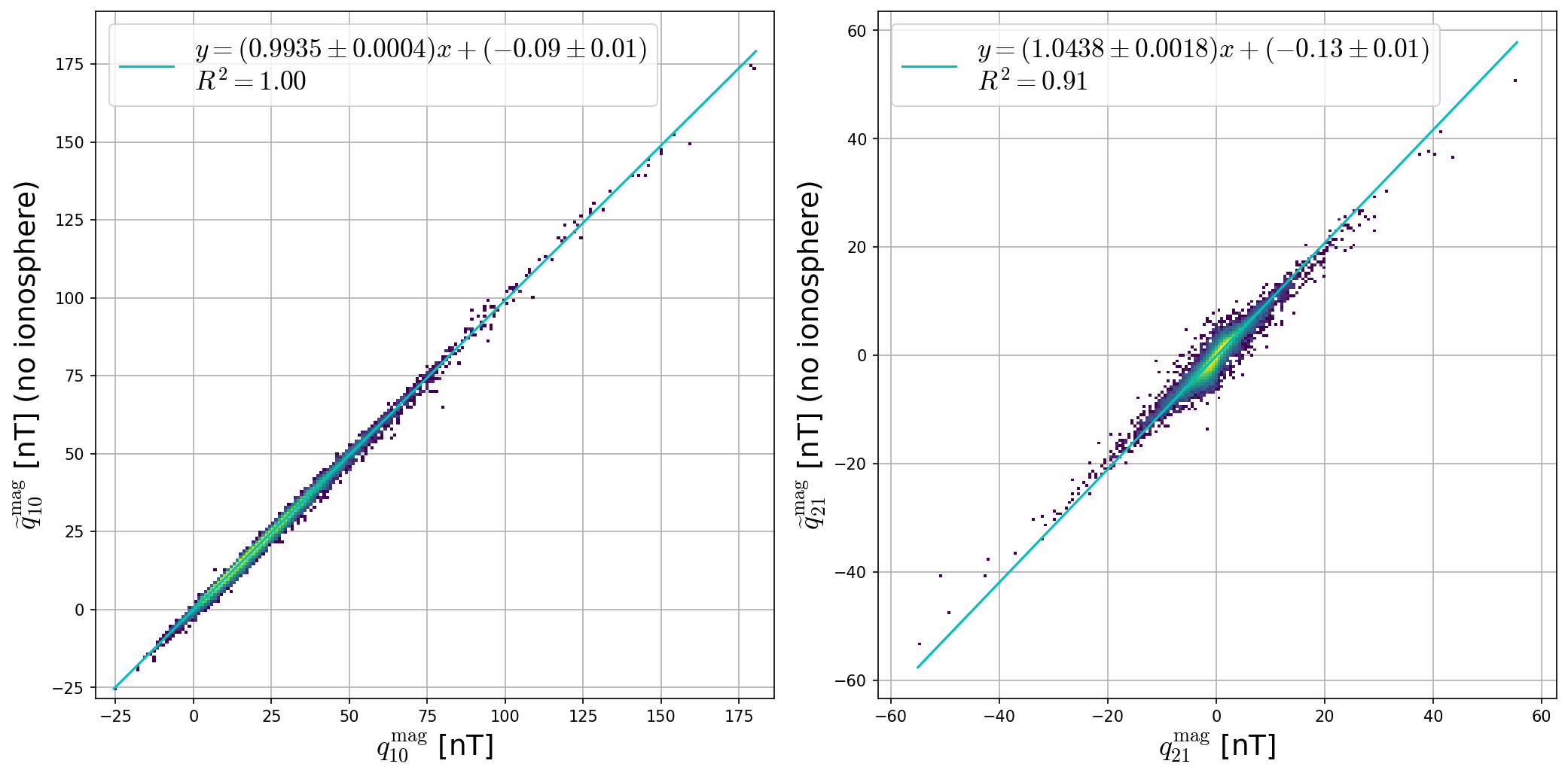}
    \caption{2-D histogram showing the distribution of the magnetospheric field Gauss coefficients estimated with and without the ionosphere layer in the model. The left panel shows the distribution for the $P_1^0$ mode, and the right panel shows the distribution for the $P_2^1$ mode. Lines show a linear fit to the distribution of points.}
    \label{fig:linfit-mag}
\end{figure}

If the ionospheric field does not strongly affect the estimate of the Gauss coefficient $f$, then the distribution of $(f_{nm}, \Tilde{f}_{nm})$ would closely follow the straight line $\Tilde{f}_{nm} = f_{nm}$. In other words, a linear model in $f_{nm}$ with a slope close to $1$ should explain most of the variance in $\Tilde{f}_{nm}$. We see that such a linear model is indeed observed for $q_{10}^\mathrm{mag}$, the first zonal harmonic of the magnetospheric field estimate (Fig. \ref{fig:linfit-mag} left panel). The correlation between the two magnetospheric field estimates deteriorates as the SH degree goes higher, e.g. for $q_{21}^\mathrm{mag}$, as shown in the right panel of Fig. \ref{fig:linfit-mag}. However, while there is a considerably larger variance and deviation from the $\Tilde{q}_{21}^\mathrm{mag} = q_{21}^\mathrm{mag}$ line, the distribution is overall largely unimodal and follows a weaker linear trend. In summary, for the joint ground observatory - satellite dataset, the introduction of the ionosphere to the model does not appear to affect strongly the estimates for the field of magnetospheric origin.

\begin{figure}
    \centering
    \includegraphics[width=\linewidth]{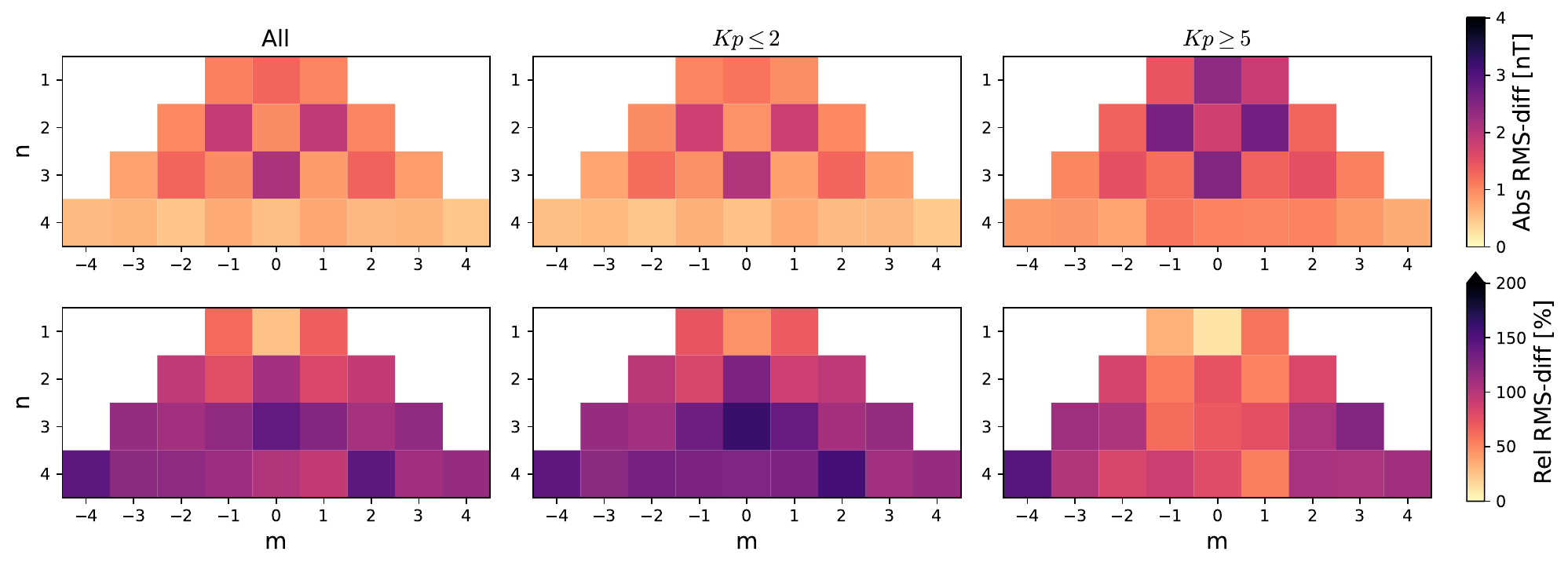}
    \caption{RMS difference for Gauss coefficients of the internally induced field estimated using the models with and without the ionosphere. The rows and columns are the same as in Fig. \ref{fig:RMS-ionos-mag}.}
    \label{fig:RMS-ionos-int}
\end{figure}

The induced field estimates, on the other hand, are much more susceptible to whether or not the ionosphere is co-modelled. This sensitivity is clearly shown in the RMS differences between the estimated made with and without the ionosphere, summarised in Fig. \ref{fig:RMS-ionos-int}. Although the absolute differences are only slightly greater than in the magnetospheric coefficients, the relatively differences are much larger. During magnetic quiet time, where the ionospheric signal is significant in proportion to the magnetosphere, the induced field coefficients are typically over $80\%$ difference for degrees $n\geq 2$, rendering the estimates with and without an ionosphere drastically different. Even the most coherent mode, $g_{10}^\mathrm{int}$, shows $> 40\%$ difference between the models with and without an ionosphere during quiet times.

\begin{figure}
    \centering
    \includegraphics[width=\linewidth]{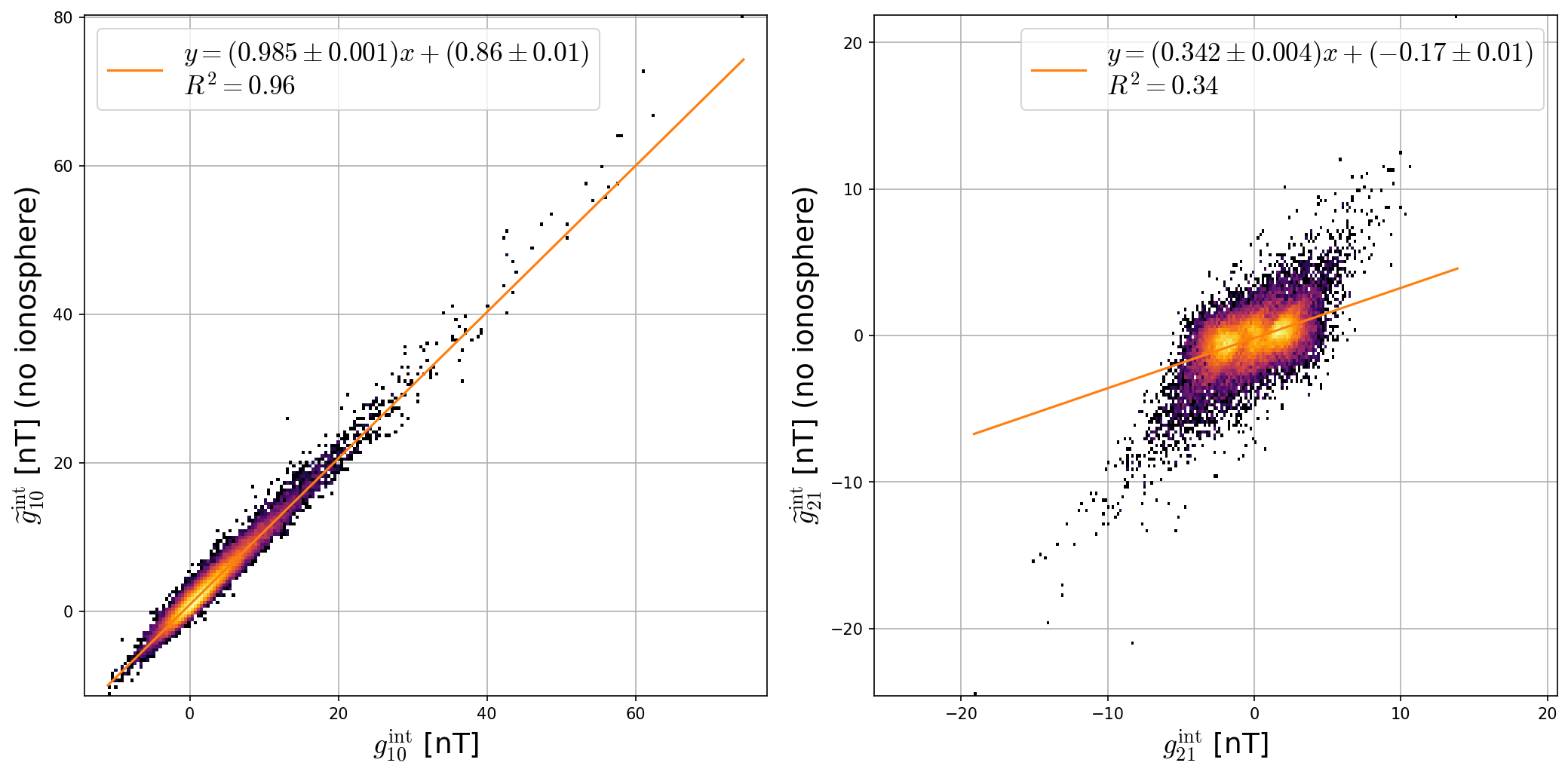}
    \caption{2-D histogram showing the distribution of the induced field Gauss coefficients estimated with and without ionosphere. The left panel shows the distribution for the $P_1^0$ mode, and the right panel shows the distribution for the $P_2^1$ mode. Lines show the linear fit to the distribution of points.}
    \label{fig:linfit-int}
\end{figure}

In the $g_{10}^\mathrm{int}$ mode, the difference between the two model estimates can be seen as the increased variability of data points around a linear trend (Fig. \ref{fig:linfit-int} left panel), compared to the same spatial mode for the magnetospheric field (Fig. \ref{fig:linfit-mag} left panel). The linear model completely fails for spatial modes such as $g_{21}^\mathrm{int}$, where the $R^2$ score decreases to a mere $0.34$. Visualisation of its distribution shows that the density is at least bimodal in the $(g_{21}^\mathrm{int}, \Tilde{g}_{21}^\mathrm{int})$ space. This is suggestive of a complex contribution to the difference in the estimated time series that cannot be described by a simple scaling in the amplitude, but must also include other effects, such as phase shifts. The phase content of the spectrum would therefore be substantially different from one estimate to another, which in turn yields considerable differences in the transfer function estimates, as will be seen in Section \ref{sec:discussions}. In summary, we observe that the introduction of the ionosphere to the model has a considerable effect on the induced field estimate in both the amplitude and phase, especially when the ionospheric field is strong.

\subsection{Time series of external and induced coefficients}\label{sec:result-tseries}

\begin{figure}
    \centering
    \begin{subfigure}[t]{\textwidth}
        \centering
        \includegraphics[width=\linewidth]{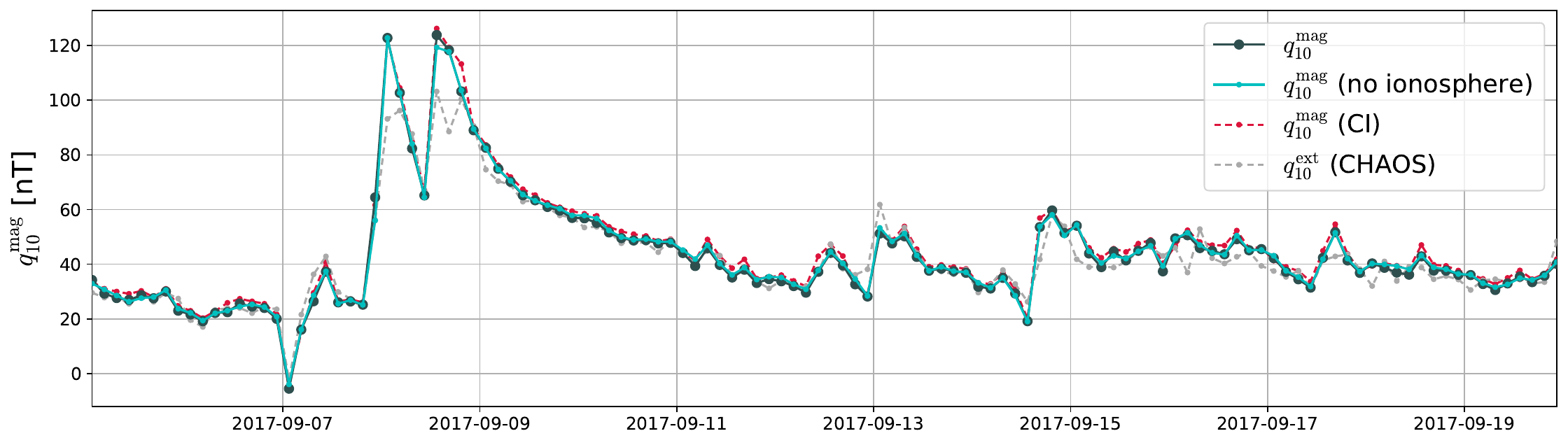}
    \end{subfigure}
    \begin{subfigure}[t]{\textwidth}
        \centering
        \includegraphics[width=\linewidth]{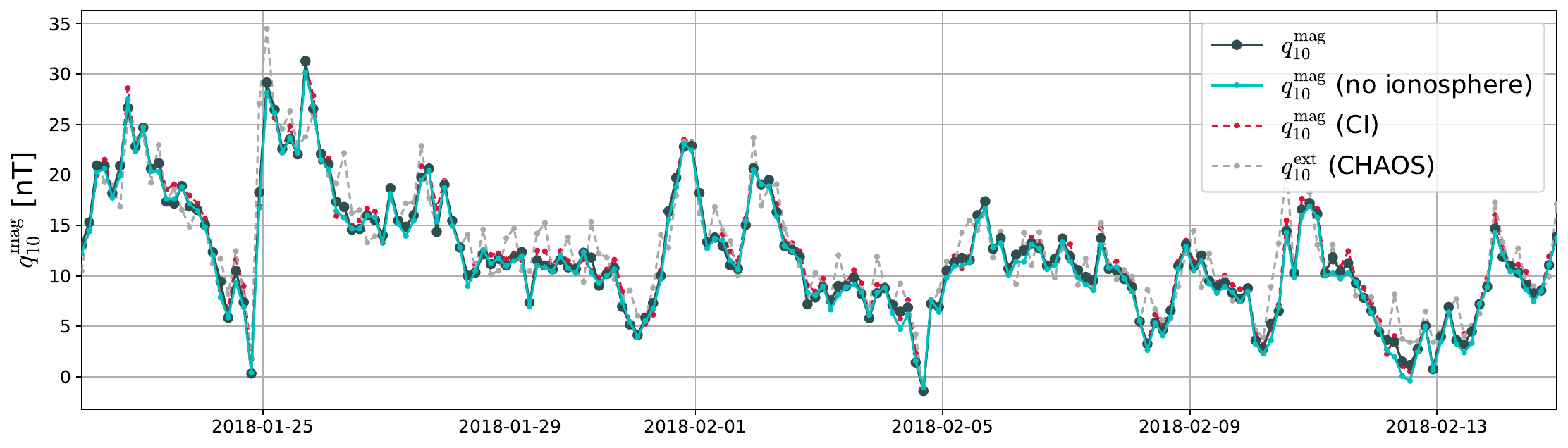}
    \end{subfigure}
    \caption{Time series of the magnetospheric Gauss coefficient $q_{10}^\mathrm{mag}(t)$ around the 2017 September geomagnetic storm (upper panel) and during magnetic quiet time, in January / February 2018 (lower panel). The plots show our model estimates (dark green lines), the estimates with no ionosphere parametrisation (cyan), the estimates from the CI model (red dashed lines) and estimates from the CHAOS-7 external field model (light gray dashed lines).}
    \label{fig:shc-tseries-mag-q10}
\end{figure}

\begin{figure}
    \centering
    \begin{subfigure}[t]{\textwidth}
        \centering
        \includegraphics[width=\linewidth]{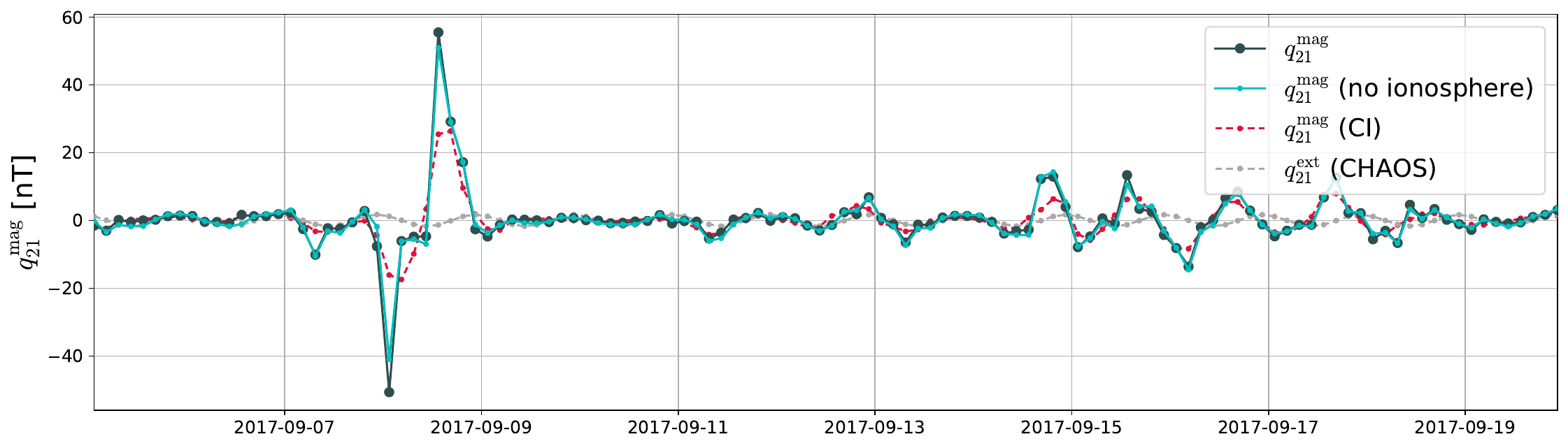}
    \end{subfigure}
    \begin{subfigure}[t]{\textwidth}
        \centering
        \includegraphics[width=\linewidth]{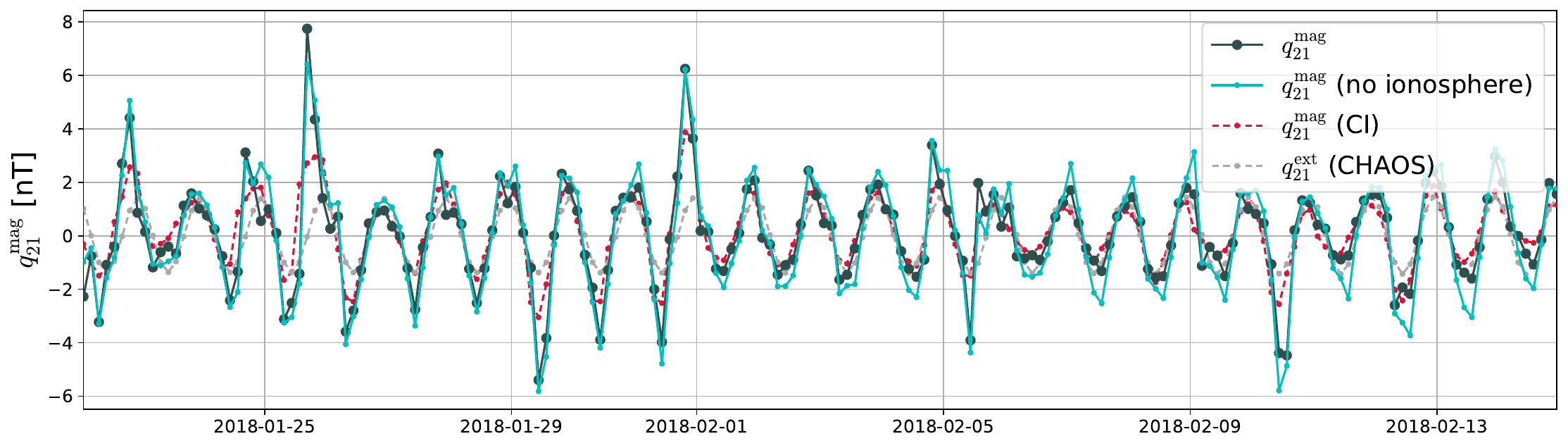}
    \end{subfigure}
    \caption{Time series of the magnetospheric Gauss coefficient $q_{21}^\mathrm{mag}(t)$ around the 2017 September geomagnetic storm (upper panel) and during magnetic quiet time, in January / February 2018 (lower panel).}
    \label{fig:shc-tseries-mag-q21}
\end{figure}

We have seen in the previous section the time-cumulative distribution of the Gauss coefficients estimated with and without the ionosphere layer. Now we investigate these coefficients further by disentangling the distribution into time series.
The time series of the estimated magnetospheric field coefficients generally show relatively small differences between different models (Figs. \ref{fig:shc-tseries-mag-q10}-\ref{fig:shc-tseries-mag-q21}), consistent with our observation on their distributions. Fig. \ref{fig:shc-tseries-mag-q10} shows time series of the estimated first zonal harmonic of the magnetospheric field ($q_{10}^\mathrm{mag}$) for a period with a geomagnetic storm and a quiet period. 
The estimates from the complete and reduced models almost coincide with each other, consistent with the low RMS difference (Fig. \ref{fig:RMS-ionos-mag}) and almost the perfect correlation between the two (Fig. \ref{fig:linfit-mag}). The close match between the model estimates with and without the ionosphere can also be seen in the $P_2^1$ mode (Fig. \ref{fig:shc-tseries-mag-q21}), where the two models exhibit nearly coincident peaks during the storm, and in-phase diurnal oscillations at quiet times. Discrepancies are however visible in the amplitudes of the $P_2^1$ estimates, especially during quiet times, which, combined with the lower energy, contribute to the aforementioned increased relative RMS difference (Fig. \ref{fig:RMS-ionos-mag}).

In addition to the same model without the ionosphere, we also compare our estimates with the external field model from the CI model \citep{sabaka_cm6_2020} and the CHAOS-7 model \citep{finlay_chaos-7_2020}. All model estimates for $q_{10}^\mathrm{mag}$ capture well the two successive geomagnetic storms (as two peaks in the time series) on 8 September 2017 \citep{tassev_analysis_2017, zhang_influence_magstorm_2019}, and are in overall good agreement during recovery phases of the geomagnetic storm and during magnetic quiet periods (Fig. \ref{fig:shc-tseries-mag-q10}). 
Discrepancies can be observed in the peak amplitude of the field coefficient in the geomagnetic storm, as well as in the oscillations in magnetic quiet times. This is anticipated considering the difference in the data used, and the different parametrisation of the external field in different models. We note that CHAOS model estimate in particular shows quasi-diurnal oscillations in the geomagnetic quiet times, which are not visible in our model estimates or CI estimates.
The discrepancies are much large for higher degrees, such as in the $q_{21}^\mathrm{mag}$ estimates (Fig. \ref{fig:shc-tseries-mag-q21}). The CHAOS external field estimation is primarily intended for quiet time on the dark side, and only allows for modes that are stationary in either the solar magnetic coordinates (SM) or the geocentric solar magnetospheric coordinates (GSM) beyond spherical harmonic degree one. As a result, it does not describe the fluctuations in the external field beyond degree one, such as the high-amplitude fluctuations for $q_{21}^\mathrm{mag}$ during a magnetic storm, which are clearly visible in our model and the CI model estimates. Compared to the CI model estimates, our model yields estimates of $q_{21}^\mathrm{mag}$ that have a larger dynamic range, especially during geomagnetic storms.

\begin{figure}
    \centering
    \begin{subfigure}[t]{\textwidth}
        \centering
        \includegraphics[width=\linewidth]{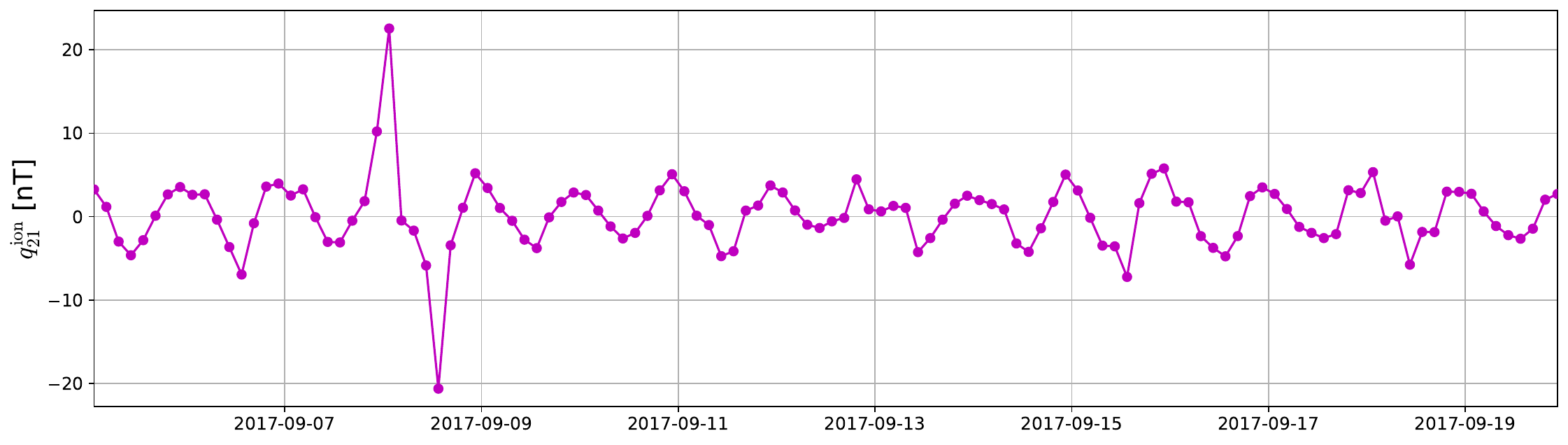}
    \end{subfigure}
    \begin{subfigure}[t]{\textwidth}
        \centering
        \includegraphics[width=\linewidth]{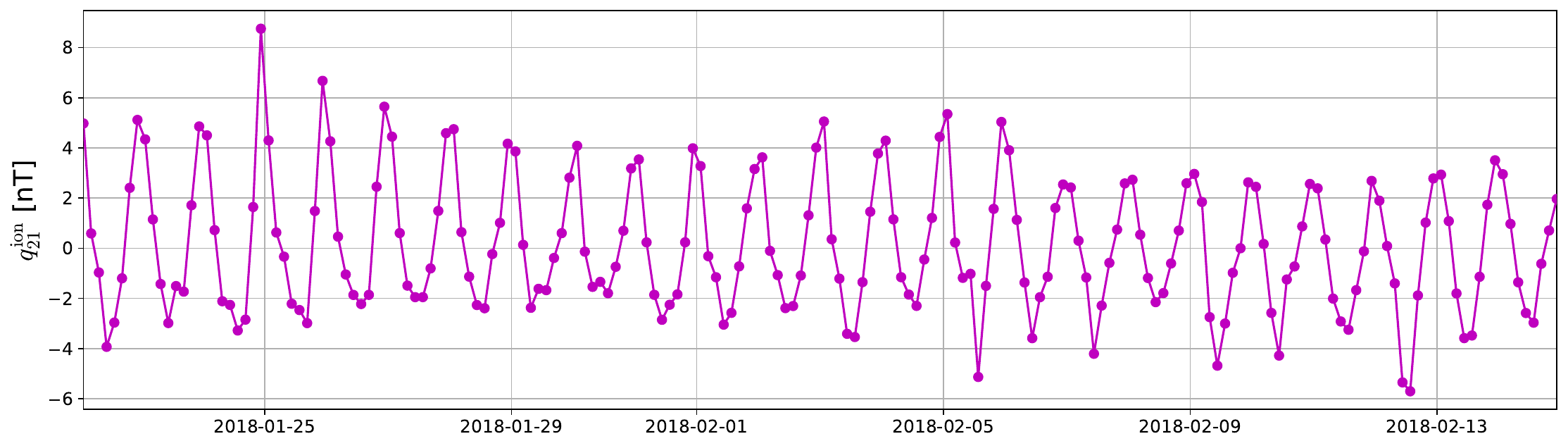}
    \end{subfigure}
    \caption{Time series of the ionospheric Gauss coefficient $q_{21}^\mathrm{ion}(t)$ around the 2017 September geomagnetic storm (upper panel) and during magnetic quiet time, in January / February 2018 (lower panel).}
    \label{fig:shc-tseries-ion-q21}
\end{figure}

In addition to the magnetospheric field, our model also simultaneously estimates the ionospheric field Gauss coefficients. The $q_{21}^\mathrm{ion}$ spatial mode (Fig. \ref{fig:shc-tseries-ion-q21}) exhibits a large amplitude in the diurnal band in the geographic frame, reflecting a standing mode in the solar-fixed frame, an important constituent of the \textit{Sq variation} \citep{malin_worldwide_1973,Schmucker1999}. The diurnal variation of $q_{21}^\mathrm{ion}$ is very clear in the reconstructed time series not only for magnetic quiet times (Fig. \ref{fig:shc-tseries-ion-q21} lower panel), but around the geomagnetic storms as well (Fig. \ref{fig:shc-tseries-ion-q21} upper panel). This suggests that the Sq variation is a robust feature that is modulated by the magnetosphere. For instance, Sq dominant structure quickly recovers after the storm's peak, although its amplitude is dwarfed by a much larger $q_{10}^\mathrm{mag}$ magnetospheric signal (Fig. \ref{fig:shc-tseries-mag-q10}). However, we observe a considerable deviation from the quiet time daily variation during the storm's main phase (Fig. \ref{fig:shc-tseries-ion-q21} upper panel). Interestingly, this deviation is anti-correlated with the magnetospheric mode $q_{21}^\mathrm{mag}$ (Fig. \ref{fig:shc-tseries-mag-q21}). This anti-correlation may represent an interaction between the ionosphere and magnetosphere current systems during the main phase of a geomagnetic storm.

\begin{figure}
    \centering
    \begin{subfigure}[t]{\textwidth}
        \centering
        \includegraphics[width=\linewidth]{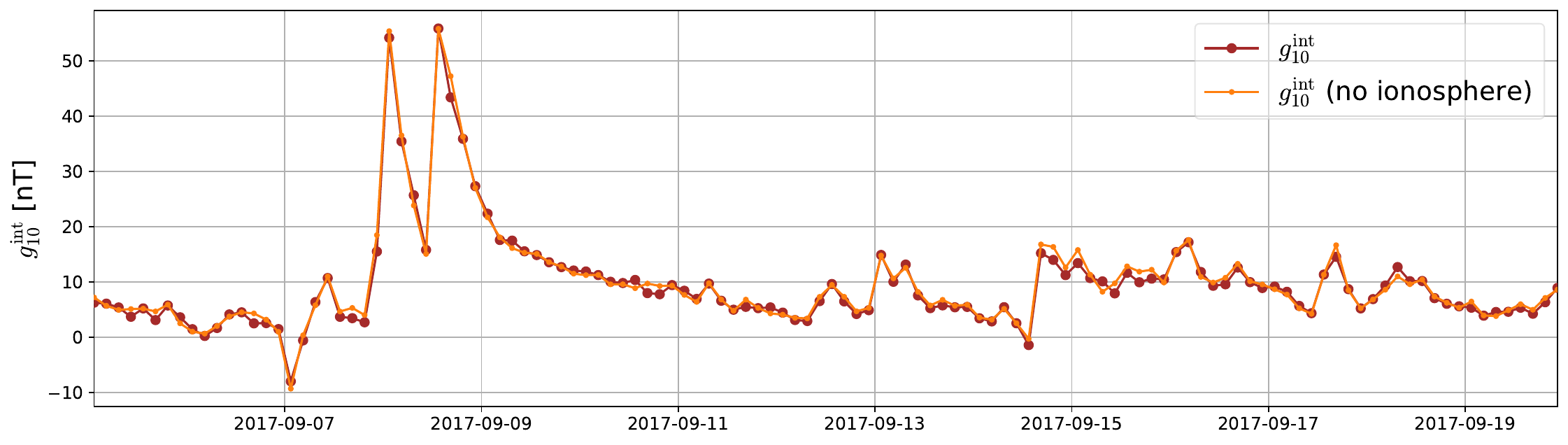}
    \end{subfigure}
    \begin{subfigure}[t]{\textwidth}
        \centering
        \includegraphics[width=\linewidth]{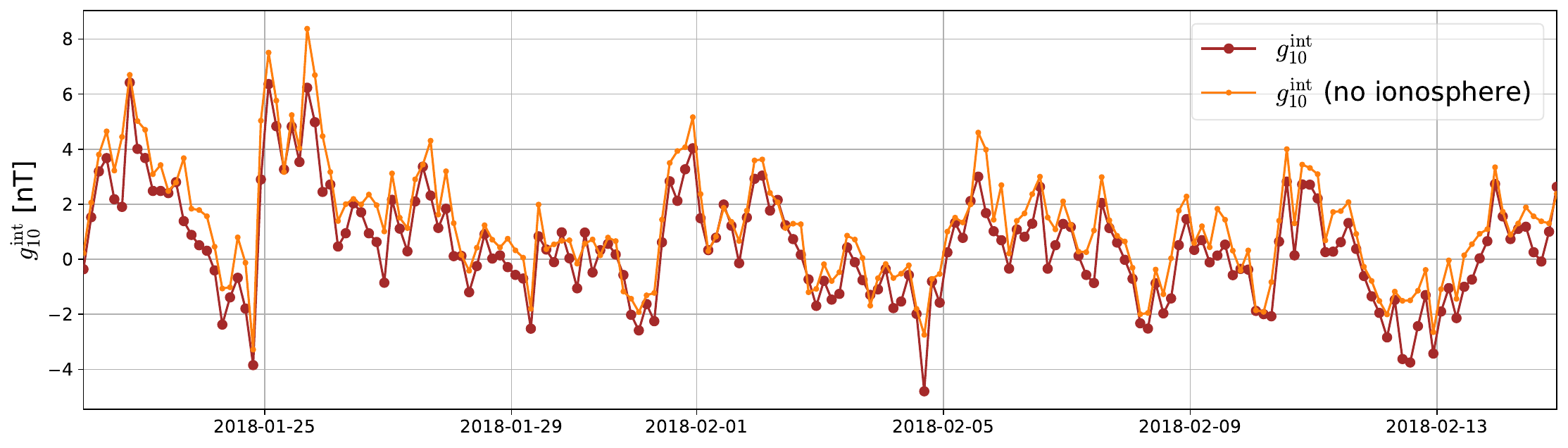}
    \end{subfigure}
    \caption{Time series of the internally induced Gauss coefficient $g_{10}^\mathrm{int}(t)$ around the 2017 September geomagnetic storm (upper panel) and during magnetic quiet time, in January / February 2018 (lower panel). The plots show our full model estimates (brown lines) and the estimates with no ionosphere parametrisation (orange).}
    \label{fig:shc-tseries-int-g10}
\end{figure}

\begin{figure}
    \centering
    \begin{subfigure}[t]{\textwidth}
        \centering
        \includegraphics[width=\linewidth]{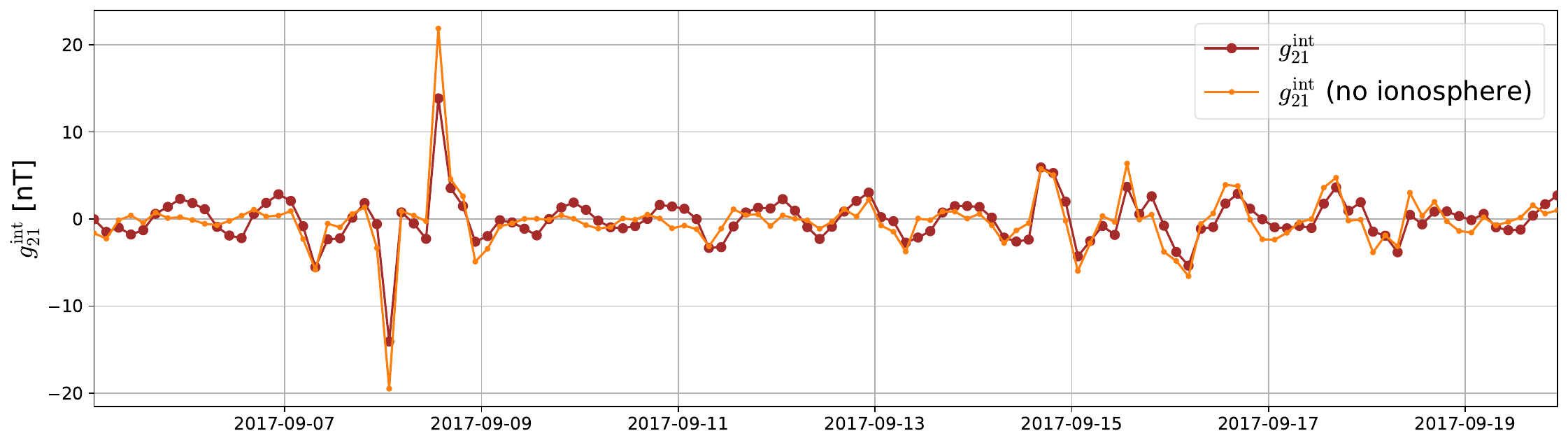}
    \end{subfigure}
    \begin{subfigure}[t]{\textwidth}
        \centering
        \includegraphics[width=\linewidth]{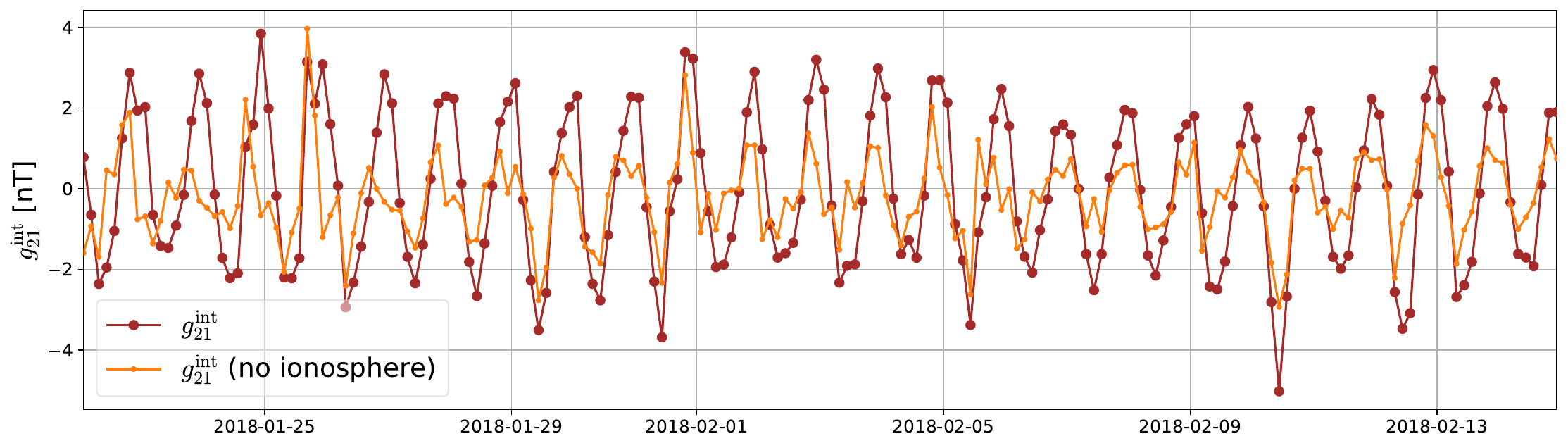}
    \end{subfigure}
    \caption{Time series of the internally induced Gauss coefficient $g_{21}^\mathrm{int}(t)$ around the 2017 September geomagnetic storm (upper panel) and during magnetic quiet time, in January / February 2018 (lower panel).}
    \label{fig:shc-tseries-int-g21}
\end{figure}

\begin{figure}
    \centering
    \includegraphics[width=\linewidth]{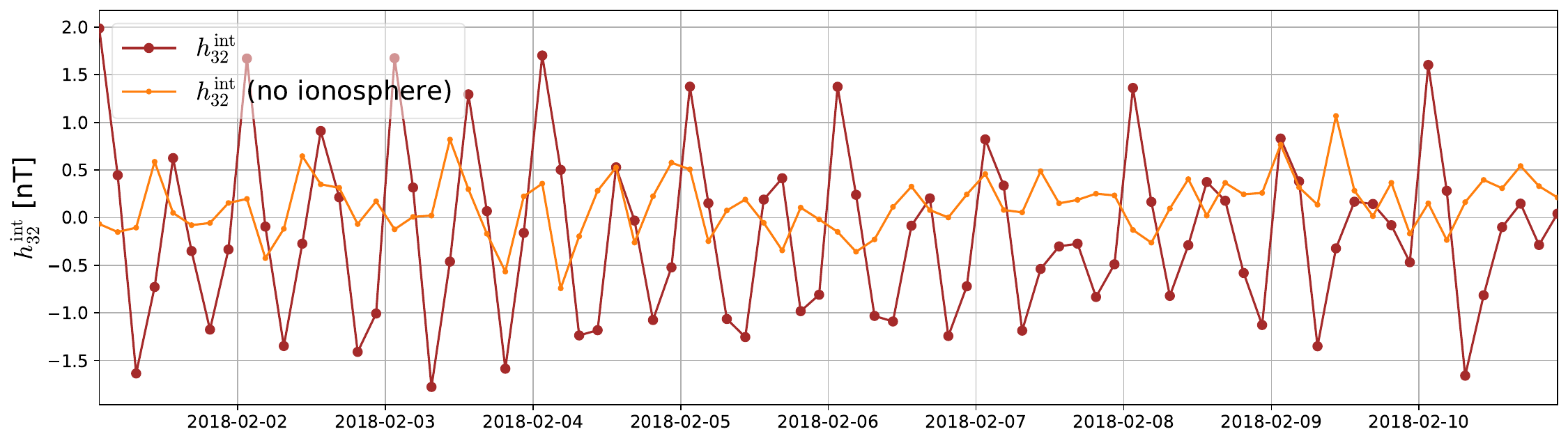}
    \caption{Time series of the internally induced Gauss coefficient $h_{32}^\mathrm{int}(t)$ during magnetic quiet time, in February 2018.}
    \label{fig:shc-tseries-int-h32}
\end{figure}

As has been observed in Figs. \ref{fig:RMS-ionos-int}-\ref{fig:linfit-int}, the induced field estimate is more susceptible to the presence of the ionospheric layer to the model. This is confirmed in the time series in Figs. \ref{fig:shc-tseries-int-g10} and \ref{fig:shc-tseries-int-g21}. 
The discrepancy is especially pronounced in spatial modes where the ionosphere is comparatively energetic, e.g. the $P_2^1$ SH mode. During the magnetic quiet time (cf. Fig. \ref{fig:shc-tseries-int-g21}, lower panel), the amplitude of the induced field estimate in the reduced model is less than 50\% of that estimated using the full model with an ionosphere. 
We interpret this systematically smaller amplitude as an imprint of the missing ionospheric field on the induced field. When the ionosphere is neglected, the internal contribution to the satellite data which should have been attributed to the ionosphere is forced into the induced coefficients.
This, in turn, may weaken the induction signal when the induced field is correlated to the ionospheric signal (Eq. \ref{eqn:model-V-repr-sat}). The difference in the model estimates with and without the ionosphere is not only in the amplitude, but also in the phase. The phase change can already be observed in $g_{21}^\mathrm{int}$, but it is even more pronounced in higher SH degrees. The semi-diurnal oscillation of $h_{32}^\mathrm{int}$, for instance, appears to be $\sim 3$ hours ahead in the model without an ionosphere compared to the full model estimate (Fig. \ref{fig:shc-tseries-int-h32}).

\subsection{Fourier spectrum of external and induced fields}\label{sec:results-spectrum}

Our 10-year time series of the SH coefficients has a theoretical bandwidth of over four decades, i.e. $f_\mathrm{Nyq}/\Delta f > 10^4$, where $f_\mathrm{Nyq}$ is the Nyquist frequency and $\Delta f$ is the frequency resolution. As such, it covers the entire period range from sub-diurnal to inter-annual periods.
In this section, we explore the frequency content of the estimated fields by deriving their amplitude spectra. The spectra are computed directly as discrete Fourier transforms of the time series of the Gauss coefficients, using a forward normalisation that preserves the time-domain amplitude of a harmonic signal in the amplitude spectrum (i.e. a sinusoidal signal with 1 nT amplitude will result in a 1 nT peak at the corresponding frequency in the amplitude spectrum).

\begin{figure}
    \centering
    \begin{subfigure}[t]{\textwidth}
        \centering
        \includegraphics[width=\linewidth]{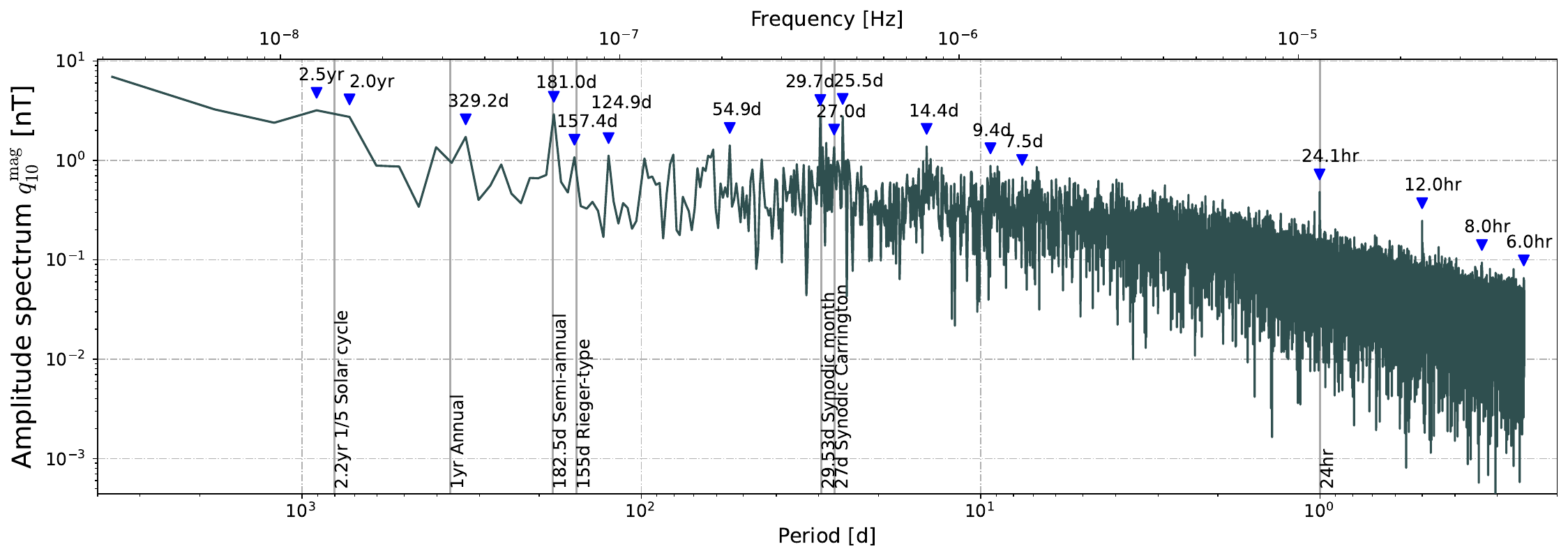}
    \end{subfigure}
    \begin{subfigure}[t]{\textwidth}
        \centering
        \includegraphics[width=\linewidth]{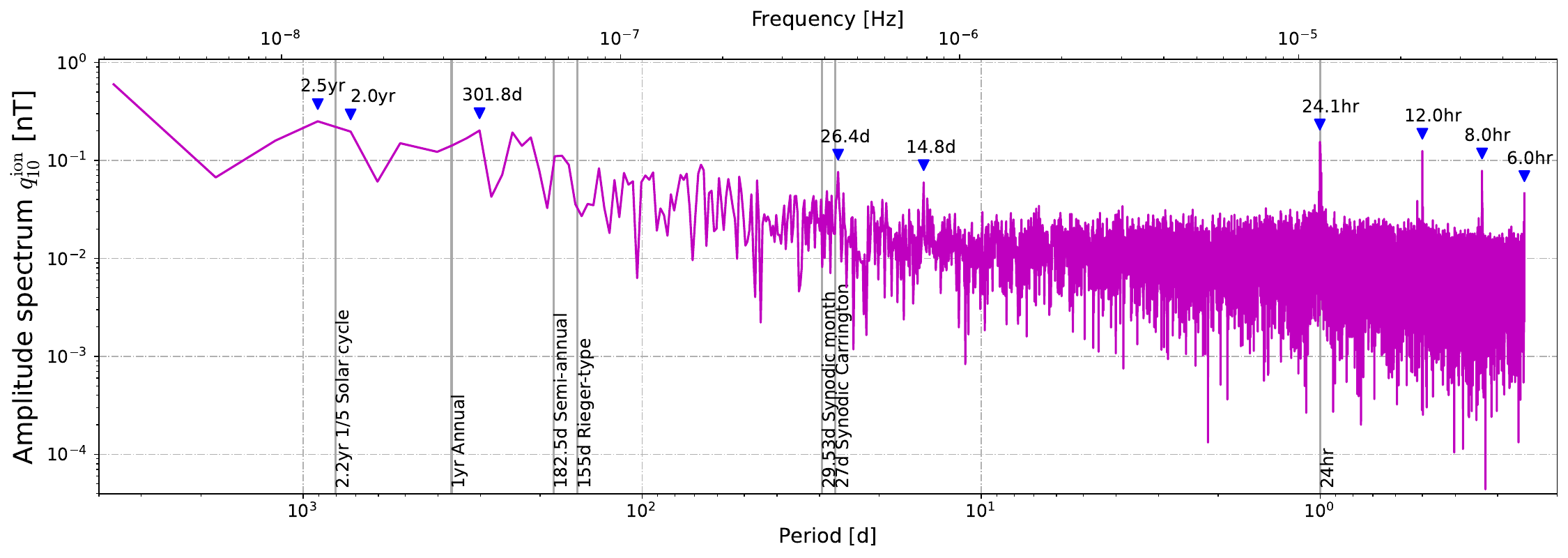}
    \end{subfigure}
    \begin{subfigure}[t]{\textwidth}
        \centering
        \includegraphics[width=\linewidth]{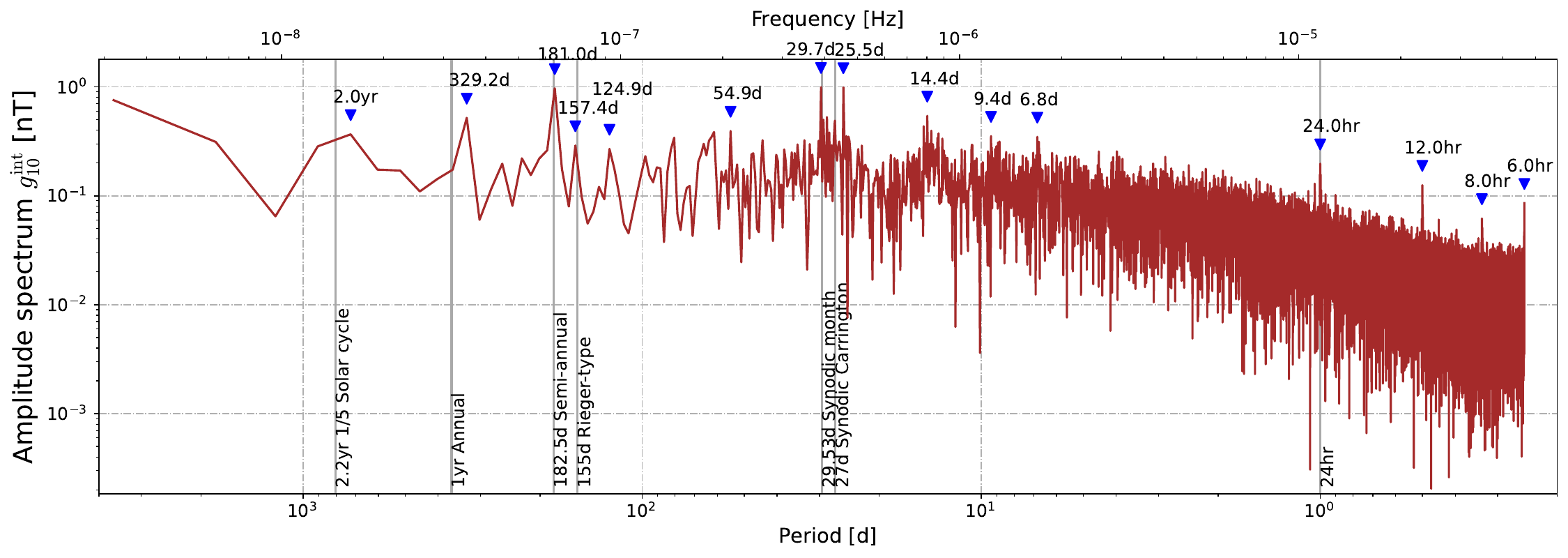}
    \end{subfigure}
    \caption{Amplitude spectra of the magnetospheric coefficient $q_{10}^\mathrm{mag}$ (upper panel), the ionospheric coefficient $q_{10}^\mathrm{ion}$ coefficient (middle panel) and the internally induced coefficient $g_{10}^\mathrm{int}$ (lower panel). Spectral peaks observed in the spectra are marked with blue triangles, with their peak periods annotated. Selected theoretical periods of the system are plotted as gray vertical lines for reference.}
    \label{fig:spec-q10-mi}
\end{figure}

The first zonal harmonic of the magnetospheric magnetic field, largely attributed to the ring current, is the most energetic contribution to the magnetic field of the external origin at periods longer than one day. The power of the spectrum for the first zonal harmonic quickly rolls off at periods shorter than one day, diurnal frequency bands, while exhibiting strong amplitude at longer periods (Fig. \ref{fig:spec-q10-mi}, upper panel). 
At longest periods, a prominent spectral peak can be observed between $2.0$ and $2.5$ years. This biennial oscillation has been previously observed in the solar magnetic field, and has been attributed to a harmonic of the solar cycle \citep{knaack_spherical_2005}. This notable feature can be similarly observed in the ionospheric signal (Fig. \ref{fig:spec-q10-mi} lower panel) and in other degree-one modes (Fig. \ref{fig:spec-q11-mi}).
The second, rather sharp peak in the magnetospheric field spectrum is at the semi-annual period (2 cycles per year). Since our estimation is conducted in the Earth-fixed frame, the semi-annual peak is likely associated with the axial tilt of the Earth and the associated seasonal variability in the Earth's frame \citep{maus2005, Olsen2005, Balasis2006}.
A secondary, wider peak can be observed at the annual period (1 cycle per year). This can be a signature of the tail current in the magnetosphere \citep{maus2005}. As the $q_{10}^\mathrm{mag}$ component experiences semi-annual changes in the solar radiation, it is further modulated by the change in the Interplanetary Magnetic Field.
Both annual and semi-annual peaks are observed not only in the first zonal harmonic component, but also in other dipole components of the magnetospheric field (Fig. \ref{fig:spec-q11-mi} upper panel). This can result from the transformation of the quasi-static magnetospheric field in GSM/SM frames to the Earth-fixed frame. In contrast, these peaks are weaker or not visible in the ionospheric spectrum (Figs. \ref{fig:spec-q10-mi} and \ref{fig:spec-q11-mi}, lower panels), suggesting that ionospheric dynamics at $n=1$ is not as strongly modulated by the orbit-related periodicities.
Note that this observation only holds for the $n=1$ spherical harmonic signals. The annual, semi-annual even quarterly peaks are clearly visible in both magnetospheric and ionospheric Gauss coefficients of $n>1$ (Figs. \ref{fig:spec-s21-mi},\ref{fig:spec-q32-mi}).

Further, we observe spectral peaks at quasi-monthly periods in the $q_{10}^\mathrm{mag}$ spectra (Fig. \ref{fig:spec-q10-mi} upper panel). These include peaks at $29.7\pm 0.2$ days and $25.5\pm 0.2$ days, and a smaller peak at $27.0\pm 0.2$ days. The uncertainty reported here is the resolution at the given peaks. This set of peaks is a robust feature that also appears in other degree-one mode spectra of the magnetospheric coefficients (Fig. \ref{fig:spec-q11-mi} upper panel). In contrast, these peaks are weaker (Fig. \ref{fig:spec-q10-mi} lower panel) or absent (Fig. \ref{fig:spec-q11-mi} lower panel) in ionospheric coefficients. We associate these with the well-known $\sim 27$-day harmonic in terrestrial magnetic field \citep{bartels_terrestrial-magnetic_1932}, stemming from mechanisms associated with the solar rotation, which has a synodic Carrington period of $\sim 27.7$ days \citep{lockwood_reconstruction_2024}. The strong peaks at $25.5$ and $29.7$ days are likely due to the variability in the solar rotation, and are consistent with the analysis of the Interplanetary Magnetic Field (IMF) in the solar cycle 24, which shows strong spectral peaks at $26$ and $30$ days \citep{chowdhury_short-term_2015}.

Additional spectral peaks are also observed at multiples or submultiples of the Carrington period. The quasi-fortnightly oscillation, observed at $14.4$ days in $q_{10}^{\mathrm{mag}}$ and $q_{11}^{\mathrm{mag}}$ and at $14.8$ days in $q_{10}^\mathrm{ion}$, is likely a 2nd harmonic of the $\sim 27$ day signal. At periods longer than the Carrington period, spectral peaks can be observed at $54.9\pm 0.8$ days (Fig. \ref{fig:spec-q10-mi} $q_{10}^\mathrm{mag}$, Fig. \ref{fig:spec-q11-mi} $q_{11}^\mathrm{mag}$), $134 \pm 5$ days (Fig. \ref{fig:spec-q11-mi}, $q_{11}^\mathrm{mag}$, $q_{11}^\mathrm{ion}$), $157\pm 7$ days ($q_{10}^\mathrm{mag}$, ibid.) and $241\pm 17$ days ($q_{11}^\mathrm{mag}$, $q_{11}^\mathrm{ion}$, ibid.). This series of spectral peaks, termed the Rieger-type periods, was first observed by \citet{rieger_154-day_1984} in the $154$-day periodic occurrence of solar flares. The periodicity was later extended to shorter periods of $\sim 130$ days and $\sim 50$ days \citep{bai_154-day_1991}, and the series of periods have since also been observed in the IMF \citep{cane_interplanetary_1998}, Coronal Mass Ejections \citep{lou_periodicities_2003} and sunspot occurrence \citep{chowdhury_short-term_2015}. Since these peaks are located close to multiples of the solar rotation period, it has long been postulated that the signals associated with these periods are sub-harmonics of a fundamental period at the solar rotation \citep{bai_154-day_1991}. It has been proposed that the mechanism underlying this signal is linked to $r$-modes or equatorial trapped Rossby-type waves in the Sun \citep{lou_rossby-type_2000}. Although the conclusion regarding the mechanism of the Rieger-type periodicity is presently not fully established, the current evidence suggests that these spectral peaks observed in our reconstruction of the magnetospheric and ionospheric fields is likely modulated by periodicities in the Sun's activity.

\begin{figure}
    \centering
    \begin{subfigure}[t]{\textwidth}
        \centering
        \includegraphics[width=\linewidth]{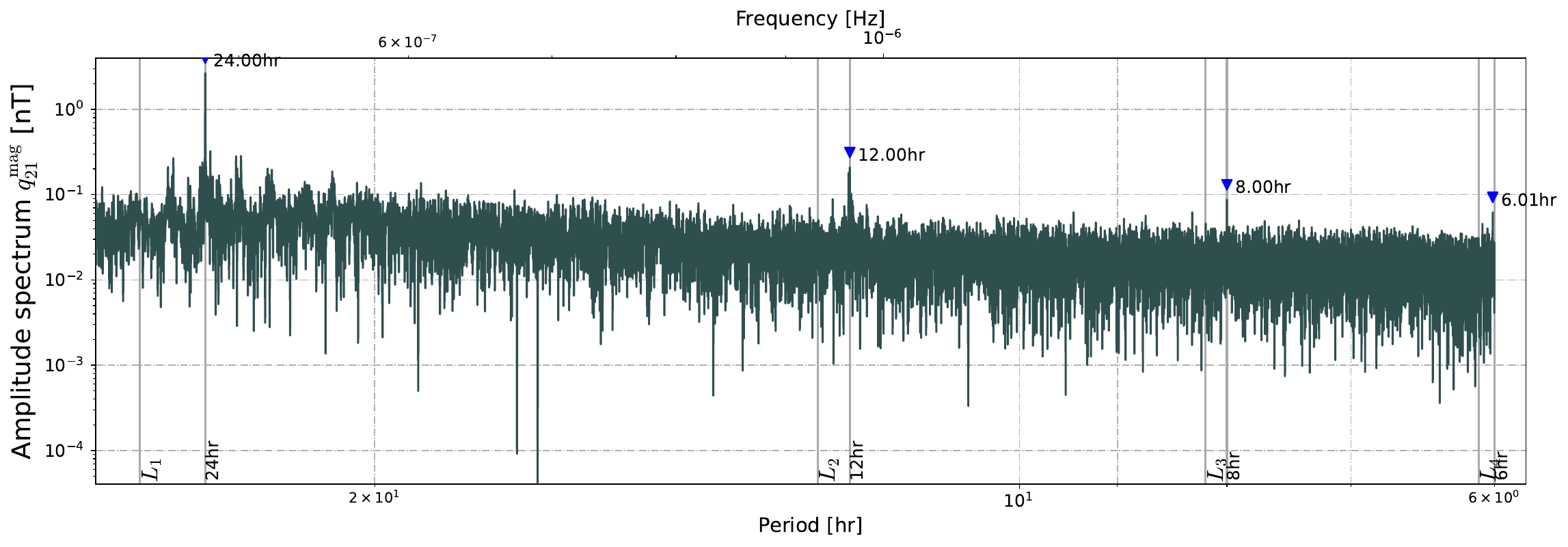}
    \end{subfigure}
    \begin{subfigure}[t]{\textwidth}
        \centering
        \includegraphics[width=\linewidth]{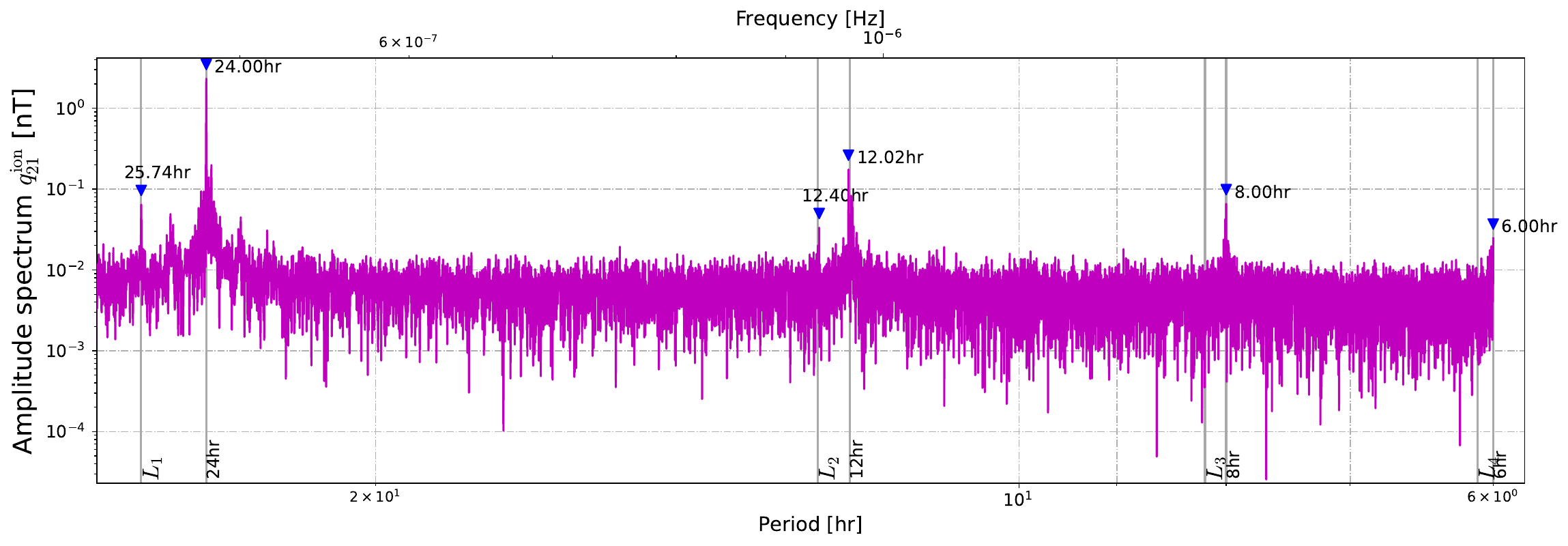}
    \end{subfigure}
    \begin{subfigure}[t]{\textwidth}
        \centering
        \includegraphics[width=\linewidth]{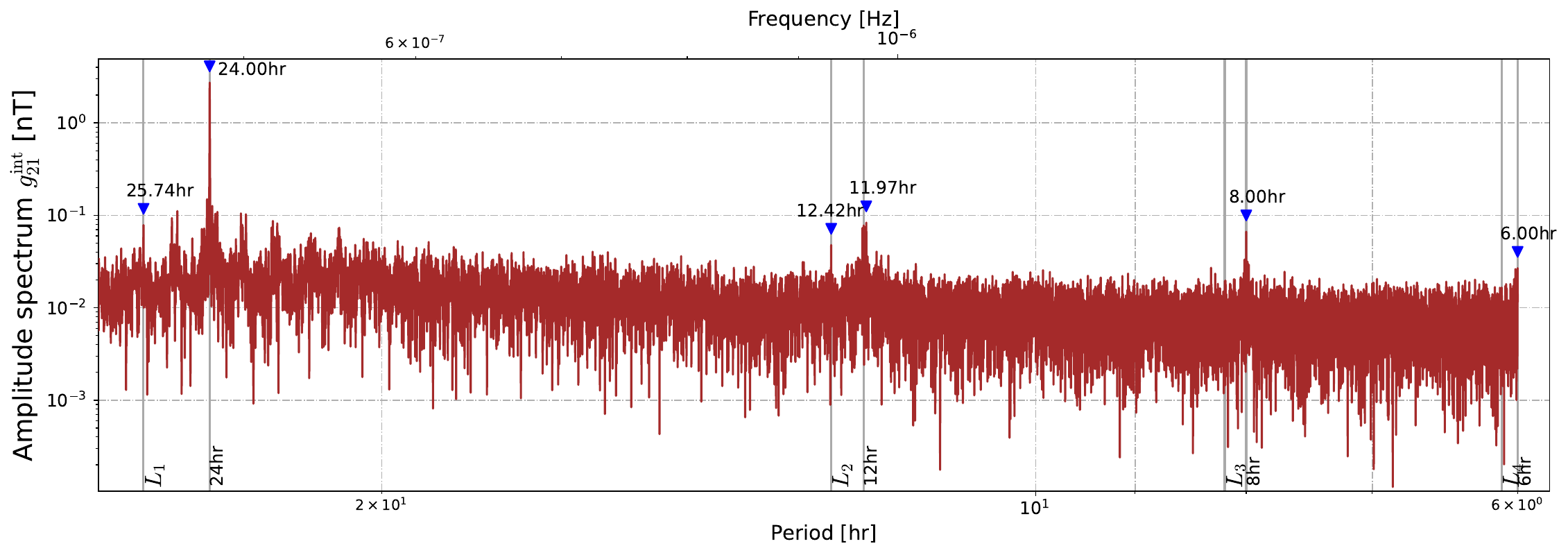}
    \end{subfigure}
    \caption{Amplitude spectra of the magnetospheric coefficient $q_{21}^\mathrm{mag}$ (upper panel), the ionospheric coefficient $q_{21}^\mathrm{ion}$ coefficient (middle panel) and the internally induced coefficient $g_{21}^\mathrm{int}$ (lower panel) around the diurnal band. Spectral peaks observed in the spectra are marked with blue triangles, with their peak periods annotated. Selected theoretical periods of the system, including the diurnal period and its harmonics, and the periods for lunar daily variations $L_p$ with $p=1,2,3,4$ are plotted as gray vertical lines for reference.}
    \label{fig:spec-q21-mi}
\end{figure}

In contrast to variations at long periods, the variations in diurnal period band are dominated by the ionospheric sources. One of the major features in the ionospheric signal in this period band is the diurnal mode and its harmonics. These are clearly visible as discrete spectral peaks at one, two, three and four cycles per day (cpd) (Fig. \ref{fig:spec-q21-mi}, lower panel). For the $P_2^1$ spatial mode, the fundamental time harmonic with a period of one day is the strongest, with an amplitude two orders of magnitudes above the average spectrum level. 
This mode, which has an azimuthal wavenumber $m=1$, and has a 1-cpd periodicity observed in the Earth-fixed geographic frame, represents a stationary $n=2$ mode in a solar-fixed frame (local-time term). This mode is predominantly influenced by the day-side structure of the Sq current system \citep{malin_worldwide_1973, winch_spherical_1981, Schmucker1999}. The same holds also for the SH coefficients $s_{32}^\mathrm{ion}$ and $s_{43}^\mathrm{ion}$ with the azimuthal wavenumbers of $m=2$ and $m=3$, respectively. They exhibit strongest spectral peaks at 2 cpd (Fig. \ref{fig:spec-s32-mi} lower panel) and 3 cpd (Fig. \ref{fig:spec-s43-mi} lower panel), respectively. These correspond to the $n=3$ and $n=4$ modes that are stationary in a solar-fixed frame. 
The diurnal periods are also visible in the spectra of magnetospheric coefficients (Figs. \ref{fig:spec-q21-mi}, \ref{fig:spec-s32-mi} and \ref{fig:spec-s43-mi} upper panels), although they have a lower signal-to-noise ratio compared to the ionosphere.

In addition to dominant solar daily variations, we also observe another set of weaker peaks at nearby periods. These periodic signals are due to lunar daily variations \citep{chapman_solar_1919, chapman1940}, hereinafter denoted by $L_p$ following the original notation by \cite{malin1970}. $L_p$ variations can be well described as harmonic signals that obey the \textit{Chapman's phase law} \citep{chapman1940, malin1970}
\begin{equation}
    L_p(t,\nu) = l_p\sin{\left( pt - 2\nu +\lambda_p\right)},
\end{equation}
where $p = 1, 2, 3 \cdots$ corresponds to diurnal, semidiurnal, terdiurnal, etc., terms with $t$ solar time in hours and $\nu$ the moon phase, measured by the hour angle between the Moon and the Sun. $l_p$ and $\lambda_p$ are the corresponding amplitude and phase values. With the Lunar synodic period of 29.53059 days, the periods of the first four lunar daily harmonics $L_p$ are 25.7435, 12.4206, 8.1847, and 6.1033 hours for $p = 1\dots 4$, respectively. These periods are marked in the power spectra of ionospheric SH coefficients in Fig. \ref{fig:spec-q21-mi} and Figs. \ref{fig:spec-s32-mi}-\ref{fig:spec-s43-mi}. Note that these peaks are not equally strong across SH coefficients owing to a non-uniform (and daily modulated) spatial structure of the E-region conductivity \citep{winch_spherical_1981}. As anticipated, peaks are strongest in the SH coefficients that correspond to local-time terms with degree $n = p + 1$ and order $m = p$ \citep{Schmucker1999}.

Although lunar daily variations are apparent in all SH coefficients describing the field of the ionospheric origin, they are absent in any of the SH coefficients describing the field of the magnetospheric origin (Figs \ref{fig:spec-q21-mi}, \ref{fig:spec-s32-mi} and \ref{fig:spec-s43-mi}). This confirms their ionospheric origin and illustrates high quality of the field separation in the present model.

The estimated internal field contains signals that originate from both the magnetospheric and ionospheric field variations. For the $P_1^0$ component, for instance, the internal coefficient $g_{10}^{\mathrm{int}}$ inherits the strong semi-annual and Rieger-type spectral peaks as well as the peak at $29.7$ days and $25.5$ days from the magnetospheric signal, whereas these peaks are absent in the ionospheric signal (Fig. \ref{fig:spec-q10-mi}). For $n>1$ in diurnal bands, the induced coefficients show clear spectral peaks at $L_p$ tidal periods, which are only observed in the ionosphere but not in the reconstructed magnetospheric signal (Fig. \ref{fig:spec-q21-mi}).
Additionally, our induced coefficients also contain oceanic tidal magnetic signals \citep{grayver_tides_2024}. However, since we use both day-side and dark-side data and adopt a low-degree spherical harmonic representation of the field, the ocean tidal contributions are superposed by signals resulting from the ionospheric lunar daily variations \citep{malin1970}.

\section{Discussion and Conclusions}\label{sec:discussions}

Compared to the previous geomagnetic field models, our new modelling approach with the ionosphere parametrisation has several distinct features. First, the magnetic fields with ionospheric and magnetospheric origins are separated. This sets our model apart from the modelling approaches where the ionospheric field is neglected \citep{lesur_second_2010, finlay_chaos-7_2020} or the magnetospheric field is removed by baseline removal and ring current corrections \citep{yamazaki_sq_2017,Guzavina2019}. Second, no temporal behaviour is prescribed on either of the external fields. This is in contrast to many other models. For instance, the temporal behaviour for magnetospheric fields beyond $P_1^0$ is prescribed to be standing modes in SM or GSM coordinates in the CHAOS model \citep{finlay_chaos-7_2020}, or the ionospheric field is assumed to consist of daily and seasonal time-harmonics in the Comprehensive Model \citep{sabaka_comprehensive_2018}. Furthermore, the secondary induced field is co-estimated alongside the external fields, rather than modelled using an existing conductivity model, as is the case in \cite{chulliat_first_2016}.

These new features make our model complementary to other existing model and in some cases more suitable for dedicated external field or mantle induction studies. Firstly, since we do not assume a subsurface electrical conductivity \textit{a priori} for modelling the induction effects, the co-estimated induced field coefficients combined with the external field coefficients can be used to probe the electrical conductivity in the interior of the Earth. Previous attempts to image the interior conductivity using joint satellite-observatory dataset tried to avoid the inconsistency between the two datasets by removing CI-based ionospheric field from the data \citep[see e.g.][]{Velimsky2021,kuvshinov_probing_2021}. This approach is limited since the ionospheric model is based on quiet-time data, and is parametrised using daily and seasonal time harmonics. For instance, we have seen in Section \ref{sec:result-tseries} that during geomagnetic storms, the ionospheric field exhibits strong non-periodic behaviour. 


When the geomagnetic field model fails to describe the ionospheric sources properly, the remnant ionospheric signal will be compensated by biasing the induced field and the magnetospheric field estimates. 
To test whether the induced and the magnetospheric fields are indeed biased if the ionospheric field is neglected, we need a physics-based metric to examine how the model with an ionosphere compares with a model without the ionosphere. 
For this purpose, we estimated the transfer functions (TFs) relating the induced and the external parts of the geomagnetic potential. 
In the frequency domain, assuming a radially symmetric electrical conductivity of the Earth, the Gauss coefficients of the induced field are related to those of the external field as
\begin{equation}\label{eqn:Q-resp}
    \iota_{nm}(\omega) = Q_n(\omega; \sigma) \, \varepsilon_{nm}(\omega).
\end{equation}
where $\omega$ is the angular frequency, and $Q_n$ is the so-called $Q$-response \citep{schmucker_magnetic_1985}, a functional of the subsurface conductivity $\sigma$. For a general 3-D subsurface conductivity, the relation above is not complete since a single external mode would induce many internal modes, all of which are coupled through a so-called Q-matrix \citep{Olsen1999Review}. As the subsurface conductivity can be approximated to the first order as a radial conductivity profile, the diagonal elements of this matrix dominate, and the relation between the induced and external coefficients can to first order be approximated by the expression above. Therefore, the equation above is still a valid test since any realistic 3-D conductivity effects will be dwarfed by much stronger source effects. 

The complex Gauss coefficients $\iota_{nm}$ and $\varepsilon_{nm}$ are the expansion coefficients of the geomagnetic potential $V$ using the complex SH basis $Y_n^m(\theta, \phi) \propto P_n^{|m|}(\cos\theta) e^{im\phi}$, and are related to their real counterparts as
\begin{equation}
\label{eqn:real2complex}
    \varepsilon_{nm} = \left\{
    \begin{aligned}
        &q_{n0},& &m=0 \\
        &\frac{1}{2} \left(q_{n|m|} - i s_{n|m|}\right),& &m > 0, \\ 
        &\frac{1}{2} \left(q_{n|m|} + i s_{n|m|}\right),& &m < 0,
    \end{aligned}\right.\qquad 
    \iota_{nm} = \left\{
    \begin{aligned}
        &g_{n0},& &m=0 \\
        &\frac{1}{2} \left(g_{n|m|} - i h_{n|m|}\right),& &m > 0, \\ 
        &\frac{1}{2} \left(g_{n|m|} + i h_{n|m|}\right),& &m < 0. 
    \end{aligned}\right.
\end{equation}
For the model with the ionosphere layer, both ionospheric and magnetospheric fields are external, and hence we use $\varepsilon_{nm} := \varepsilon_{nm}^\mathrm{ion} + \varepsilon_{nm}^\mathrm{mag}$ for $\varepsilon_{nm}$ in (\ref{eqn:Q-resp}). The internal coefficients are simply the Gauss coefficients for the induced field, thus we have $\iota_{nm}:=\iota_{nm}^\mathrm{int}$. For the model without ionosphere, the magnetospheric and the induced fields serve as the external and the internal fields, respectively.

\begin{figure}
    \centering
    \includegraphics[width=\linewidth]{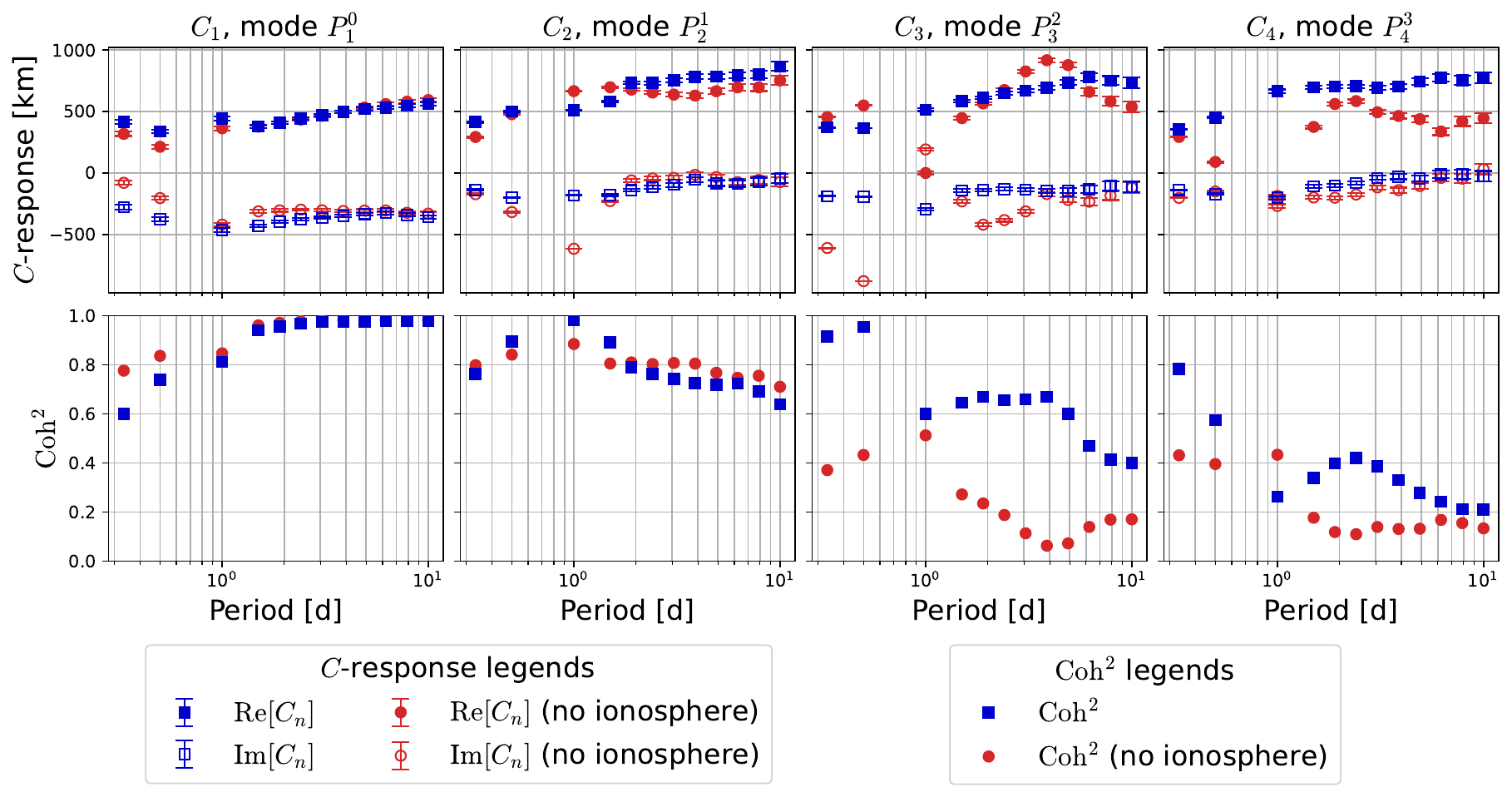}
    \caption{$C_n$-responses (upper panel) and their squared coherences (lower panel) at periods between $8$ hours and $10$ days. The columns show the $C_1$, $C_2$, $C_3$ and $C_4$ responses estimated from the SH coefficients described by the $P_1^0$, $P_2^1$, $P_3^2$ and $P_4^3$ modes, respectively.}
    \label{fig:Cn-10d}
\end{figure}

For a radially symmetric electrical conductivity of the Earth, the $Q$-response is only dependent on the SH degree $n$, but is independent of the order $m$. Therefore, an estimation of $Q_n$ can be conducted using any spherical harmonic mode of the same degree. The results are further converted to the global $C$-response \citep{schmucker_magnetic_1985,Olsen1999Review} as
\begin{equation}
    C_n(\omega) = \frac{a}{n+1} \frac{1 - \frac{n+1}{n} Q_n(\omega)}{1 + Q_n(\omega)}.
\end{equation}
The real part of the $C_n$-response is a depth of the maximum induced current, thus approximating the sounding depth at a given frequency \citep{weidelt1972inverse}. 
The value of $C_n$ at $\omega$ is estimated following the general method outlined in \citet{Olsen1998}. First, the time series of the complex Gauss coefficients $\varepsilon_{nm}$ and $\iota_{nm}$ (obtained from real counterparts using Eq. \ref{eqn:real2complex}) are split into segments with length $3T$, where $T = 2\pi/\omega$ is the period. Second, the short-time Fourier transform of the time series with Hamming time window is computed for every segment $\tau_j$ ($j=1,2,\cdots N_\tau$), generating the spectrum points within the segment, denoted as $\iota_{nm}(\tau_j, \omega)$ and $\varepsilon_{nm}(\tau_j, \omega)$. The final transfer function is estimated by performing a robust regression on the linear system (\ref{eqn:Q-resp}), with $\varepsilon_{nm}(\tau_j, \omega)$ and $\iota_{nm}(\tau_j, \omega)$ estimated in the previous step.
The resulting $C_n$-responses from the field estimates of the model with and without the ionosphere are visualised in Fig. \ref{fig:Cn-10d} at $9$ log-spaced periods between $1.5$ days and $10$ days, as well as at the diurnal ($24$ hours), semidiurnal ($12$ hours) and terdiurnal ($8$ hours) periods. 
It is also possible to estimate these transfer functions at longer periods, but the effect of the ionosphere becomes smaller at periods longer than 20-30 days.  
To evaluate the linear correlation between the internal and external fields, we also computed the squared coherence ($\mathrm{Coh}^2$) for each spherical harmonic mode and each frequency, defined as
\begin{equation}
    \mathrm{Coh}^2 = \frac{\left|\sum_{j=1}^{N_\tau} \iota^*_{nm}(\tau_j, \omega) \, \varepsilon_{nm}(\tau_j, \omega)\right|^2}{\sum_{j=1}^{N_\tau}\left|\iota_{nm}(\tau_j, \omega)\right|^2 \sum_{j=1}^{N_\tau}\left|\varepsilon_{nm}(\tau_j, \omega)\right|^2}.
\end{equation}

The TFs for the model with the ionosphere have generally higher squared coherence, especially for higher SH degress where the contribution of the ionosphere is expected to be larger. For instance, $C_3$ estimated from the $P_3^2$ mode boasts a squared coherence of $\mathrm{Coh}^2 = 0.95$ at the dominant period of $12$hrs, a drastic improvement over the mere $\mathrm{Coh}^2 = 0.42$ for the model without the ionosphere; $C_4$ estimated from the $P_4^3$ mode has $\mathrm{Coh}^2 = 0.78$ at the period of $8$hrs, compared to $\mathrm{Coh}^2=0.43$ for the model that neglects ionosphere (Fig. \ref{fig:Cn-10d}). Similar but somewhat smaller advantages can also be observed for $C_2$ estimated from the $P_2^1$ mode around the diurnal period band, and across almost the entire period range for $C_3$ and $C_4$, albeit both have lower coherences at longer periods. For the $C_1$ estimated from the first zonal SH $P_1^0$, we note that the transfer functions from both models deteriorate as they approach the diurnal band compared to longer periods, where the two estimates are nearly identical in terms of coherence.

Besides improved coherences, $C_n$-response values appear more physical when the ionosphere is modelled. In higher degree responses ($C_2$, $C_3$ and $C_4$), the $C$-responses derived using the reduced model without the ionosphere feature spurious fluctuations and non-monotonic behaviour, occasionally changing the sign in the real parts (e.g. $C_3$ and $C_4$ at diurnal periods). This behaviour is not physical since global $C_n$-responses are strictly monotonic for a 1-D model and can deviate from this in 3-D settings, but such deviations are smooth and mild for realistic 3-D models \citep{Puethe2014, Grayver2021}. In contrast, these strong fluctuations and unrealistic values of the $C_n$-responses are completely absent in the TFs computed from the model with the ionosphere.

The difference in the estimated transfer functions and their coherence shows that a lack of proper treatment of the ionosphere in the model results in transfer functions that are in some cases significantly incoherent and physically inconsistent. Fortunately, this does not affect the $C_1$-response computed from the first zonal harmonic at periods longer than one day, rendering previous global reference models \citep[e.g.][]{Grayver2017, kuvshinov_probing_2021} still valid. The situation is different for the 3-D mantle inversions. Previous studies \citep{Velimsky2021,kuvshinov_probing_2021} used external and internal coefficients derived from satellite and ground observations whereby the ionospheric effect was not explicitly modelled. The authors subtracted a model of the ionosphere (together with the corresponding induced field) with a prescribed periodic behaviour, constructed using quiet-time magnetic data as a part of the CI model \citep{sabaka_comprehensive_2018}. Failing to co-model ionospheric sources or extrapolating quiet-time ionosphere models to the entire data set will likely result in source effects being propagated to 3-D conductivity models. To illustrate this, we also first subtracted a CI-derived ionospheric model (with its induced field) from our dataset and then carried out the internal-external field separation. The resulting magnetospheric spectra show lunar tidal peaks (Fig. \ref{fig:spec-ms4-MIOres}), implying the ionospheric signal may have leaked into the magnetospheric estimates. At the same time, there is almost no improvement and even occasional deterioration in the coherence of the estimated TFs (Fig. \ref{fig:Cn-10d-MIOres}) compared to the internal-external field separation without the ionosphere on the original dataset (Fig. \ref{fig:Cn-10d}). Therefore, explicitly co-estimating the ionosphere is necessary to achieve coherent and physical transfer functions for probing the 3-D subsurface conductivity of the Earth with satellite and ground data.

Besides new opportunities for induction studies, our model also enables a more flexible and versatile characterisation of the external magnetic field sources. Thanks to the relaxed assumptions on temporal behaviour of the fields and the environmental setting, our model is capable of exploiting data across all local times and during all magnetic conditions. To get a glimpse of the spacial structure and the temporal behaviour of the estimated external field, we visualise the results in terms of the streamfunction of an equivalent current, defined as 
\begin{equation}\label{eqn:Psi-expansion}
    \Psi(\mathbf{r}, t) = - \frac{a}{\mu_0} \sum_{nm} \frac{2n+1}{n+1} \left(\frac{a + h}{a}\right)^n \left[q_{nm}(t) \cos m\phi + s_{nm}(t)\sin\phi \right] P_n^m(\cos\theta).
\end{equation}
The electric current is then linked to the streamfunction via
\begin{equation}
    \mathbf{j}(\mathbf{r}, t) = - \delta (r - (a+h)) \, \hat{\mathbf{e}}_r \times \nabla_H \Psi(\mathbf{r}, t)
\end{equation}
where $\hat{\mathbf{e}}_r$ is the unit vector in the radial direction, and $\nabla_H$ is the surface gradient. On the surface of a sphere with radius $b$, the surface gradient can be explicitly written as $\nabla_H = \hat{\mathbf{e}}_\theta b^{-1} \partial_\theta + \hat{\mathbf{e}}_\phi (b\sin\theta)^{-1} \partial_\phi$. This formulation, describing purely toroidal electric currents concentrated in an infinitely thin layer at a radius $a+h$, is consistent with the current sheet approximation for the ionospheric currents (Section \ref{sec:method}). Therefore, the ionospheric equivalent current streamfunction $\Psi^{\mathrm{ion}}$ can be calculated using the expansion (\ref{eqn:Psi-expansion}) and $q_{nm}^\mathrm{ion}(t)$, $s_{nm}^\mathrm{ion}(t)$.

\begin{figure}
    \centering
    \includegraphics[width=0.9\linewidth]{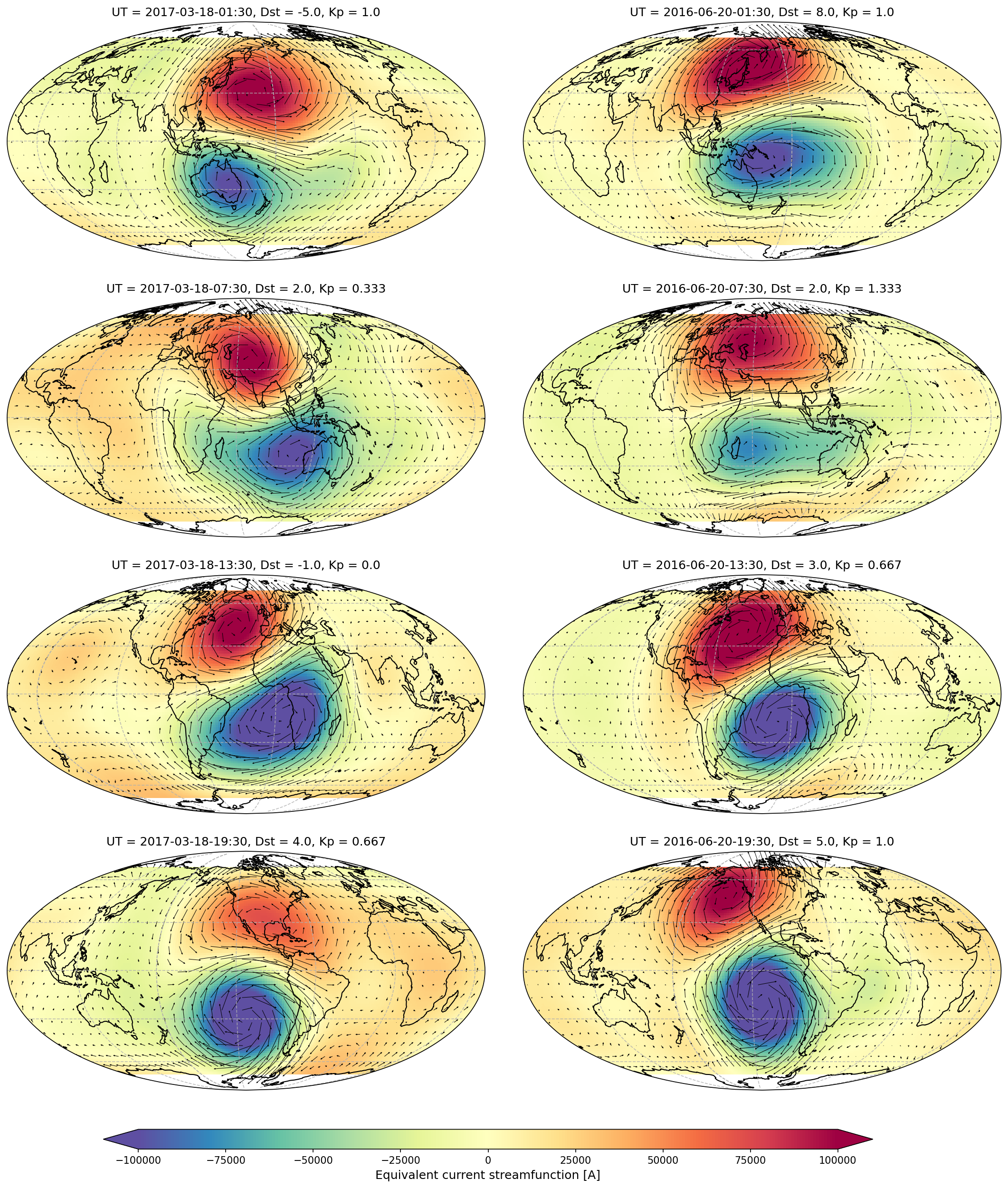}
    \caption{Streamfunctions of the equivalent current for the ionospheric field ($\Psi^{\mathrm{ion}}$) on magnetic quiet days close to a spring equinox (left column) and a summer solstice (right column). The streamfunction is cut off above an absolute latitude of $70$ degrees to filter out the values in high latitudes where our dataset and parametrisation have limited resolution. The central longitude for each subplot is at the local noon.}
    \label{fig:Psi-equinox-solstice-quiet}
\end{figure}

During geomagnetic quiet times, our model recovers the dominant Sq current system in the ionosphere, as is shown in Fig. \ref{fig:Psi-equinox-solstice-quiet}. The equivalent current of the Sq current system features one large-scale counter-clockwise vortex in the northern hemisphere, and one large-scale clockwise vortex in the southern hemisphere. Near the summer solstice, the foci of the Sq currents in the northern hemisphere are observed to consistently lag behind in local time compared to the southern foci, consistent with the previous observations on the seasonal variation in the local times of the focus positions \citep{yamazaki_sq_2017}. Near the spring equinox, the northern and southern foci of the Sq currents alternate to lead in local times on a semi-diurnal basis. These seasonal changes of the Sq system are also reflected by the annual, semi-annual and seasonal periodicities in the ionospheric Gauss coefficient spectra at spherical harmonic degree $n > 1$ (Figs. \ref{fig:spec-s21-mi} and \ref{fig:spec-q32-mi} lower panels).

\begin{figure}
    \centering
    \includegraphics[width=0.9\linewidth]{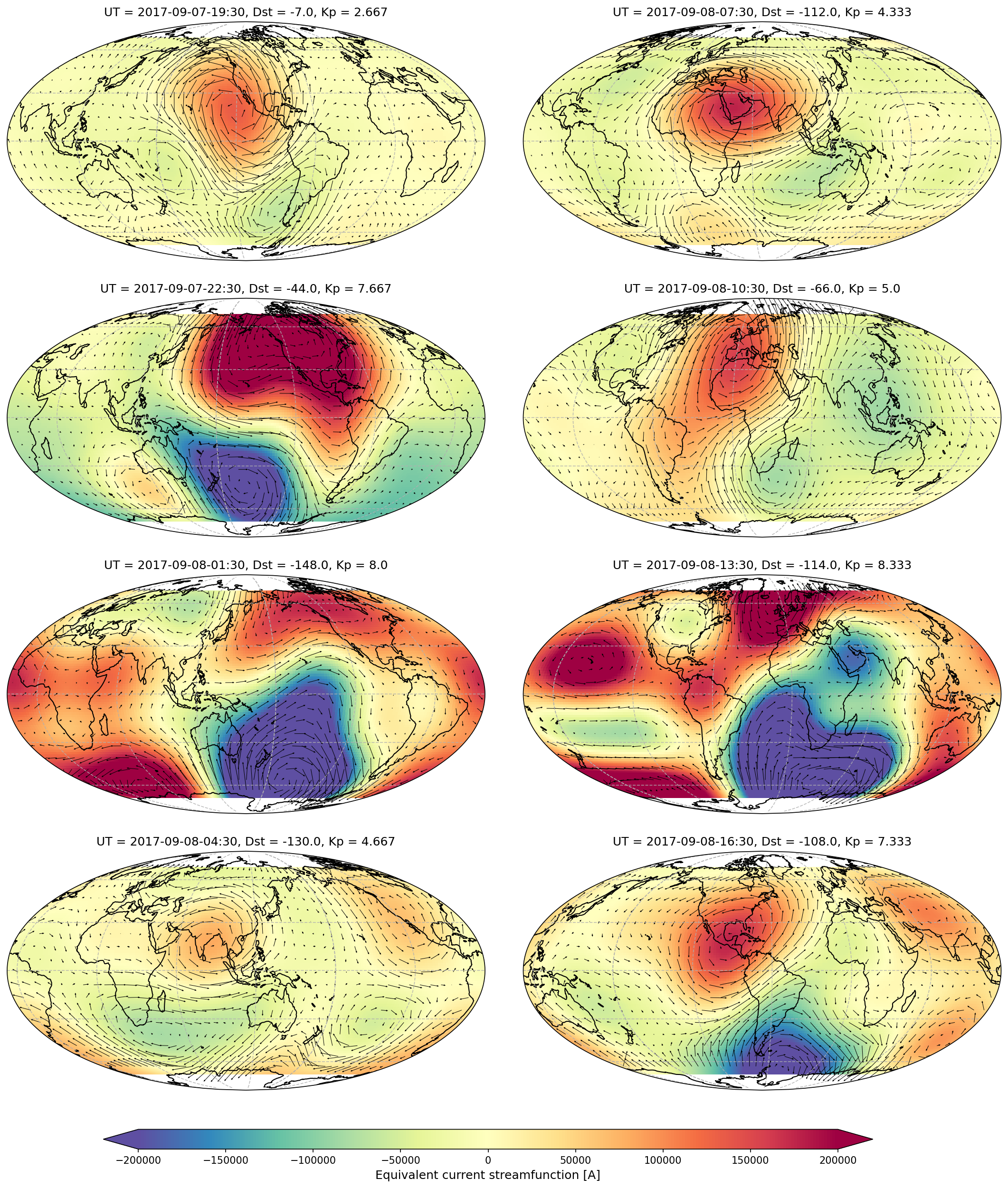}
    \caption{Streamfunction of the equivalent current for the ionospheric field ($\Psi^{\mathrm{ion}}$) during a geomagnetic storm on Sept. 8, 2017. The same cutoff latitude of $70$ degrees is used. Note that the maximum value of the color scale is extended to twice of that for the quiet day plot (Fig. \ref{fig:Psi-equinox-solstice-quiet}) to accommodate higher amplitudes of the streamfunction during the storm. Nevertheless, the strong equivalent currents saturate the colour scales when $Kp > 7$.}
    \label{fig:Psi-2017-09-storm}
\end{figure}

During geomagnetic storms, the estimated ionospheric current deviates considerably from a quiet Sq current, exemplified by the ionospheric equivalent current reconstructed during the geomagnetic storm on Sept. 8, 2017 (Fig. \ref{fig:Psi-2017-09-storm}). Compared to the Sq current during quiet times, the equivalent sheet shows two distinct features during geomagnetic storms. First, the day-night dichotomy, ubiquitous during geomagnetic quiet times, is no longer present. Instead, the sheet current shows strong current amplitude across all local times (see UT=1:30 and 13:30 on 8 September). Second, the structurally robust vortices centred at mid-latitudes in the northern and southern hemispheres give way to multiple transient, rapid changing vortices centred at high latitudes (see UT=22:30, 1:30, 13:30, 16:30 on 8 September, when $Kp \geq 7$). This may imply the energy input from the magnetosphere in high latitudes, which then propagates to lower latitudes by virtue of electric field penetration \citep{yamazaki_sq_2017}. Interestingly, we see that in between the time bins with $Kp \geq 7$, the Sq current system still seems to persist, observable as the two vortices in the northern and southern hemispheres. 

Our model provides a continuous description of the mid-latitude ionospheric and magnetospheric fields for the period 2014-2023, covering all local times and magnetic conditions. The temporal resolution of the presented model is three hours, indicating that dynamics on a time scale that is similar to or less than three hours cannot be well captured. This is especially pronounced in phases of geomagnetic storms with fast Dst variations, where we observe higher misfits of the model ($R^2 \sim 0.7$ for UT=22:30, 01:30, 13:30, Fig. \ref{fig:Psi-2017-09-storm}). In comparison, the misfits during other times tend to be lower ($R^2 > 0.85$ for all other snapshots). In addition, the spatial parametrisation of the current model is up to SH degree and order $4$ for induced and magnetospheric fields, and up to degree and order $5$ for ionospheric field. Localised features, such as equatorial electrojets and the polar current system, cannot be described within our parametrisation and hence have been avoided in data selection. The (in-)ability to capture certain temporal and spatial variations of the magnetic fields is controlled by a trade-off between space and time coverage of the data, which is ultimately dictated by the distribution of the observations.

In light of these limitations, our framework can benefit both from the incorporation of additional data, as well as dedicated parametrisation for localised current systems. The new low-orbit geomagnetic satellites, such as the Macau Science Satellites (MSS) and a planned ESA NanoMagSat mission, will significantly improve the spatial (especially local-time) coverage, allowing us to increase the spatio-temporal resolution of the model. Dedicated parametrisation of polar electrojets based on loop or point currents can be integrated into our model to improve the fit at polar regions.
Thanks to its flexibility, the model can be updated on the fly in real time, thus making it particularly suitable for working with low-time-lag datasets, such as the Swarm FAST data, in support of space weather operations and hazard assessment.

\begin{acknowledgments}

J.M. is grateful for funding from the European Research Council (agreement No. 833848-UEMHP) under the Horizon 2020 program, and the funding from Swiss National Science Foundation (grant No. 219247) under the MINT 2023 call. AG was supported by the Heisenberg Grant from the German Research Foundation, Deutsche Forschungsgemeinschaft (Project No. 465486300) and ESA Swarm DISC project No. 4000109587.


The author contributions are as follows. J.M.: conceptualisation (equal), data curation (lead), methodology (lead), investigation (equal), resources (equal), software (lead), visualisation (lead), writing - original draft (lead), writing - review \& editing (equal). A.G.: conceptualisation (equal), methodology (supporting), investigation (equal), project administration (lead), resources (equal), visualisation (supporting), writing - original draft (supporting), writing - review \& editing (equal).

\end{acknowledgments}

\section*{DATA AVAILABILITY STATEMENT}

The input Swarm, CryoSat-2, and Grace-FO data are part of the Swarm product and are available at \texttt{https://swarmhandbook.earth.esa.int/catalogue/}. The input ground observatory data are from INTERMAGNET and WDC data, accessed via the VirES platform (\texttt{https://vires.services/}). The CI core and lithospheric models subtracted from the data and the CI magnetospheric model used for comparison are available at \texttt{https://swarmhandbook.earth.esa.int/catalogue/}. The CHAOS-7.18 model used for comparison is available at \texttt{https://spacecenter.dk/files/magnetic-models/CHAOS-7}.
Time series of estimated ionospheric and magnetospheric Gauss coefficients are available on Zenodo, at \texttt{https://dx.doi.org/10.5281/zenodo.14787587}.

\bibliography{references}

\appendix

\section{Non-unique least squares estimation of a piece-wise constant field and observatory bias from observatory data}\label{sec:app-bias-obs}

Given only observatory data, a piece-wise constant field along with the observatory bias cannot be uniquely determined in the least squares sense. This has been noted already in \citet{langel_new_methods_1982}. Although not complicated, there is no reference to our knowledge where a clear mathematical proof or physical explanation is given. We thus document our mathematical proof here, followed by a physical explanation.

In the $i$-th time bin ($i=1,2,\cdots N_t$ where $N_t$ is the total number of time bins), the data $\mathbf{d}_i\in \mathbb{R}^{N_i}$ measured at ground observatories are linearly related to the field coefficients $\mathbf{x}_i \in \mathbb{R}^{M_0}$ via the design matrix $\mathbf{G}_i \in \mathbb{R}^{N_i\times M_0}$, combined with a contribution from the observatory bias
\begin{equation}
    \mathbf{d}_i \simeq \mathbf{G}_i \mathbf{x}_i + \mathbf{b}_i.
\end{equation}
Here $\mathbf{b}_i \in \mathbb{R}^{N_i}$ represents the observatory bias corresponding to the measurements in this time bin. We use $N_i$ and $M_0$ to denote the dimensionality of the data and field coefficients, respectively. The dimensionality of the field coefficients $M_0$ is assumed to be uniform for all time bins. Since the observatory biases are static biases, the bias term in the $i$-th time bin is simply sampled from the collective observatory biases according to the measurements available in this time bin, and hence can be related to the vector representing the collective observatory biases $\mathbf{b} \in \mathbb{R}^{M_b}$ by a sampling matrix $\mathbf{S}_i \in \mathbb{R}^{N_i\times M_b}$. Therefore, we have
\begin{equation}
    \mathbf{d}_i \simeq \mathbf{G}_i \mathbf{x}_i + \mathbf{S}_i \mathbf{b} = \begin{pmatrix} \mathbf{G}_i & \mathbf{S}_i \end{pmatrix}
    \begin{pmatrix} \mathbf{x}_i \\ \mathbf{b} \end{pmatrix}.
\end{equation}
Since the bias $\mathbf{b}$ is shared across all time bins, one cannot solve each time window separately, but has to solve the full system, written as
\begin{equation}
    \mathbf{d} = \begin{pmatrix} \mathbf{d}_1 \\ \vdots \\ \mathbf{d}_{N_t} \end{pmatrix} \simeq \begin{pmatrix} 
    \mathbf{G}_1 & & \mathbf{0} & \mathbf{S}_1 \\
    & \ddots & & \vdots \\
    \mathbf{0} & & \mathbf{G}_{N_t} & \mathbf{S}_{N_t} \\
    \end{pmatrix}
    \begin{pmatrix} \mathbf{x}_1 \\ \vdots \\ \mathbf{x}_{N_t} \\ \mathbf{b} \end{pmatrix} = \widetilde{\mathbf{G}} \widetilde{\mathbf{x}}.
\end{equation}
where $\widetilde{\mathbf{G}} \in \mathbb{R}^{N\times M}$ is the full design matrix for the problem, with $N = \sum_{i=1}^{N_t} N_i$ and $M = {N_t} M_0 + M_b$. 
$\widetilde{\mathbf{x}} \in \mathbb{R}^M$ is the augmented model vector. Solving the least squares problem above is mathematically equivalent to solving the normal equation
\begin{equation}
    \widetilde{\mathbf{G}}^\top \widetilde{\mathbf{G}} \widetilde{\mathbf{x}} = 
    \begin{pmatrix}
        \mathbf{G}_1^\top \mathbf{G}_1 & & \mathbf{0} & \mathbf{G}_1^\top \mathbf{S}_1 \\ 
        & \ddots & & \vdots \\ 
        \mathbf{0} & & \mathbf{G}_{N_t}^\top \mathbf{G}_{N_t} & \mathbf{G}_{N_t}^\top \mathbf{S}_{N_t} \\ 
        \mathbf{S}_1^\top \mathbf{G}_1 & \cdots & \mathbf{S}_{N_t}^\top \mathbf{G}_{N_t} & \sum_i \mathbf{S}_i^\top \mathbf{S}_i
    \end{pmatrix} \widetilde{\mathbf{x}}
    = \widetilde{\mathbf{G}}^\top \mathbf{d}.
\end{equation}
We now want to calculate the rank or the nullity of the matrix $\widetilde{\mathbf{G}}^\top \widetilde{\mathbf{G}} \in \mathbb{R}^{M\times M}$. We assume that the submatrices $\mathbf{G}_i^\top \mathbf{G}_i$ are all invertible (i.e. the field coefficients can be estimated time-bin-wise if there is no bias on the measurements).
In this case, we can apply the Guttman rank additivity formula
\begin{equation}
    \mathrm{rank} 
    \begin{pmatrix} \mathbf{A} & \mathbf{B} \\ \mathbf{C} & \mathbf{D} \end{pmatrix} = \mathrm{rank}(\mathbf{A}) + \mathrm{rank}(\mathbf{D} - \mathbf{C} \mathbf{A}^{-1} \mathbf{B})
\end{equation}
iteratively to the $\widetilde{\mathbf{G}}^\top \widetilde{\mathbf{G}}$ matrix. The rank reduces to
\begin{equation}
\begin{aligned}
    \mathrm{rank}\left(\widetilde{\mathbf{G}}^\top \widetilde{\mathbf{G}}\right) &= 
    \mathrm{rank}\left(\mathbf{G}_1^\top \mathbf{G}_1\right) 
    + \mathrm{rank} \begin{pmatrix}
        \mathbf{G}_2^\top \mathbf{G}_2 & & \mathbf{0} & \mathbf{G}_2^\top \mathbf{S}_2 \\ 
        & \ddots & & \vdots \\ 
        \mathbf{0} & & \mathbf{G}_{N_t}^\top \mathbf{G}_{N_t} & \mathbf{G}_{N_t}^\top \mathbf{S}_{N_t} \\ 
        \mathbf{S}_2^\top \mathbf{G}_2 & \cdots & \mathbf{S}_{N_t}^\top \mathbf{G}_{N_t} & \sum_i \mathbf{S}_i^\top \mathbf{S}_i - \mathbf{S}_1^\top \mathbf{G}_1 \mathbf{G}_1^+ \mathbf{S}_1
    \end{pmatrix} \\ 
    &= \sum_{i=1}^{N_t} \mathrm{rank}\left(\mathbf{G}_i^\top \mathbf{G}_i\right) + \mathrm{rank} \left(\sum_{i=1}^{N_t} \mathbf{S}_i^\top \mathbf{P}_{\mathbf{G}_i}^\perp \mathbf{S}_i\right)
\end{aligned}
\end{equation}
where $\mathbf{A}^+ = (\mathbf{A}^\top \mathbf{A})^{-1} \mathbf{A}^\top$ is the Moore-Penrose pseudoinverse of matrix $\mathbf{A}$, and $\mathbf{P}_{\mathbf{A}}^\perp = \mathbf{I} - \mathbf{G} \mathbf{G}^+$ is the projection operator onto the orthogonal complement of the image of $\mathbf{A}$.
Under our assumption, $\mathbf{G}_i$ is full column rank and $\mathrm{rank}(\mathbf{G}_i^\top \mathbf{G}_i) = M_0$. Applying the rank-nullity theorem, we have
\begin{equation}\label{eqn:nullity-intermediate}
\begin{aligned}
    &\mathrm{dim} \, \mathrm{ker}\left(\widetilde{\mathbf{G}}^\top \widetilde{\mathbf{G}}\right) = 
    \mathrm{dim}\left(\widetilde{\mathbf{G}}^\top \widetilde{\mathbf{G}}\right) 
    - \mathrm{rank} \left(\widetilde{\mathbf{G}}^\top \widetilde{\mathbf{G}}\right) \\ 
    &= \mathrm{dim}\left(\widetilde{\mathbf{G}}^\top \widetilde{\mathbf{G}}\right) - \sum_{i=1}^{N_t} \mathrm{rank}\left(\mathbf{G}_i^\top \mathbf{G}_i\right) 
    - \mathrm{dim} \left(\sum_{i=1}^{N_t} \mathbf{S}_i^\top \mathbf{P}_{\mathbf{G}_i}^\perp \mathbf{S}_i\right)
    + \mathrm{dim}\, \mathrm{ker} \left(\sum_{i=1}^{N_t} \mathbf{S}_i^\top \mathbf{P}_{\mathbf{G}_i}^\perp \mathbf{S}_i\right)\\ 
    &= M - {N_t} M_0 - M_b + \mathrm{dim}\, \mathrm{ker} \left(\sum_{i=1}^{N_t} \mathbf{S}_i^\top \mathbf{P}_{\mathbf{G}_i}^\perp \mathbf{S}_i\right) = \mathrm{dim}\, \mathrm{ker} \left(\sum_{i=1}^{N_t} \mathbf{S}_i^\top \mathbf{P}_{\mathbf{G}_i}^\perp \mathbf{S}_i\right)
\end{aligned}
\end{equation}
Therefore, the dimension of the nullspace of the original linear system can be identified with that of the matrix $\sum_i \mathbf{S}_i^\top \mathbf{P}_{\mathbf{G}_i}^\perp \mathbf{S}_i$.
In a piece-wise constant field model with observatory data, not only does $\mathbf{S}_i$ sample the biases in each time bin, but the same sampling matrix also samples the design matrix. In other words, the design matrices in each time bin are sampled from the rows of a design matrix $\mathbf{G}_0 \in \mathbb{R}^{M_b\times M_0}$, whose rows contain the collective design vectors for all observatories, hence $\mathbf{G}_i = \mathbf{S}_i \mathbf{G}_0$.
The whole column space of $\mathbf{G}_0$ hence maps to zero under the matrix $\sum_i \mathbf{S}_i^\top \mathbf{P}_{\mathbf{G}_i}^\perp \mathbf{S}_i$
\begin{equation}
    \left(\sum_{i=1}^{N_t} \mathbf{S}_i^\top \mathbf{P}_{\mathbf{G}_i}^\perp \mathbf{S}_i\right) \mathbf{G}_0 = \sum_{i=1}^{N_t} \mathbf{S}_i^\top \mathbf{P}_{\mathbf{G}_i}^\perp (\mathbf{S}_i \mathbf{G}_0) = \sum_{i=1}^{N_t} \mathbf{S}_i^\top \mathbf{P}_{\mathbf{G}_i}^\perp \mathbf{G}_i = \mathbf{0}
\end{equation}
since $\mathbf{P}_{\mathbf{A}}^\perp \mathbf{A}\equiv \mathbf{0}$, $\forall \mathbf{A}$. Therefore, we arrive at
\begin{equation}
    \mathrm{dim} \, \mathrm{ker}\left(\widetilde{\mathbf{G}}^\top \widetilde{\mathbf{G}}\right) = \mathrm{dim}\, \mathrm{ker} \left(\sum_{i=1}^{N_t} \mathbf{S}_i^\top \mathbf{P}_{\mathbf{G}_i}^\perp \mathbf{S}_i\right) \geq \mathrm{rank}(\mathbf{G}_0) = M_0.
\end{equation}
The full linear system thus is singular, with a nullspace of dimension $M_0$, the same dimension as the field coefficients. There is no unique least squares solution to this problem.

From a physical point of view, introducing the observatory bias in the regular spherical harmonic analysis allows one to embed an arbitrary time-invariant field that can be represented with the field coefficients in the bias. If a set of field coefficients solves the original least squares problem, adding an arbitrary constant vector $\delta \mathbf{x}$ to the field coefficients in all time bins uniformly also solves the problem, as long as the observatory biases are adjusted accordingly to compensate for the static field. Therefore there is no uniqueness in the solution.

\section{Non-unique least squares estimation of a piece-wise constant field and observatory bias from ground and satellite data}\label{sec:app-bias-full}

The observatory bias along with the field coefficients can be uniquely determined in a least squares sense if both satellite and ground observatory data are used, provided that the two datasets share the same internal and external contributions in the geomagnetic field parametrisation. 
However, since we introduced an extra ionospheric field in our complete model which contributes to the fields measured by observatories and satellites differently, the solution remains non-unique. Here we also provide a mathematical and a physical explanation for this non-uniqueness.

The least squares problem in this scenario can be formulated in the same way as in the last section. However, the setup of the submatrices are different from the previous section due to the different dataset and field parametrisation. In this scenario, we have both the data from ground observatories, and the data from satellites, which are unaffected by the observatory bias. Therefore, the data vectors, the design matrices and the sampling matrices have row divisions
\begin{equation}
    \mathbf{d}_i = \begin{pmatrix} \mathbf{d}_{i,\mathrm{obs}} \\ \mathbf{d}_{i,\mathrm{sat}} \end{pmatrix}, \quad 
    \mathbf{G}_i = \begin{pmatrix} \mathbf{G}_{i,\mathrm{obs}} \\ \mathbf{G}_{i,\mathrm{sat}} \end{pmatrix}, \quad 
    \mathbf{S}_i = \begin{pmatrix} \mathbf{S}'_{i} \\ \mathbf{0} \end{pmatrix}.
\end{equation}
On the other hand, in our complete model, the field coefficients are separated into internal, ionospheric and magnetospheric parts. The field coefficient vector and the design matrix has the following row and column divisions,
\begin{equation}
    \mathbf{x}_i = \begin{pmatrix} \mathbf{x}_i^{\mathrm{int}} \\ \mathbf{x}_i^{\mathrm{ion}} \\ \mathbf{x}_i^{\mathrm{mag}} \end{pmatrix}, \quad 
    \mathbf{G}_i = \begin{pmatrix} \mathbf{G}_{i,\mathrm{obs}} \\ \mathbf{G}_{i,\mathrm{sat}} \end{pmatrix} = \begin{pmatrix}
        \mathbf{G}_{i,\mathrm{obs}}^{\mathrm{int}} & \mathbf{G}_{i,\mathrm{obs}}^{\mathrm{ion}} & \mathbf{G}_{i,\mathrm{obs}}^{\mathrm{mag}} \\ 
        \mathbf{G}_{i,\mathrm{sat}}^{\mathrm{int}} & \mathbf{G}_{i,\mathrm{sat}}^{\mathrm{ion}} & \mathbf{G}_{i,\mathrm{sat}}^{\mathrm{mag}}
    \end{pmatrix}.
\end{equation}
For simplicity, we shall only discuss the case where the three fields are parametrised in the same set of spherical harmonics, hence $\mathbf{x}_i^{\mathrm{int}}, \mathbf{x}_i^{\mathrm{ion}}, \mathbf{x}_i^{\mathrm{mag}} \in \mathbb{R}^{M_0/3}$. A similar conclusion holds for a non-uniform truncation of the fields, but the notation will be much more complicated.

Following Eq. (\ref{eqn:model-V-repr-obs}), we have $\mathbf{G}_{i,\mathrm{obs}}^{\mathrm{ion}} = \mathbf{G}_{i,\mathrm{obs}}^{\mathrm{mag}} = \mathbf{G}_{i,\mathrm{obs}}^{\mathrm{ext}}$, i.e. the design matrix linking the ionospheric and magnetospheric coefficients to the ground observations are the same, and are simply the design matrix for external coefficients $\mathbf{G}_{i,\mathrm{obs}}^{\mathrm{ext}}$ in conventional internal-external field separation. These sub-matrices corresponding to the ground observations are also sampled by the sampling matrix from the conglomerate design matrices, i.e. 
\begin{equation}
    \mathbf{G}_{i,\mathrm{obs}}^\mathrm{int} = \mathbf{S}_i' \mathbf{G}_{0,\mathrm{obs}}^\mathrm{int},\quad 
    \mathbf{G}_{i,\mathrm{obs}}^\mathrm{ext} = \mathbf{S}_i' \mathbf{G}_{0,\mathrm{obs}}^\mathrm{ext}.
\end{equation}
On the other hand, following Eq. (\ref{eqn:model-V-repr-sat}), we have $\mathbf{G}_{i,\mathrm{sat}}^\mathrm{mag} = \mathbf{G}_{i,\mathrm{sat}}^\mathrm{ext}$, but $\mathbf{G}_{i,\mathrm{sat}}^\mathrm{ion} = \mathbf{G}_{i,\mathrm{sat}}^\mathrm{int} \bm{\Lambda}$. $\bm{\Lambda}$ is diagonal matrix responsible for changing the internal coefficients with ionospheric origins observed at satellite altitudes into the external coefficients with ionospheric origins observed on the ground, as described by Eq. (\ref{eqn:gaussc-ionos-link-sheet}). In the end, the sub-matrix $\mathbf{G}_i$ takes the form
\begin{equation}\label{eqn:Gk-divided}
    \mathbf{G}_i = \begin{pmatrix}
        \mathbf{S}_i' \mathbf{G}_{0,\mathrm{obs}}^{\mathrm{int}} & \mathbf{S}_i' \mathbf{G}_{0,\mathrm{obs}}^{\mathrm{ext}} & \mathbf{S}_i' \mathbf{G}_{0,\mathrm{obs}}^{\mathrm{ext}} \\ 
        \mathbf{G}_{k,\mathrm{sat}}^{\mathrm{int}} & \mathbf{G}_{k,\mathrm{sat}}^{\mathrm{int}} \bm{\Lambda} & \mathbf{G}_{k,\mathrm{sat}}^{\mathrm{ext}}
    \end{pmatrix}.
\end{equation}

The mathematical formulation of this proof follows the same line as that in Appendix \ref{sec:app-bias-obs}. The dimension of the nullspace of the original linear system can still be identified with the that of the matrix $\sum_i \mathbf{S}_i^\top \mathbf{P}_{\mathbf{G}_i}^\perp \mathbf{S}_i$ (Eq. \ref{eqn:nullity-intermediate}). The underlying assumption here is that the field coefficients can still be determined uniquely in a least squares sense in the absence of observatory bias, i.e. $\mathbf{G}_i^\top \mathbf{G}_i$ invertible. 
A corollary is that $\mathbf{G}_{0,\mathrm{obs}}^\mathrm{int}$ and $\mathbf{G}_{0,\mathrm{obs}}^\mathrm{ext}$ must both be full column rank, and their column spaces should be mutually linearly independent. Therefore, we can introduce a full column rank matrix $\mathbf{G}' = \mathbf{G}_{0,\mathrm{obs}}^\mathrm{int} \bm{\Lambda} - \mathbf{G}_{0,\mathrm{obs}}^\mathrm{ext}$. The action of the sampling matrix $\mathbf{S}_i$ on $\mathbf{G}'$ lies within the image of $\mathbf{G}_i$, as
\begin{equation}
    \mathbf{S}_i \mathbf{G}' = \begin{pmatrix} \mathbf{S}_i' \\ \mathbf{0} \end{pmatrix} ( \mathbf{G}_{0,\mathrm{obs}}^\mathrm{int} \bm{\Lambda} - \mathbf{G}_{0,\mathrm{obs}}^\mathrm{ext}) = \begin{pmatrix} \mathbf{S}_i' \mathbf{G}_{0,\mathrm{obs}}^\mathrm{int} \bm{\Lambda} - \mathbf{S}'_i \mathbf{G}_{0,\mathrm{obs}}^\mathrm{ext} \\ \mathbf{0} \end{pmatrix} 
    = \mathbf{G}_i \begin{pmatrix} \bm{\Lambda} \\ - \mathbf{I} \\ \mathbf{0} \end{pmatrix}
\end{equation}
where we used Eq. (\ref{eqn:Gk-divided}) in the last step. As a result, the image of $\mathbf{G}'$ lies in the nullspace of $\sum_i \mathbf{S}_i^\top \mathbf{P}_{\mathbf{G}_i}^\perp \mathbf{S}_i$,
\begin{equation}
    \left(\sum_{i=1}^{N_t} \mathbf{S}_i^\top \mathbf{P}_{\mathbf{G}_i}^\perp \mathbf{S}_i\right) \mathbf{G}' = \sum_{i=1}^{N_t} \mathbf{S}_i^\top \mathbf{P}_{\mathbf{G}_i}^\perp \mathbf{G}_i \begin{pmatrix} \bm{\Lambda} \\ - \mathbf{I} \\ \mathbf{0} \end{pmatrix} = \mathbf{0}.
\end{equation}
The nullity of the full system can be calculated via
\begin{equation}
    \mathrm{dim} \, \mathrm{ker}\left(\widetilde{\mathbf{G}}^\top \widetilde{\mathbf{G}}\right) = \mathrm{dim}\, \mathrm{ker} \left(\sum_{i=1}^{N_t} \mathbf{S}_i^\top \mathbf{P}_{\mathbf{G}_i}^\perp \mathbf{S}_i\right) \geq \mathrm{rank}(\mathbf{G}') = \frac{1}{3} M_0
\end{equation}
which concludes our proof that the problem has no unique least squares solution.

From a physical point of view, when the complete model is augmented with observatory bias, an arbitrary time-invariant field can be embedded in the induced field, which can always be compensated by the ionospheric field and the observatory bias. If a set of field coefficients solves the original least squares problem, we can add an arbitrary vector $\delta \mathbf{x}^\mathrm{int}$ to the induced coefficients in every time bin. An additional ionospheric field vector $\delta \mathbf{x}^\mathrm{ion} = - \bm{\Lambda}^{-1} \delta \mathbf{x}^\mathrm{int}$ can be added so that they cancel out in the satellite data. The residual in the observatory data can be compensated by adjusting the observatory biases accordingly. The adjusted model vectors will provide the exact same data fit as the original model vectors, and hence the solution is non-unique.

\section{Solving block-diagonal linear system with intercept}\label{sec:app-bias-diag}

In spherical harmonic analysis of a piece-wise constant geomagnetic field given vector magnetic field measurements, the least squares problem that arises takes the form
\begin{equation}
    \begin{pmatrix}
        \mathbf{G}_1 & & \mathbf{0} \\
        & \ddots & \\
        \mathbf{0} & & \mathbf{G}_{N_t} 
    \end{pmatrix}
    \begin{pmatrix} \mathbf{x}_1 \\ \vdots \\ \mathbf{x}_{N_t} \end{pmatrix} \simeq \begin{pmatrix} \mathbf{d}_1 \\ \vdots \\ \mathbf{d}_{N_t} \end{pmatrix}
\end{equation}
which we denote using the shorthand notation $\mathbf{G} \mathbf{x} \simeq \mathbf{d}$. Since the matrix $\mathbf{G}$ is block diagonal, the field coefficients for different time bins naturally decouple, and the least squares problem can be solved for each time bin separately. The computational and memory cost is therefore bounded by the summation of the costs for solving sub-systems $\mathbf{G}_i \mathbf{x}_i \simeq \mathbf{d}_i$, without involving the action of the full matrix $\mathbf{G}$.

In our estimation of the observatory bias, the least squares problem that arises takes the form
\begin{equation}
    \begin{pmatrix} \mathbf{G} & \mathbf{S} \end{pmatrix} \begin{pmatrix} \mathbf{x} \\ \mathbf{b} \end{pmatrix} = 
    \begin{pmatrix} 
    \mathbf{G}_1 & & \mathbf{0} & \mathbf{S}_1 \\
    & \ddots & & \vdots \\
    \mathbf{0} & & \mathbf{G}_{N_t} & \mathbf{S}_{N_t} \\
    \end{pmatrix}
    \begin{pmatrix} \mathbf{x}_1 \\ \vdots \\ \mathbf{x}_{N_t} \\ \mathbf{b} \end{pmatrix} 
    \simeq \begin{pmatrix} \mathbf{d}_1 \\ \vdots \\ \mathbf{d}_{N_t} \end{pmatrix}.
\end{equation}
This linear system has been described in Appendix \ref{sec:app-bias-obs}, and is denoted $\widetilde{\mathbf{G}} \widetilde{\mathbf{x}} \simeq \mathbf{d}$. Although the left part of the matrix is still block diagonal, the right sub-matrix couples all the sub-systems together by sampling the contribution from the observatory biases in every sub-system, and prohibits solving each sub-system separately.
Naive implementation might result in solving the full system, incurring large computational and memory costs.

We present here an algorithm that reduces the full system into solving a series of smaller sub-systems by virtue of matrix partitioning. The algorithm consists of two steps. In the first step, the bias $\mathbf{b}$ is estimated. After removing the bias contribution from the data, the $\mathbf{x}_i$ vectors can then be estimated using the block-diagonal linear least squares solver in the second step. 
To begin with, solving the aforementioned linear least squares is equivalent mathematically to solving the normal equation
\begin{equation}
    \widetilde{\mathbf{G}}^\top \widetilde{\mathbf{G}} \widetilde{\mathbf{x}} = 
    \begin{pmatrix}
        \mathbf{G}^\top \mathbf{G} & \mathbf{G}^\top \mathbf{S} \\ 
        \mathbf{S}^\top \mathbf{G} & \mathbf{S}^\top \mathbf{S}
    \end{pmatrix} \begin{pmatrix} \mathbf{x} \\ \mathbf{b} \end{pmatrix}
    = \begin{pmatrix} \mathbf{G}^\top \mathbf{d} \\ \mathbf{S}^\top \mathbf{d} \end{pmatrix} = \widetilde{\mathbf{G}}^\top \mathbf{d}.
\end{equation}
The solution can be formally written as $\widetilde{\mathbf{x}} = (\widetilde{\mathbf{G}}^\top \widetilde{\mathbf{G}})^{-1} \widetilde{\mathbf{G}}^\top \mathbf{d}$, where we assume that the matrix $\widetilde{\mathbf{G}}^\top \widetilde{\mathbf{G}}$ is invertible. Assuming $\mathbf{G}^\top \mathbf{G}$ is also invertible, we can use the first row block to express $\mathbf{x}$ as 
\begin{equation}
    \mathbf{x} = (\mathbf{G}^\top \mathbf{G})^{-1} \mathbf{G}^\top (\mathbf{d} - \mathbf{S} \mathbf{b}) = \mathbf{G}^+ (\mathbf{d} - \mathbf{S} \mathbf{b})
\end{equation}
which, when inserted into the second row block, yields
\begin{equation}
\begin{gathered}
    \left(\mathbf{S}^\top\mathbf{S} - \mathbf{S}^\top \mathbf{G} \mathbf{G}^+ \mathbf{S}\right)\mathbf{b} = \mathbf{S}^\top \mathbf{d} - \mathbf{S}^\top \mathbf{G} \mathbf{G}^+ \mathbf{d} \\ 
    \left(\mathbf{S}^\top \mathbf{P}_{\mathbf{G}}^\perp \mathbf{S}\right) \mathbf{b} = \mathbf{S}^\top \mathbf{P}_{\mathbf{G}}^\perp \mathbf{d}.
\end{gathered}
\end{equation}
It can be shown using Guttman rank additivity theorem that if $\widetilde{\mathbf{G}}^\top \widetilde{\mathbf{G}}$ and $\mathbf{G}^\top \mathbf{G}$ are both invertible, then $\mathbf{S}^\top \mathbf{P}_{\mathbf{G}}^\perp \mathbf{S}$ must be invertible as well. Therefore, we arrive at the solution of $\mathbf{b}$
\begin{equation}
    \mathbf{b} = \left(\mathbf{S}^\top \mathbf{P}_{\mathbf{G}}^\perp \mathbf{S}\right)^{-1} \mathbf{S}^\top \mathbf{P}_{\mathbf{G}}^\perp \mathbf{d}.
\end{equation}
Thanks to the block-diagonality of the $\mathbf{G}$ matrix, the matrix-matrix and matrix-vector multiplications in the equation above can be drastically simplified. The seemingly high-dimensional multiplications turn out to be merely a summation of low-dimensional products; the solution of $\mathbf{b}$ therefore reads
\begin{equation}
    \mathbf{b} = \left(\sum_{i=1}^{N_t} \mathbf{S}_i^\top \mathbf{P}_{\mathbf{G}_i}^\perp \mathbf{S}_i\right)^{-1} \sum_{i=1}^{N_t} \mathbf{S}_i^\top \mathbf{P}_{\mathbf{G}_i}^\perp \mathbf{d}_i.
\end{equation}
The bias estimation problem is reduced to collecting the matrix-matrix and matrix-vector products in each sub-system, and then performing one low-dimensional linear solve. This completes the first step.

In the second step, the contribution from $\mathbf{b}$ is removed from each sub-system. The remaining sub-systems can be solved separately as they are now decoupled. 

This algorithm does not require the action or the decomposition of the full matrix, but merely requires matrix multiplications of low-dimensional sub-matrices. As such, the algorithm provides a computation- and memory-efficient way to solve such block-diagonal systems with bias contribution.

\end{document}


\maketitle

\renewcommand\thefigure{S\arabic{figure}}
\renewcommand\thetable{S\arabic{table}}

\begin{figure}
    \centering
    \includegraphics[width=\linewidth]{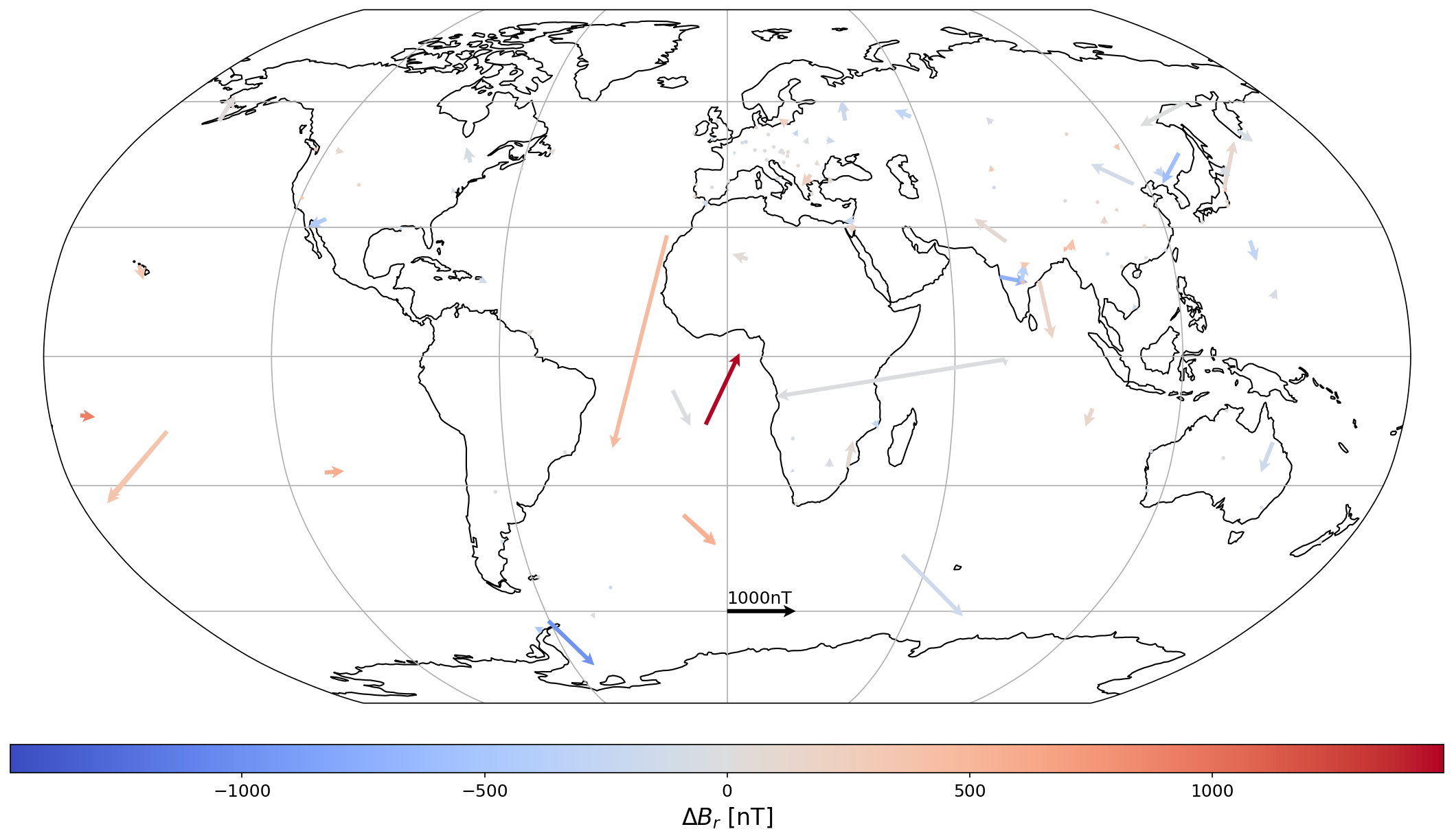}
    \caption{Geographical distribution of the estimated observatory vector bias. The vectors on the map show the horizontal components of the bias, while the radial components are represented by the colour. For the scale, a black arrow at ($60^\circ$ south, $0^\circ$) shows an east bias of $1000$nT.}
    \label{fig:obs-bias-est-map}
\end{figure}

\begin{figure}
    \centering
    \includegraphics[width=\linewidth]{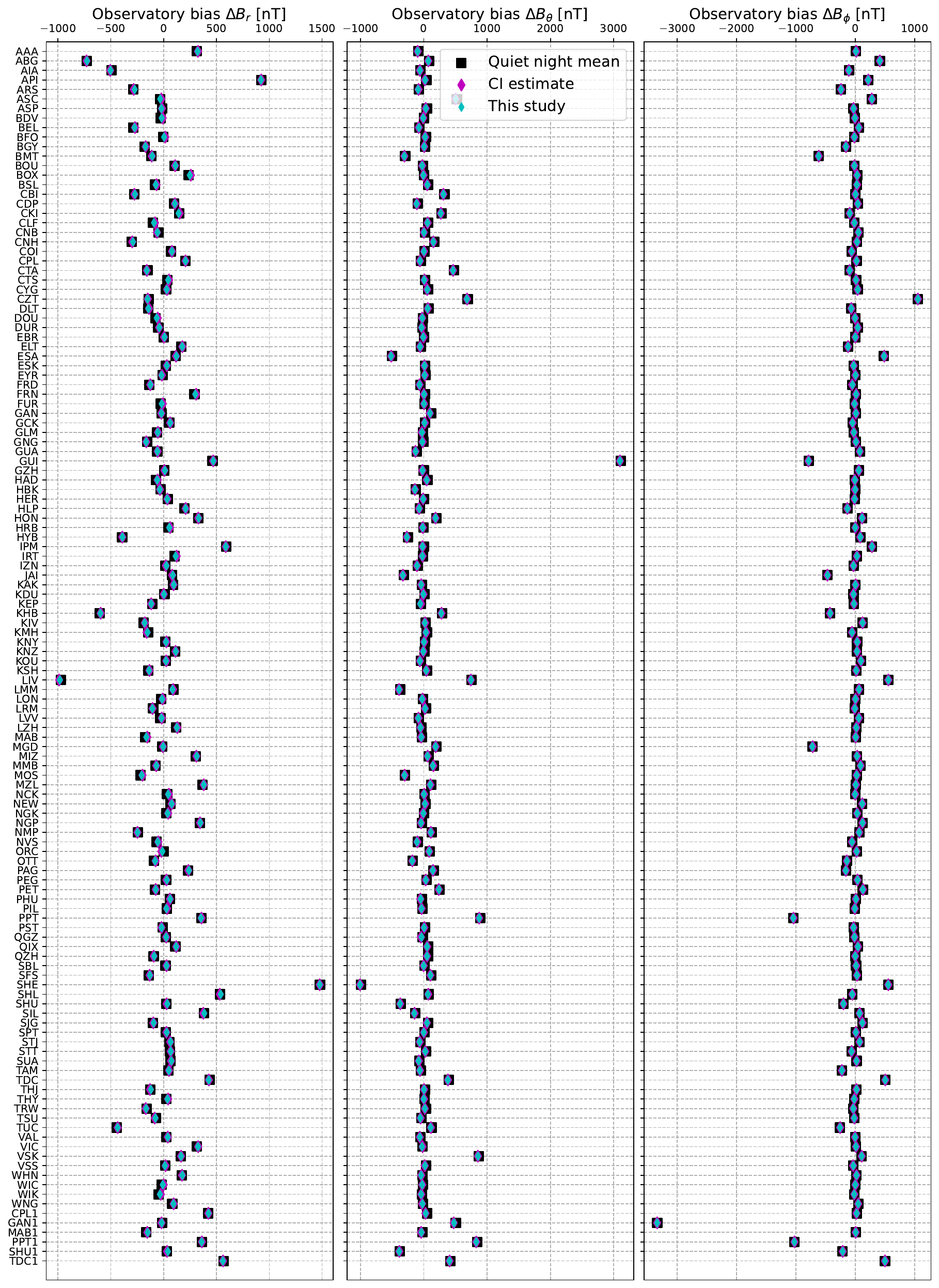}
    \caption{Estimated observatory vector bias (cyan), the quiet night mean (black) and CI estimates (magenta).}
    \label{fig:obs-bias-est}
\end{figure}

\begin{figure}
    \centering
    \includegraphics[width=\linewidth]{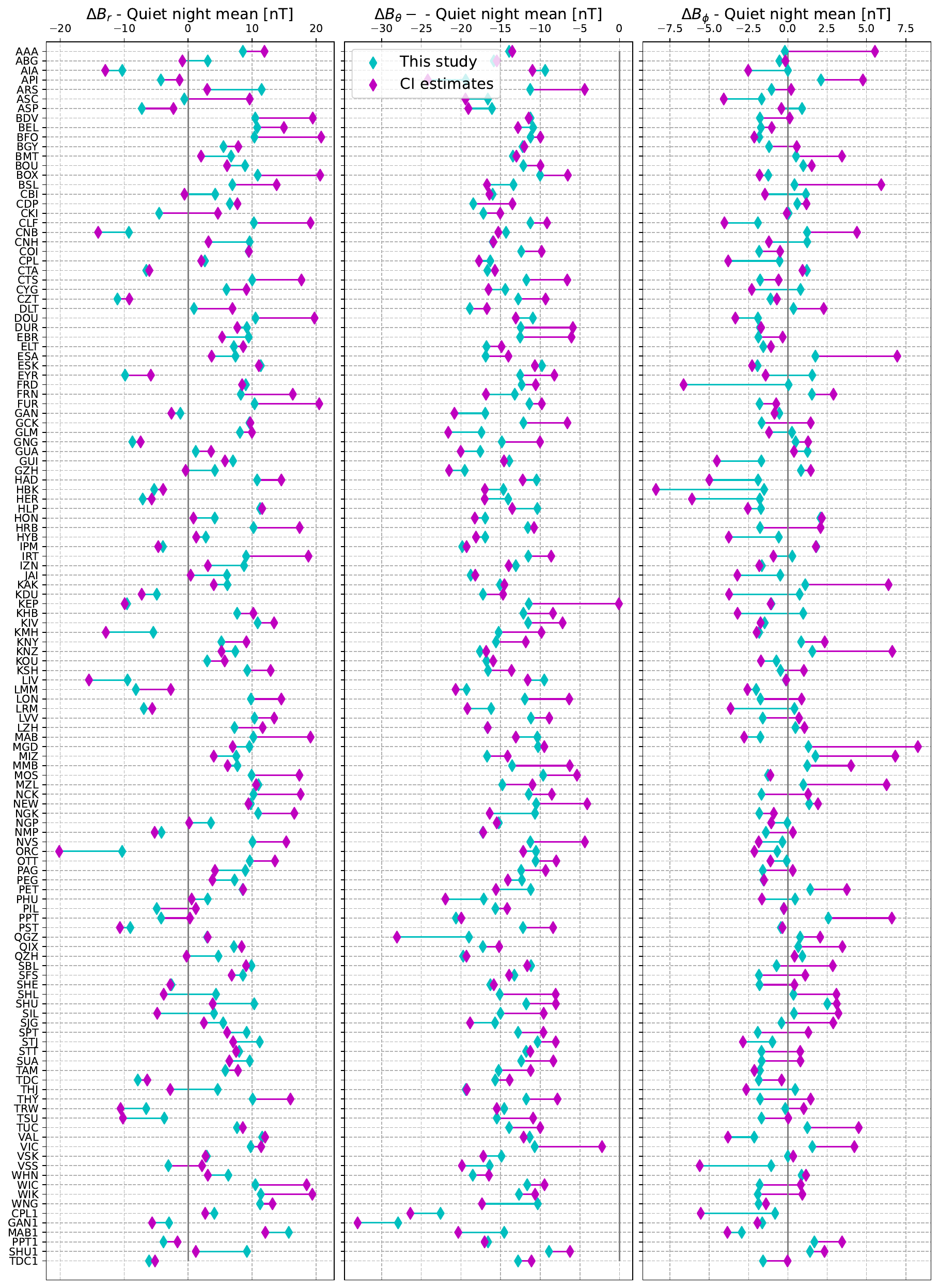}
    \caption{Difference between the estimated observatory bias and the quiet night mean.}
    \label{fig:obs-bias-est-diffmean}
\end{figure}

\begin{table}
    \centering
    \caption{Root-mean-square differences (RMSD) between bias values from different datasets.}
    \label{tab:bias-rmsd}
    \begin{tabular}{lrrr}
    \toprule
    Bias sets & RMSD $\Delta B_r$ [nT] & RMSD $\Delta B_\theta$ [nT] & RMSD $\Delta B_\phi$ [nT] \\
    \midrule
    This study - Quiet night mean & 8.07 & 14.55 & 1.38 \\
    CI bias - Quiet night mean & 10.16 & 14.35 & 3.05 \\
    This study - CI bias & 4.83 & 3.54 & 2.60 \\
    \bottomrule
    \end{tabular}
\end{table}

\clearpage

\begin{figure}
    \centering
    \begin{subfigure}[t]{\textwidth}
        \centering
        \includegraphics[width=\linewidth]{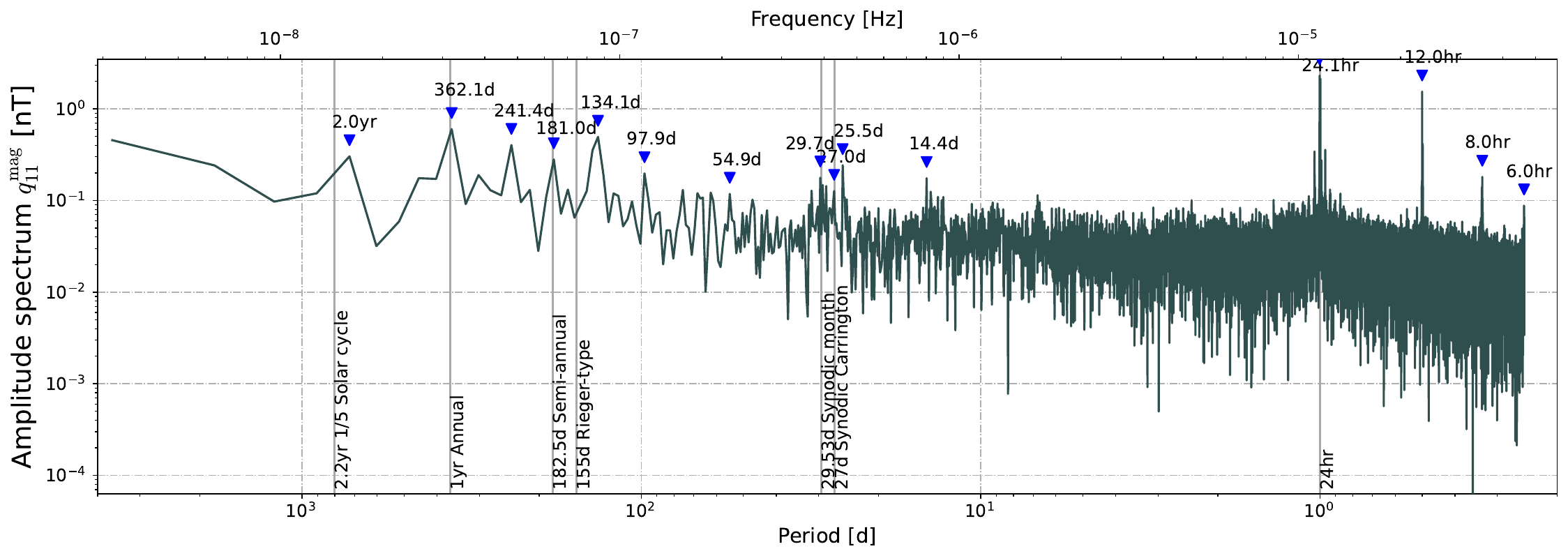}
    \end{subfigure}
    \begin{subfigure}[t]{\textwidth}
        \centering
        \includegraphics[width=\linewidth]{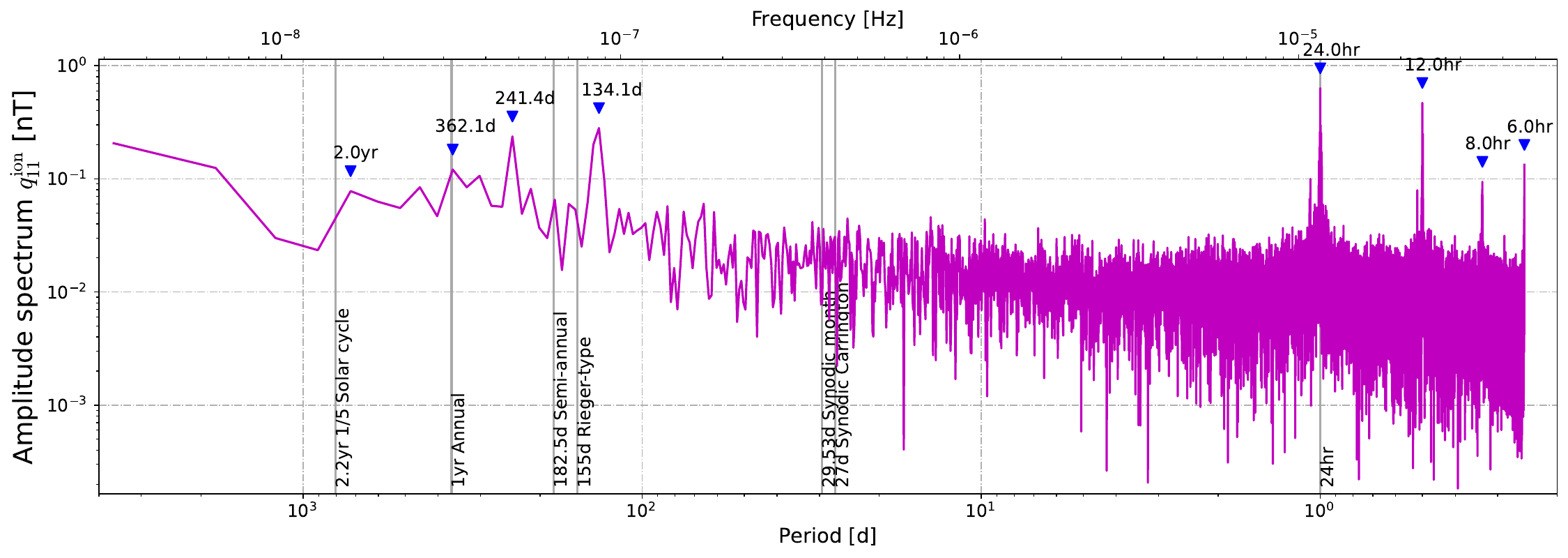}
    \end{subfigure}
    \begin{subfigure}[t]{\textwidth}
        \centering
        \includegraphics[width=\linewidth]{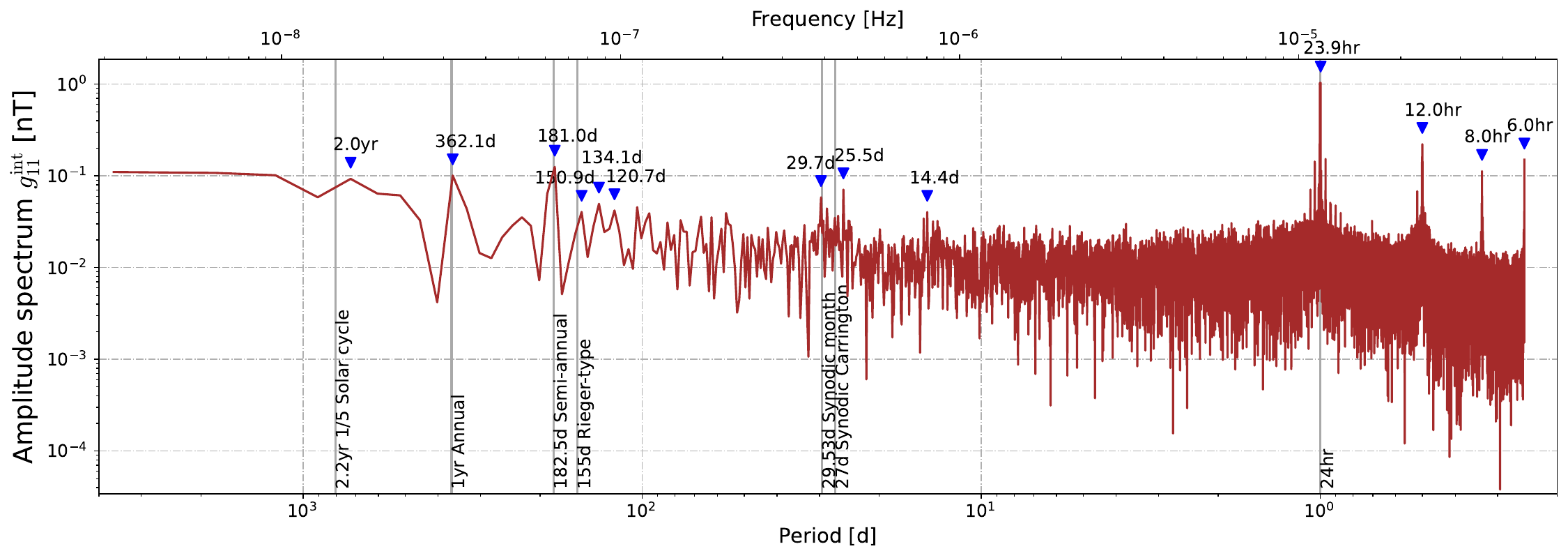}
    \end{subfigure}
    \caption{Amplitude spectra of the magnetospheric coefficient $q_{11}^\mathrm{mag}$ (upper panel), the ionospheric coefficient $q_{11}^\mathrm{ion}$ (middle panel) and the internally induced coefficient $g_{11}^\mathrm{int}$ (lower panel).}
    \label{fig:spec-q11-mi}
\end{figure}

\begin{figure}
    \centering
    \begin{subfigure}[t]{\textwidth}
        \centering
        \includegraphics[width=\linewidth]{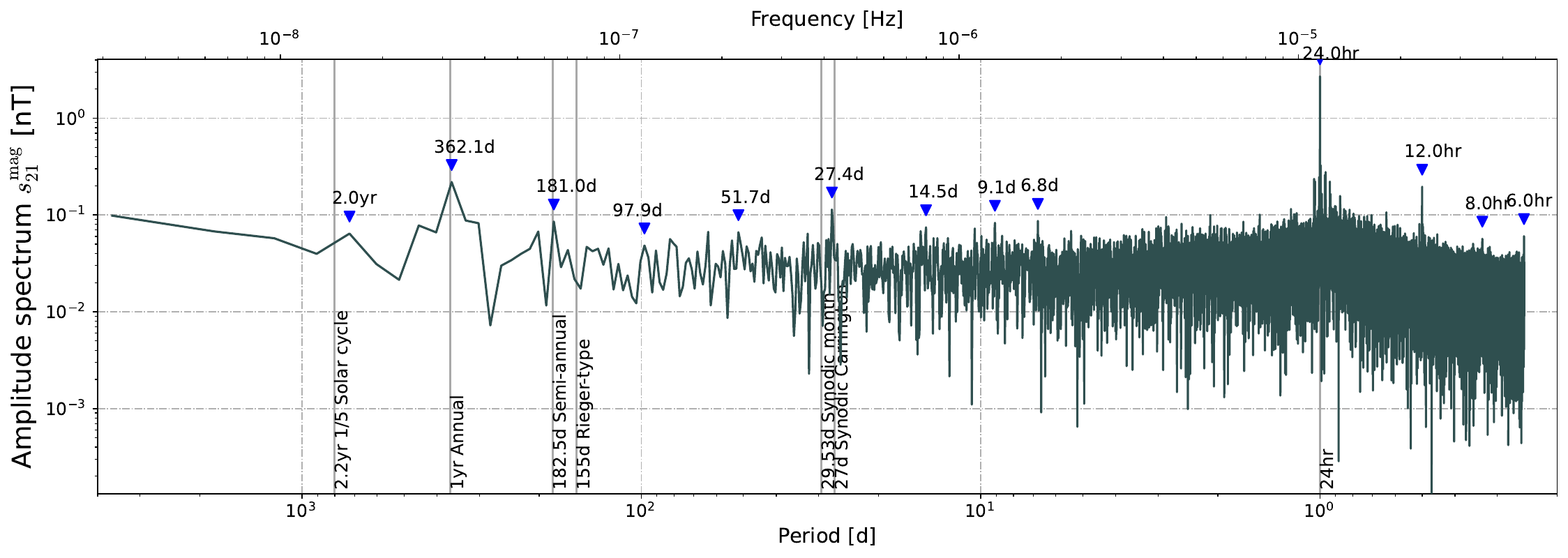}
    \end{subfigure}
    \begin{subfigure}[t]{\textwidth}
        \centering
        \includegraphics[width=\linewidth]{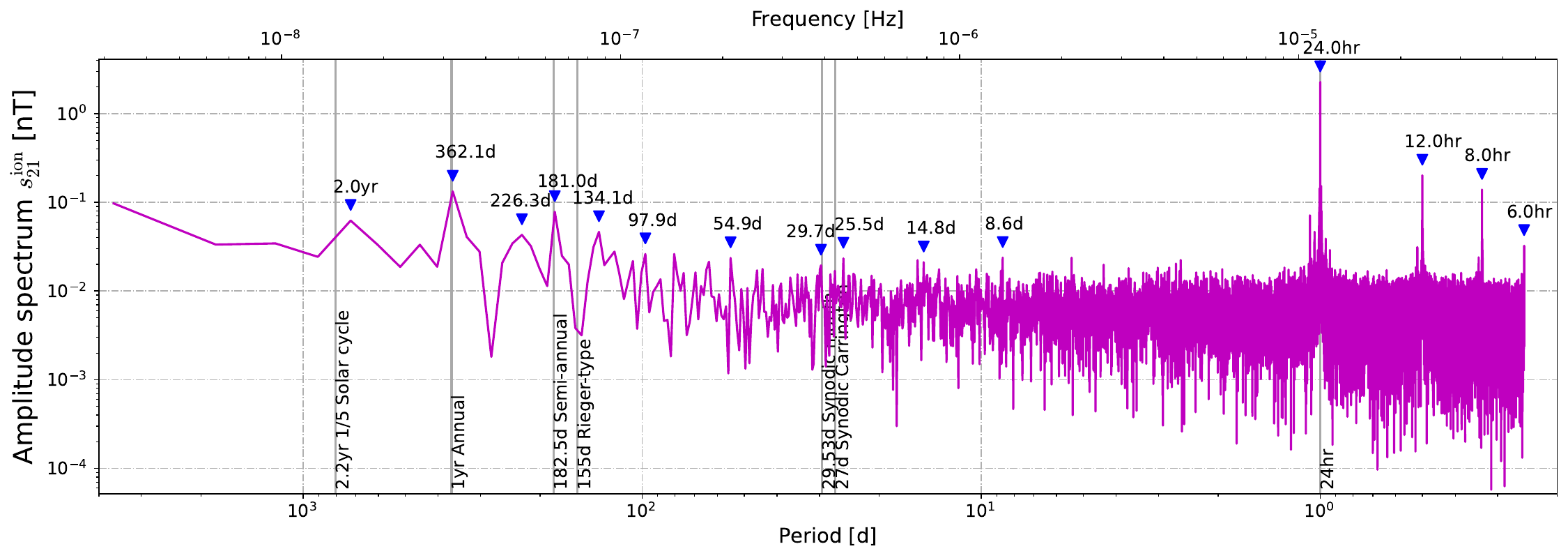}
    \end{subfigure}
    \begin{subfigure}[t]{\textwidth}
        \centering
        \includegraphics[width=\linewidth]{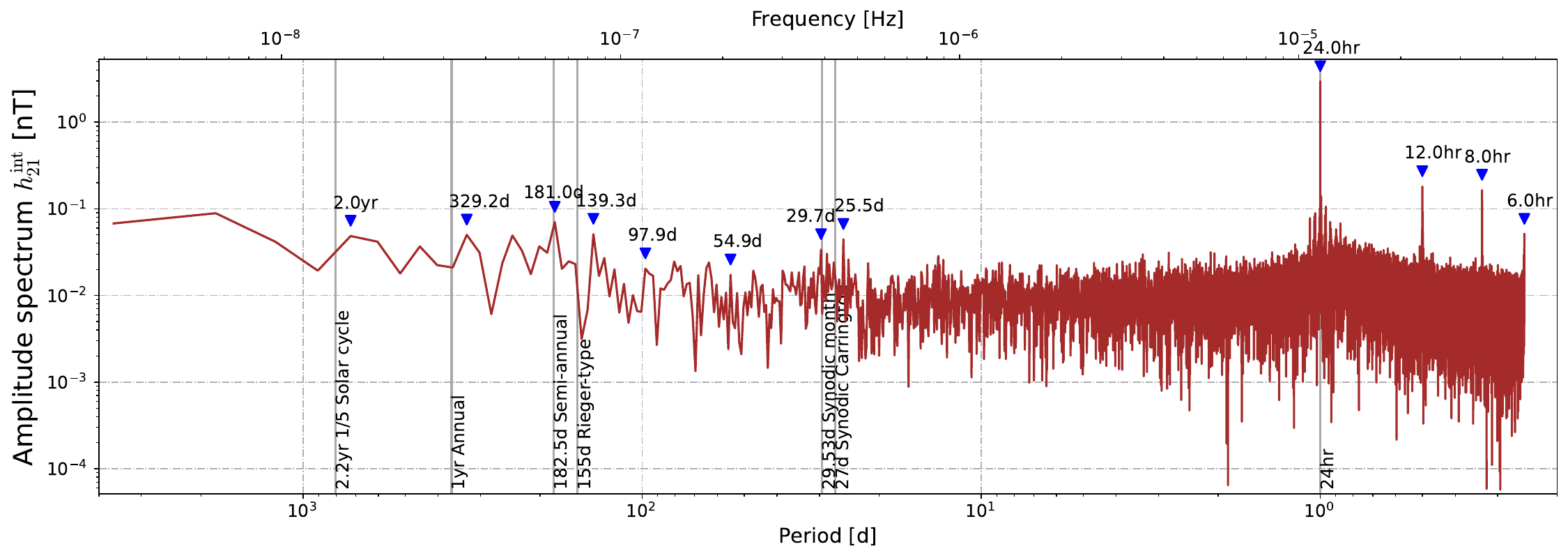}
    \end{subfigure}
    \caption{Amplitude spectra of the magnetospheric coefficient $s_{21}^\mathrm{mag}$ (upper panel), the ionospheric coefficient $s_{21}^\mathrm{ion}$ (middle panel) and the internally induced coefficient $h_{21}^\mathrm{int}$ (lower panel).}
    \label{fig:spec-s21-mi}
\end{figure}

\begin{figure}
    \centering
    \begin{subfigure}[t]{\textwidth}
        \centering
        \includegraphics[width=\linewidth]{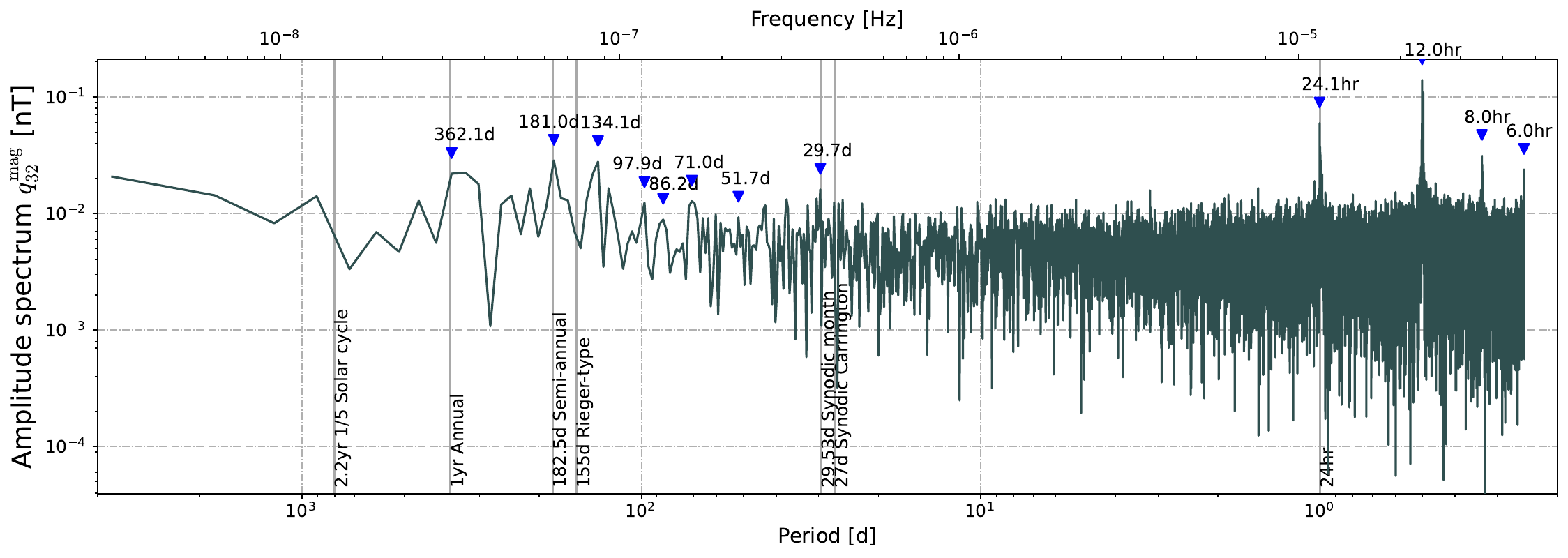}
    \end{subfigure}
    \begin{subfigure}[t]{\textwidth}
        \centering
        \includegraphics[width=\linewidth]{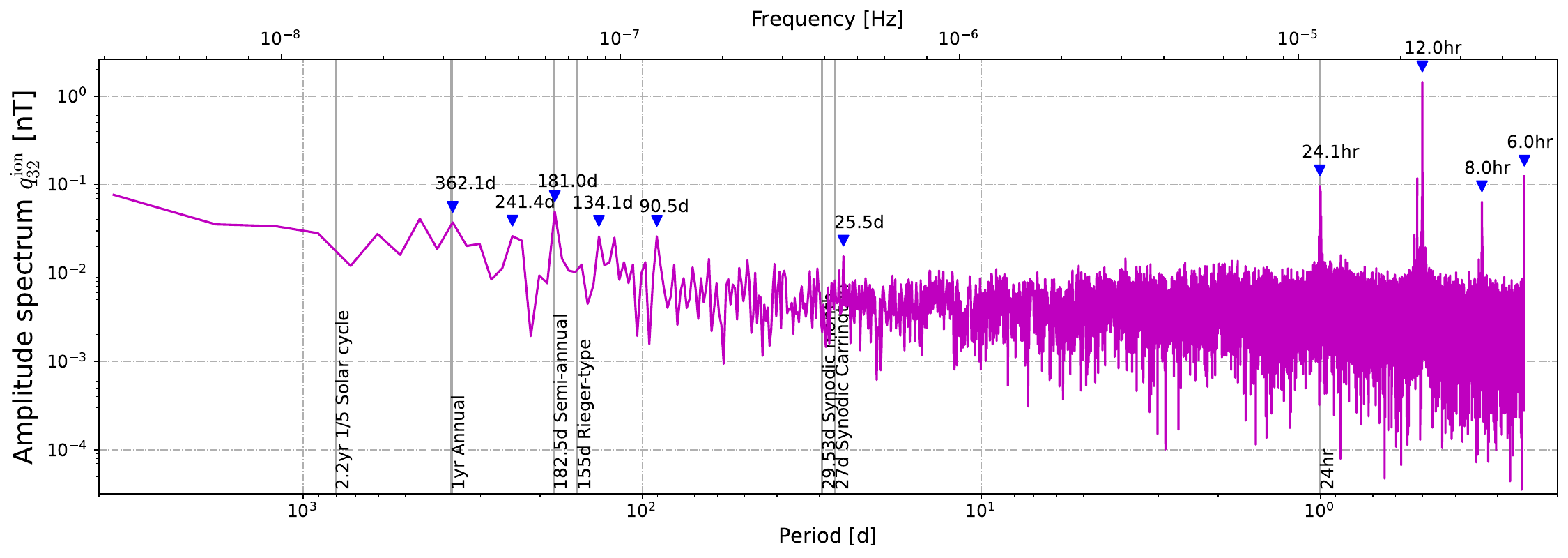}
    \end{subfigure}
    \begin{subfigure}[t]{\textwidth}
        \centering
        \includegraphics[width=\linewidth]{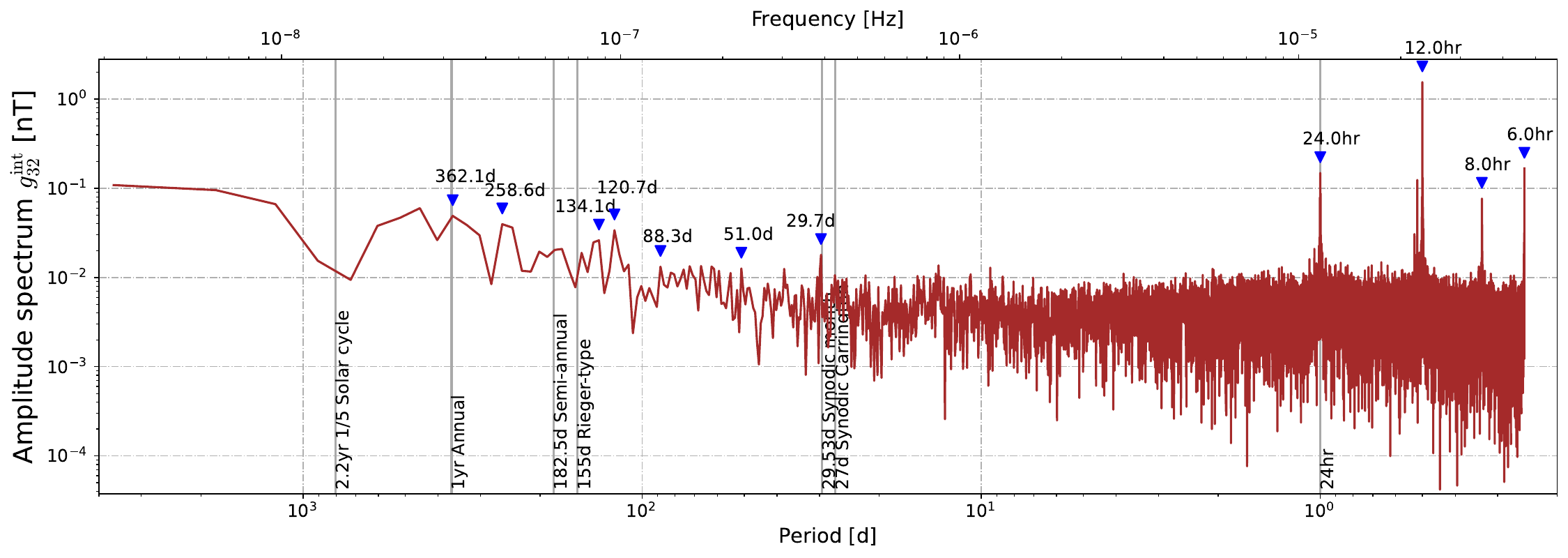}
    \end{subfigure}
    \caption{Amplitude spectra of the magnetospheric coefficient $q_{32}^\mathrm{mag}$ (upper panel), the ionospheric coefficient $q_{32}^\mathrm{ion}$ (middle panel) and the internally induced coefficient $g_{32}^\mathrm{int}$ (lower panel).}
    \label{fig:spec-q32-mi}
\end{figure}

\begin{figure}
    \centering
    \begin{subfigure}[t]{\textwidth}
        \centering
        \includegraphics[width=\linewidth]{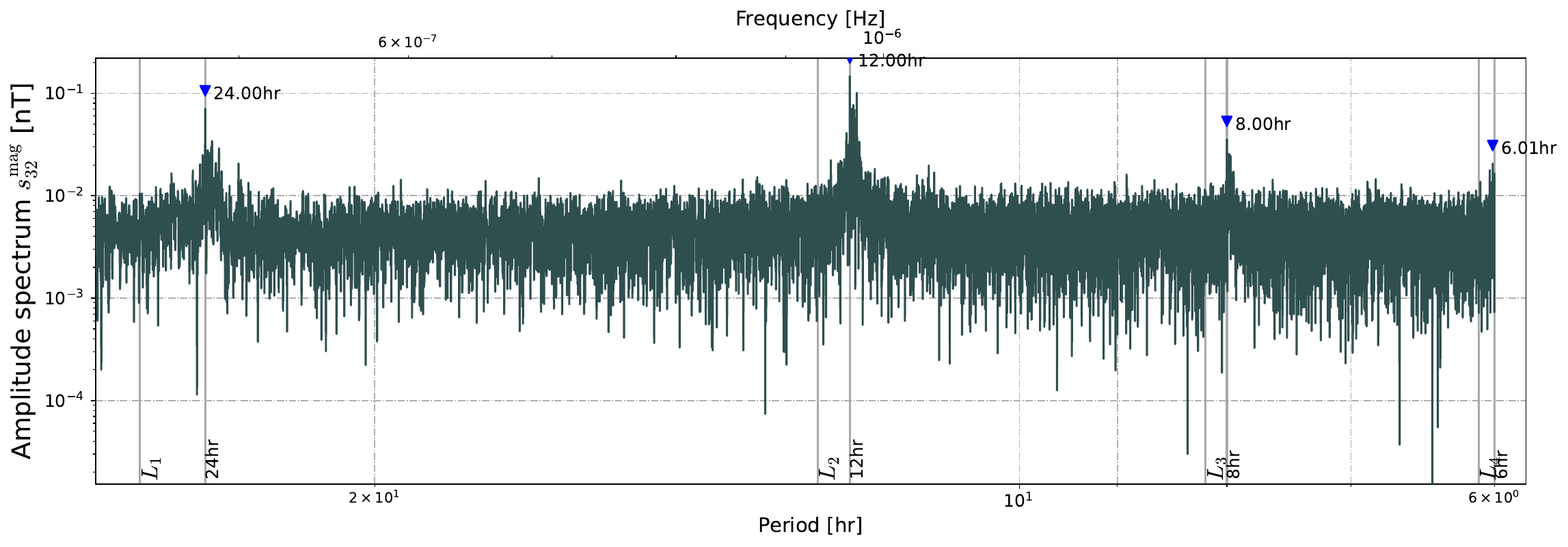}
    \end{subfigure}
    \begin{subfigure}[t]{\textwidth}
        \centering
        \includegraphics[width=\linewidth]{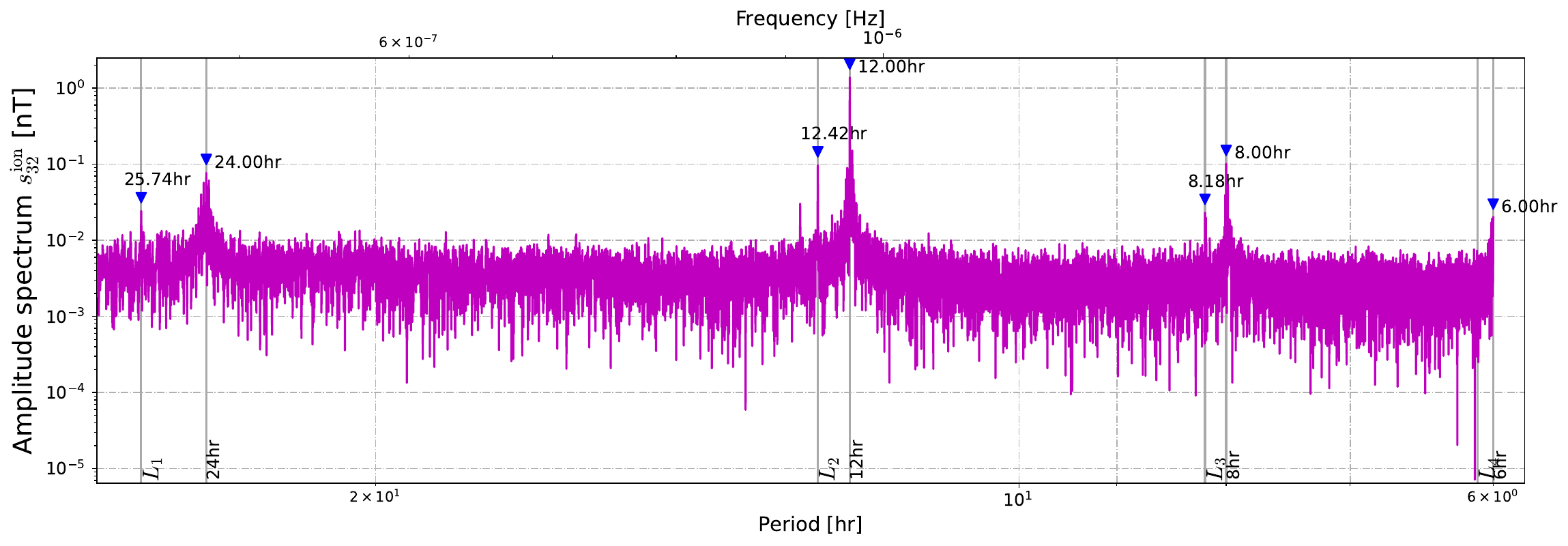}
    \end{subfigure}
    \begin{subfigure}[t]{\textwidth}
        \centering
        \includegraphics[width=\linewidth]{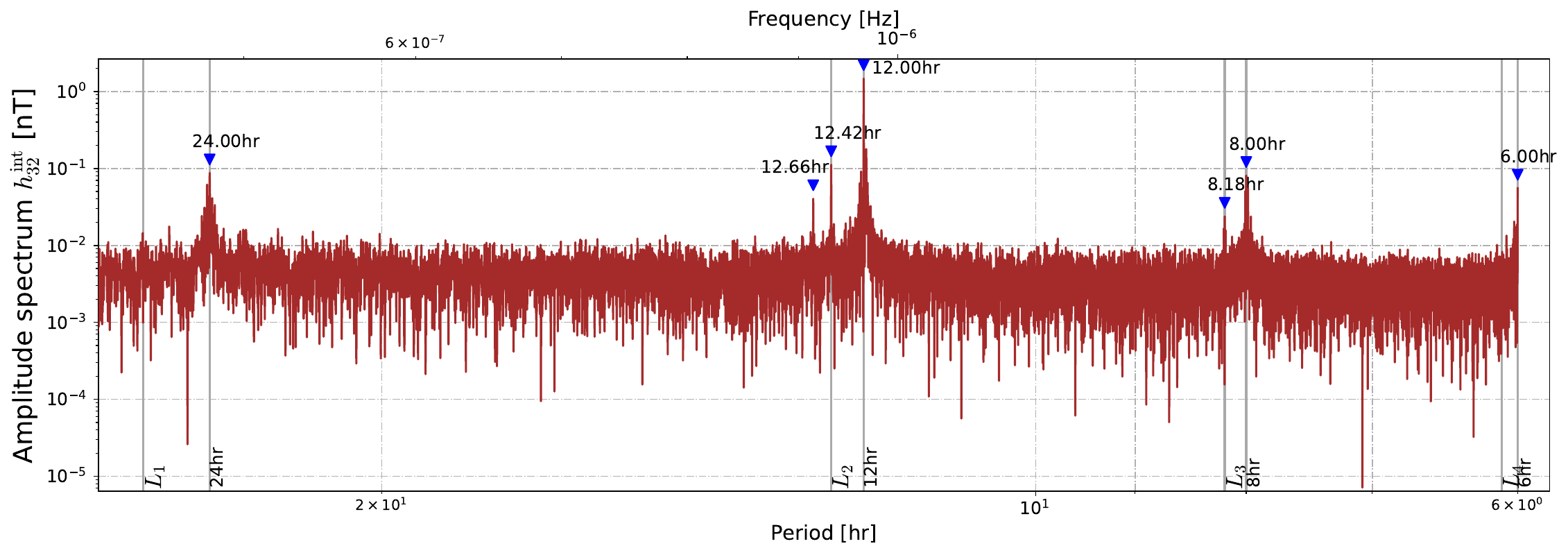}
    \end{subfigure}
    \caption{Amplitude spectra of the magnetospheric coefficient $s_{32}^\mathrm{mag}$ (upper panel), the ionospheric coefficient $s_{32}^\mathrm{ion}$ (middle panel) and the internally induced coefficient $h_{32}^\mathrm{int}$ (lower panel) around the diurnal band.}
    \label{fig:spec-s32-mi}
\end{figure}

\begin{figure}
    \centering
    \begin{subfigure}[t]{\textwidth}
        \centering
        \includegraphics[width=\linewidth]{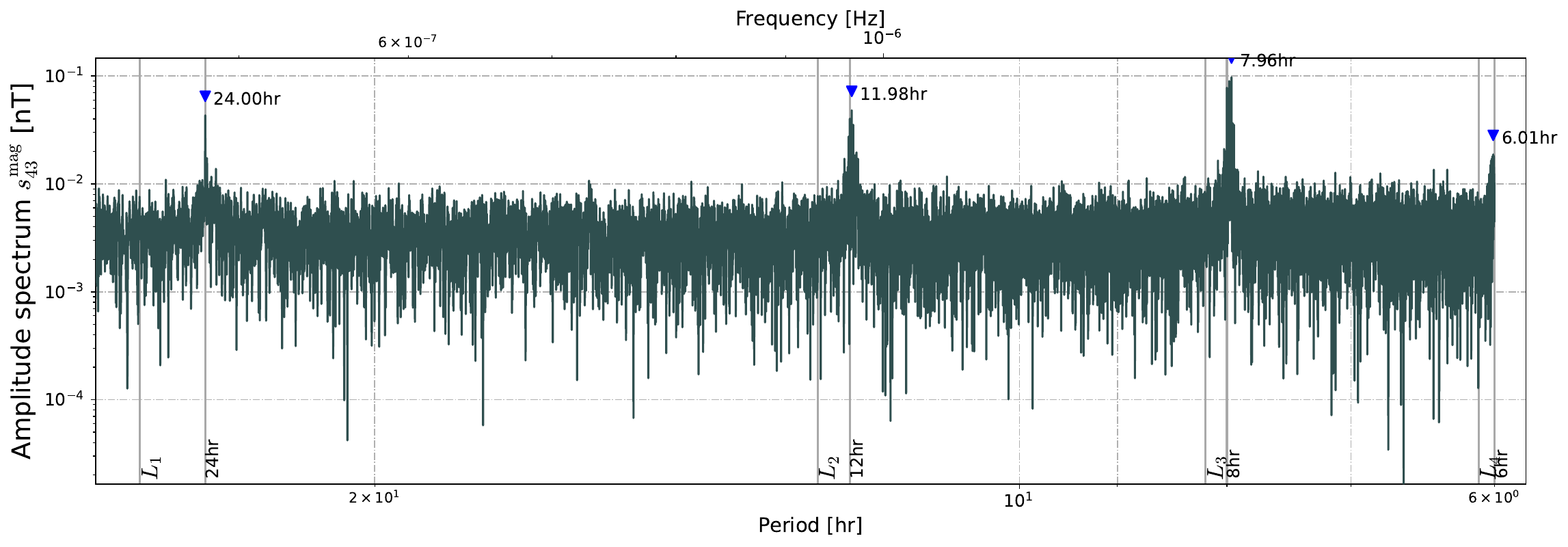}
    \end{subfigure}
    \begin{subfigure}[t]{\textwidth}
        \centering
        \includegraphics[width=\linewidth]{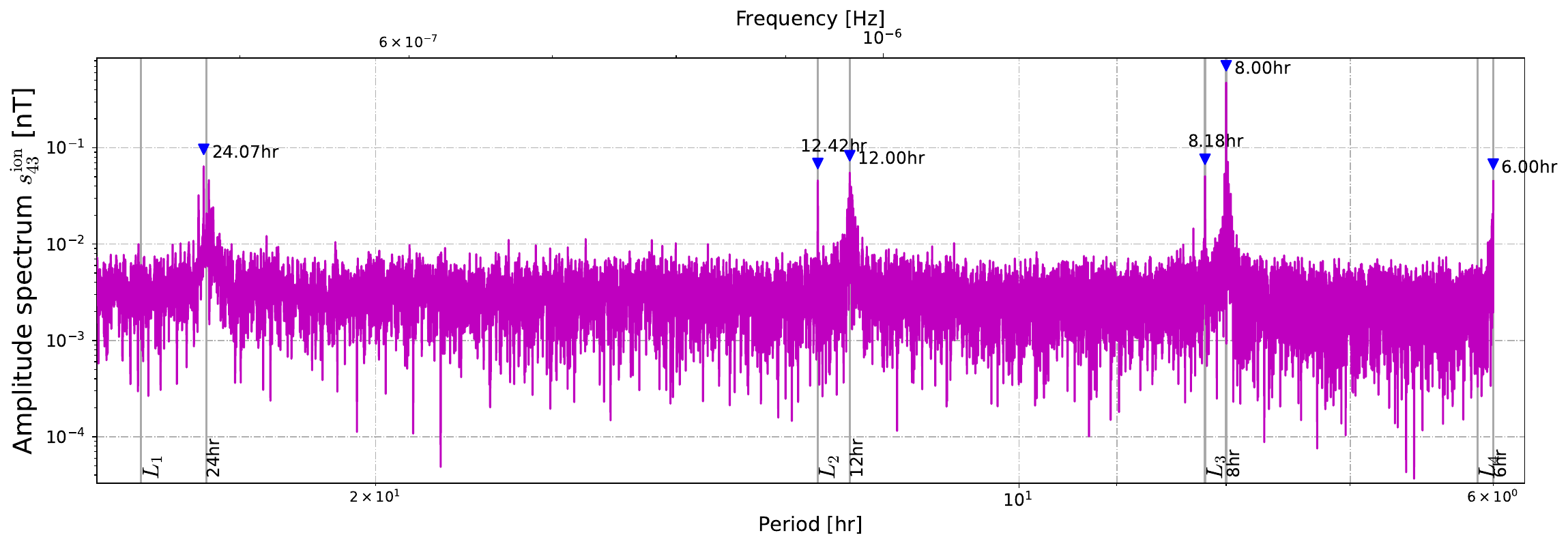}
    \end{subfigure}
    \begin{subfigure}[t]{\textwidth}
        \centering
        \includegraphics[width=\linewidth]{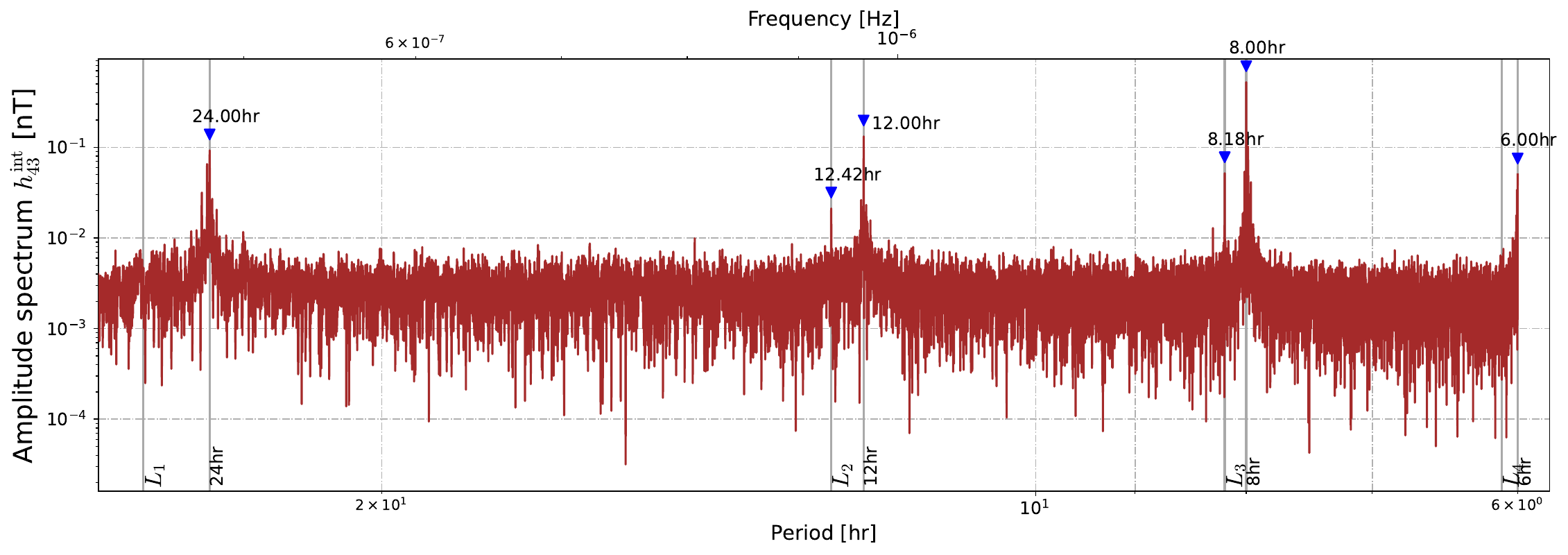}
    \end{subfigure}
    \caption{Amplitude spectra of the magnetospheric coefficient $s_{43}^\mathrm{mag}$ (upper panel), the ionospheric coefficient $s_{43}^\mathrm{ion}$ (middle panel) and the internally induced coefficient $h_{43}^\mathrm{int}$ (lower panel) around the diurnal band.}
    \label{fig:spec-s43-mi}
\end{figure}

\begin{figure}
    \centering
    \begin{subfigure}[t]{\textwidth}
        \centering
        \includegraphics[width=\linewidth]{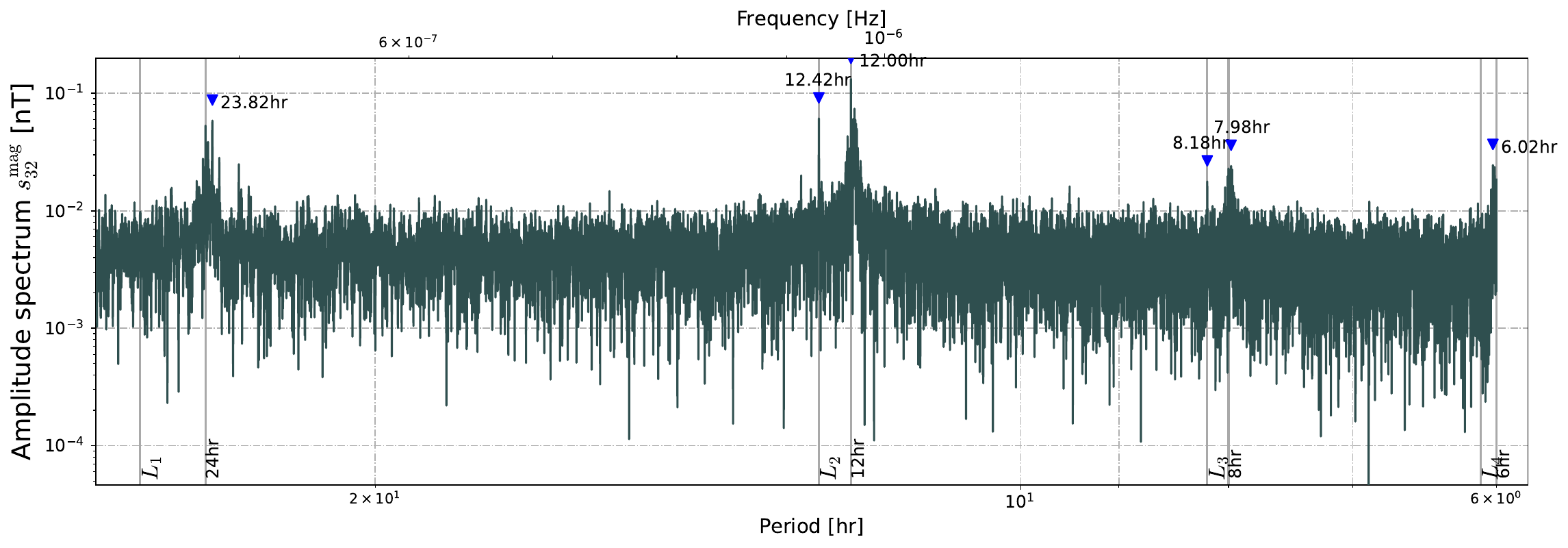}
    \end{subfigure}
    \begin{subfigure}[t]{\textwidth}
        \centering
        \includegraphics[width=\linewidth]{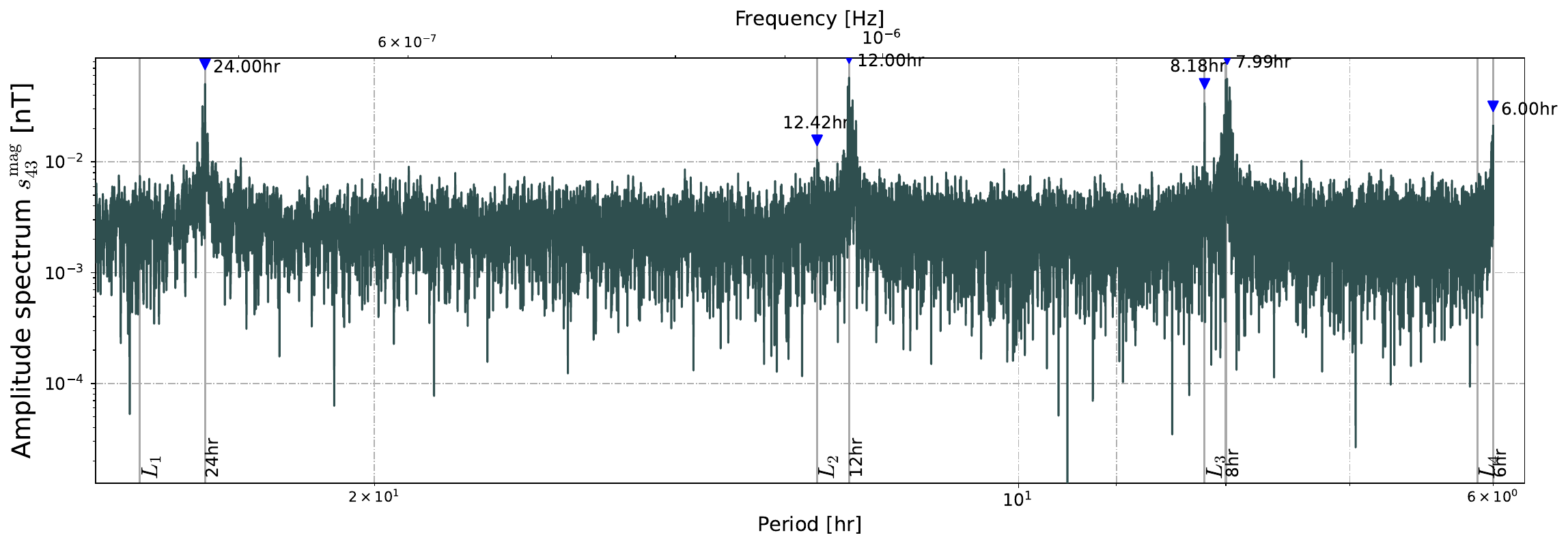}
    \end{subfigure}
    \caption{Amplitude spectra around the diurnal band of the magnetospheric coefficient $s_{32}^\mathrm{mag}$ (upper panel) and $s_{43}^\mathrm{mag}$ (lower panel) calculated from a dataset with CI ionospheric model (along with its induced counterpart) subtracted. The magnetospheric field spectra now contain peaks at lunar tidal periods, which were only observed in the ionospheric signal as $Lp$ but not in the magnetospheric signal in our model.}
    \label{fig:spec-ms4-MIOres}
\end{figure}

\begin{figure}
    \centering
    \includegraphics[width=\linewidth]{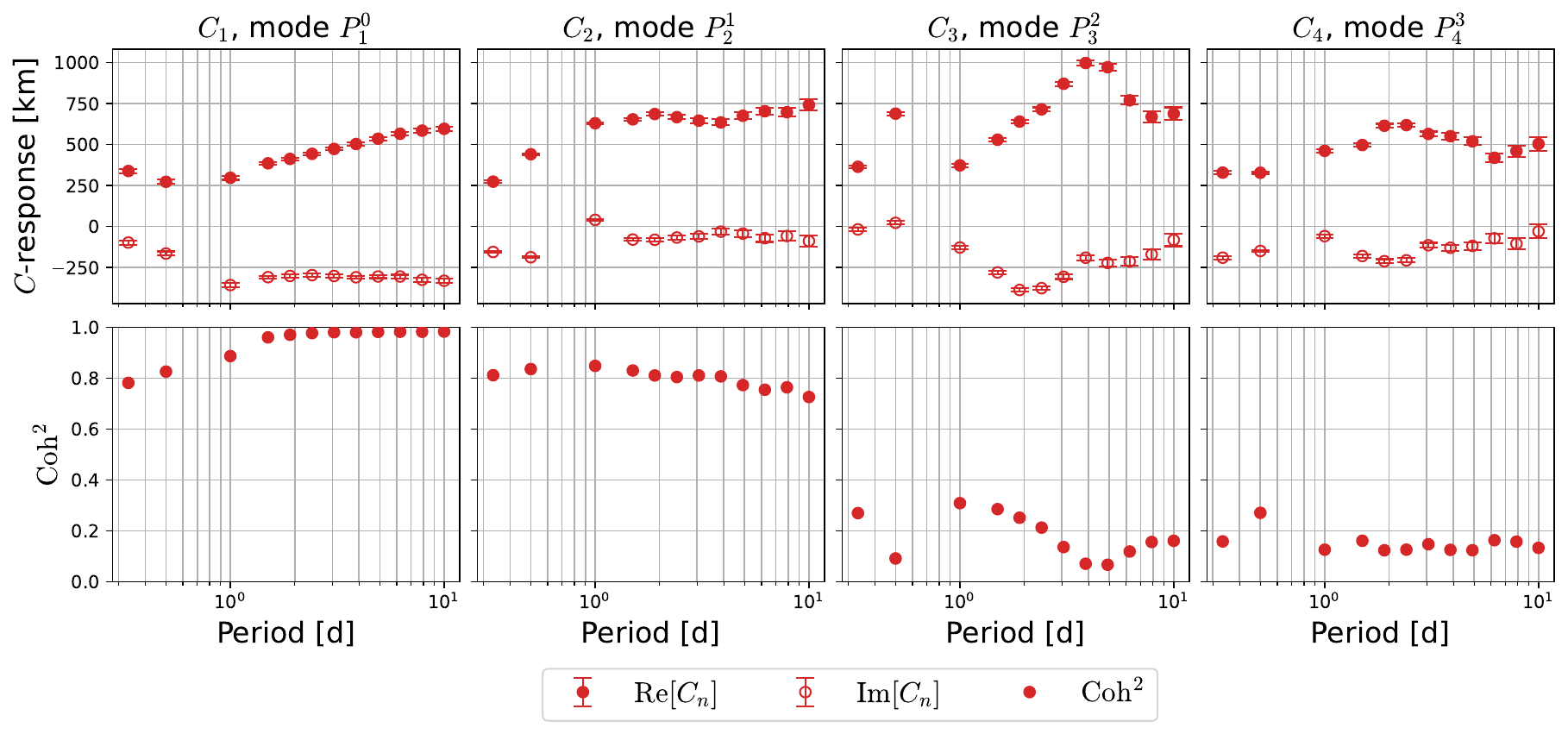}
    \caption{$C_n$-responses (upper panel) and their squared coherences (lower panel) at periods between $8$ hours and $10$ days, calculated from a dataset with CI ionospheric model (along with its induced counterpart) subtracted. The columns show the $C_1$, $C_2$, $C_3$ and $C_4$ responses estimated from the SH coefficients described by the $P_1^0$, $P_2^1$, $P_3^2$ and $P_4^3$ modes, respectively.}
    \label{fig:Cn-10d-MIOres}
\end{figure}